\documentclass[english,11pt,reqno]{article}

\usepackage{latexsym, amsmath, amssymb,amsthm, natbib,verbatim, xy}
\xyoption{all}

\usepackage[dvipsnames]{xcolor}
\usepackage{tgpagella}
\usepackage{mathpazo}
\usepackage{float}
\usepackage[labelfont=bf]{caption}
\usepackage[left=1in,top=1in,right=1in,bottom=1in]{geometry}
\usepackage{setspace,color}
\usepackage{graphicx}
\usepackage{enumerate}
\usepackage[multiple]{footmisc}
\usepackage{hyperref}
\usepackage{fancyhdr}
\usepackage{subfig}
\usepackage{rotating}
\usepackage{etoolbox}
\patchcmd{\section}{\scshape}{\bfseries}{}{}

\newcommand{\fig}[1]{{\color{black} #1}}

\newtheorem{theorem}{Theorem}
\newtheorem{corollary}{Corollary}
\newtheorem{definition}{Definition}
\newtheorem{lemma}{Lemma}
\newtheorem{proposition}{Proposition}

\theoremstyle{definition}

\newtheorem{example}{Example}

\def\cala{\mathcal{A}} 
  
\def\calx{\mathcal{X}}

\def\calc{\mathcal{C}} 
\def\cals{\mathcal{S}}

\def\calq{\mathcal{Q}}
\def\calx{\mathcal{X}} 
 
\def\calp{\mathcal{P}} 

\def\i{\mathsf{i}}
\def\b{\mathsf{b}}
\def\t{\mathsf{t}}

\renewenvironment{quote}{%
   \list{}{%
     \leftmargin1.5cm   
     \rightmargin\leftmargin
   }
   \item\relax
}
{\endlist}

\begin{document}

\title{Mechanism Design meets Priority Design:\\
Redesigning the US Army's Branching Process\thanks{All opinions expressed in this manuscript are those of the authors and do not represent the opinions of the United States Military Academy (USMA), United States Cadet Command, the United States Army, or the Department of Defense. We are grateful for excellent research assistance from Kate Bradley and Robert Upton. Eryn Heying provided superb help with research administration. The Army's Office of Economic and Manpower Analysis provided administrative branching data for
this project to Kyle Greenberg as part of a restricted use agreement with USMA and MIT that specifies that data can only be stored, accessed, and analyzed within USMA's information system.  Any parties interested in accessing this data must make a direct application to USMA.  We are grateful to Scott Kominers for helpful conversations.  Pathak
acknowledges support from the National Science Foundation for this project.}}
\author{Kyle Greenberg \and Parag A. Pathak  \and Tayfun S\"{o}nmez\thanks{Greenberg: Department of Social Sciences, United States Military Academy, email: kyle.greenberg@westpoint.edu.
Pathak: Department of Economics, MIT and NBER, email: ppathak@mit.edu, S\"{o}nmez: Department of Economics, Boston College, email: sonmezt@bc.edu.}}

\date{June 2021}

\maketitle

\begin{abstract}
Army cadets obtain occupations through a centralized process.  Three objectives – increasing retention, aligning talent, and enhancing trust – have guided reforms to this process since 2006.  West Point’s mechanism for the Class of 2020 exacerbated challenges implementing Army policy aims.   We formulate these desiderata as axioms and study their implications theoretically and with administrative data.   We show that the Army’s objectives not only determine an allocation mechanism, but also a specific priority policy, a uniqueness result that integrates mechanism and priority design. These results led to a re-design of the mechanism, now adopted at both West Point and ROTC.   \\[.5cm]
\end{abstract}
\thispagestyle{empty}

\newpage


\section{Introduction}
Each year, the US Army assigns thousands of graduating cadets from the United States Military Academy (USMA) at West Point and the Reserve Officer Training Corps (ROTC) to their first job in a military occupation, or branch, through centralized systems.  Combined, the West Point and ROTC branching systems determine the branch placements for 70 percent of newly commissioned Army officers \citep{dodpoprep2018}. In 2006, the US Army created a ``market-based'' system for branch assignments with the goal of increasing officer retention \citep{colarusso/lyle/wardynski:10}.  The system, known as the \textit{Branch-of-Choice} or \textit{BRADSO program}, gives cadets heightened priority for a fraction of a branch's positions  
if they express a willingness to BRADSO, or extend the length of their service commitment.\footnote{ADSO is short for Active Duty Service Obligation.  BRADSO stands for Branch of Choice Active Duty Service Obligation. BRADSO slots are 25\% of total branch allocations at USMA from the Class of 2006 through 2020 and 35\% for the Class of 2021, and either 50\% or 60\% of total branch allocations at ROTC depending on the graduating class. USMA and ROTC cadets receive branches through separate centralized branching systems.}  

Since the allocation problem involves both branch assignment and length of service commitment, the Army's branching system is a natural application of the matching with contracts framework developed by \cite{kelso/crawford:82} and \cite{hatfield/milgrom:05}.  
In that framework, a centralized mechanism assigns both positions and contractual terms.
However, the Army's mechanism, hereafter USMA-2006, was designed while the matching with contracts
model was still being developed and the original formulation in \cite{hatfield/milgrom:05} did not directly apply to the Army's problem.  
Subsequent research by \cite{hatfield/kojima:10} broadened the framework in a way that allows it to
apply to the Army's problem.\footnote{Further elaboration is provided by \cite{echenique:12}, \cite{schlegel:15}, and \cite{jagadeesan:19}.} Building on this research,
 \citet{sonmez/switzer:13} proposed that the Army use the cumulative offer mechanism to assign cadets to branches.  
While this proposal had desirable theoretical properties, it required a more complex strategy space  in which
cadets have to rank branches and terms jointly.  Under the USMA-2006 mechanism, 
cadets only rank branches and separately indicate their willingness to BRADSO for 
any branch.  The Army considered the existing strategy space more manageable than a more complex alternative. 
In addition, \cite{sonmez/switzer:13} showed that the Nash equilibrium outcome of the USMA-2006 mechanism was equivalent to the outcome of the cumulative
offer mechanism if cadet preferences took a particular form, where willingness to BRADSO is secondary to rankings of branches. 
Seeing the proximity between USMA-2006 and the proposal, the Army decided to keep the simpler strategy space and maintain the USMA-2006 mechanism.  

In 2012, the US Army introduced Talent-Based Branching to develop a ``talent market'' where additional information 
about each cadet influences the priority a cadet receives at a branch \citep{colarusso2016starting}. 
In the branch assignment process, prioritization at each branch has long been based on the order-of-merit list (OML), 
a composite of a cadet's academic, physical, and military performance scores.  Talent-Based Branching was
introduced to allow branches and cadets to better align their interests and fit for one another.  Under Talent-Based Branching, 
branches prioritize cadets into one of three tiers: high, medium, and low. These ratings of cadets were originally a pilot initiative, 
but for the Class of 2020, the US Army decided to use these ratings to adjust the underlying OML-based prioritization, 
constructing priorities at each branch first by the tier and then by the OML within the tier.  

The desire to use branching to improve 
talent alignment created a new objective for the Branch-of-Choice program beyond retention.
Since the decision to integrate cadet ratings into the mechanism took place under an abbreviated timeline, the US Army maintained the same strategy
space for the mechanism as in previous years, 
and devised the USMA-2020 mechanism to accommodate heterogenous branch priorities.  
In their design, the Army created two less-than-ideal theoretical possibilities in the USMA-2020 mechanism. 
First, a cadet could be charged BRADSO under the USMA-2020 mechanism even if she does not need heightened priority to receive a position at that branch.
While this was also possible under USMA-2006, it was nearly four times as common under USMA-2020. 
 Second, under USMA-2020, a cadet's willingness to BRADSO for a branch can improve priorities even for regular positions.
Surveys of cadets showed that these aspects potentially undermined trust in the branching system, 
and led the Army to reconsider the cumulative offer mechanism, despite its more complex strategy space. 
At that point, the Army established a partnership with market designers.

This paper reports on the design of a new branching system for the Class of 2021, COM-BRADSO, based on the cumulative offer mechanism
together with a choice rule for each branch that reflects the Army's dual objectives of retention and talent alignment.
We develop a model that integrates priority design with mechanism design.  Our main formal result is that the 
Army's objectives, when formulated through intuitive axioms, uniquely give
us the cumulative offer mechanism together with a choice rule, endogenous in our  setting.  
In developing this result, we provide direct evidence of the relevance of these axioms in the design.  To the best of our knowledge, our main result is the first joint characterization of the cumulative offer mechanism 
along with a specific choice rule that is induced by the central planner's policy objectives.\footnote{\cite{hirata/kasuya:17} 
and \cite{hatfield/kominers/westkamp:21}  provide characterizations of the cumulative offer mechanism for fixed choice 
rules that satisfy various technical conditions. Our main result differs from theirs in the endogeneity of the choice rule that emerges in our characterization.}

A second contribution of this paper is to provide a formal analysis of the USMA-2020 mechanism.  Our analysis shows how issues related to the lack of incentive compatibility
became more pressing with the USMA-2020 mechanism, leading the  Army to abandon this mechanism.  
We illustrate the issues using a single-branch model and by characterizing Nash equilibria of the game induced by the USMA-2020 mechanism.   
This characterization for a complete information environment 
and an example on the Bayesian equilibria of the same game for an incomplete information environment
support our argument  that the structure of incentives under the USMA-2020 mechanism is highly complex.  
We complement this theoretical analysis with field evidence on the performance of the USMA-2020 mechanism. 
Taken together, this analysis provides insight into why the Army adopted COM-BRADSO after using USMA-2020.

Finally, as part of the design, the US Army also considered policies to affect the balance between talent alignment and retention.  
To do so, the Army
considered two  policy levers: increasing the number of BRADSO-eligible positions  and making the BRADSO policy more effective.  
Based on the tools developed in this paper, the Army decided to use a more effective BRASDO policy for the Class of 2021 than it
used for the Class of 2020. 
We establish  comparative static results about
these policy levers and show how each increases the total number of BRADSOs collected using data from the Class of 2021.

Aside from our specific application, our paper offers two additional lessons for market design.  A longstanding folk-wisdom about the matching with contracts framework is that its applicability may be limited because it is too complex for participants to submit rich information on preferences over positions and contractual terms.\footnote{For instance, \cite{crawford:08}  proposes that a flexible-salary match based on \cite{kelso/crawford:82} is a natural way to incorporate wages into the National Residency Matching Program.  He argues that participants would ``be willing to bear the additional reporting costs to reap the benefits of improved allocation.''  Communication costs of mechanisms are old theme in mechanism design, including \citet{hurwicz:77}, \citet{mount/reiter:74}, and \citet{segal:07}.} Indeed, this was a major reason USMA did not adopt a cumulative offer mechanism as proposed in \cite{sonmez/switzer:13}.  We show that while not all cadets used the flexibility of the richer strategy space, many valued this option and several took advantage of it when submitting preferences.  And because the new mechanism utilizes this more detailed information, it is able to avoid failures due to its unavailability.  Therefore, our application shows possibilities for using more complicated strategy spaces in the field.\footnote{A related analogy is two-sided matching with couples.  \cite{roth:84} shows that previous systems that elicited preferences from couples did not succeed in eliciting preferences over pairs of jobs. Subsequent reforms changed the strategy-space to include such information.}   Second, field evidence on the failures of incentive and equity properties of existing allocation mechanisms is important for making the case to change mechanisms.\footnote{A growing literature has shown that a major cost of a manipulable mechanism is that some participants may not strategize while others may make mistakes by not strategizing optimally.  Studies including \cite{apsr:06}, \cite{pathak/sonmez:08}, \cite{pathak/sonmez:13}, and \cite{budish/cantillon:12} relate field evidence on incentives to arguments about changing mechanisms.} The adoption of the USMA-2020 mechanism led to a dramatic increase in the prevalence of failures due to the mechanism's lack of incentive compatibility. This field evidence laid the foundation for the new mechanism as West Point leadership decided that incentivizing cadets to misreport their true preferences degraded cadets' trust in each other and the Army.

The rest of this paper is organized as follows.  The next section introduces model and additional background on BRADSO policies.  
Section \ref{sec:bradso} provides details on quasi-direct mechanisms, including the mechanism the US Army used starting in 2006.   
Section \ref{sec:usma2020} describes the mechanism used by USMA in 2020, and uses data on cadet preferences and 
branch priorities to measure issues related to incentive compatibility and the accommodation of 
the Army's objectives between these two mechanisms.  
Focusing on the simpler case of a single branch, 
Section \ref{sec:singlebranch} illustrates the complexity of the game induced by that USMA-2020 mechanism.   
In this section,  we also  present an alternative and intuitive formulation of our proposed mechanism,  
and relate its outcome to the Nash equilibrium outcome of the  USMA-2020 mechanism.  
Section \ref{sec:2020reform} extends our analysis to the general multiple branch case, presents our 
main result characterizing COM-BRADSO, and describes some design issues with the new mechanism.  
The last section concludes.  All proofs are contained in \fig{Appendix \ref{proofs}}.

\section{Model}\label{sec:model}

There is a finite set of cadets $I$ and  a finite set of branches $B$.  There are $q_b$ identical positions
at any given branch $b \in B$, and a total of $\sum_{b\in B} q_b$ positions across all branches.   
Each cadet is in need of at most one position,
and she can be assigned one at any branch
either at a \textbf{base cost} of $t^0$ years of mandatory service, 
or at an \textbf{increased cost} of $t^+$ years  through
a  \textbf{BRADSO} program.
Let $T = \{t^0,t^+\}$ denote the set of possible mandatory service lengths. 
For any branch $b\in B$, at most $q^+_b$ of its positions can be assigned at the increased cost of $t^+$.  
We refer these positions as \textbf{BRADSO-eligible} positions. 
For any branch $b\in B$,
let $q^0_b = (q_b-q^+_b)$ denote the number  of remaining positions which can only be assigned at the base cost of $t^0$. 

\subsection{Cadet Preferences and Branch Baseline Priorities}

Each cadet has a  strict preference relation on branch-cost pairs and remaining unmatched, represented by a 
linear order on $B \times T \cup \{\emptyset\}$.
We assume that,  at any branch $b\in B$,  each cadet $i\in I$ strictly
prefers a position at the base cost $t^0$ to  one at the increased cost $t^+$. 
Let $\calq$ denote the set of linear orders  on $B \times T \cup \{\emptyset\}$  identified by this assumption. Therefore,   
for any $i \in I$,\, $\succ_i\;\in\calq$, and $b \in B$, 
\[ (b,t^0) \; \succ_i  (b,t^+).
\]
For any strict preference relation $ \succ_i \, \in \calq$, let $\succeq_i$ denote the resulting weak preference relation. 

In parts of our analysis, cadet preferences over branches (alone) and remaining unmatched will also be useful. 
In these preferences, each branch is evaluated at its base cost $t^0$. 
Let $\calp$ be the set of linear orders on $B  \cup \{\emptyset\}$. 
Here, for any $i \in I$, $P_i\in\calp$, and  $b, b' \in B$,
\[ b \; P_i \; b'  \] 
means that branch $b$ at base cost $t^0$ is strictly preferred by cadet $i$ to branch $b'$ at base cost $t^0$.

Let $\Pi$ denote the set of all linear orders on the set of cadets $I$.
Each branch $b \in B$ has a strict priority order $\pi_b \in \Pi$ on the set of cadets $I$. 
We refer $\pi_b$ as the \textbf{baseline priority order} at branch $b$. 

\subsection{BRADSO Policy}

For any branch $b\in B$, in addition to the baseline priority order $\pi_b$ (which represents the 
``baseline claims'' of cadets for positions at branch $b$), cadets' willingness to serve the
increased cost $t^+$ for a position at branch $b$ may also affect the allocation of positions at this branch. 

Given a branch $b\in B$ and a baseline priority order $\pi_b \in\Pi$, a \textbf{BRADSO policy}  
is a linear order $\omega^+_b$  on $I\times T$  with the following two properties:
\begin{enumerate}
\item for any $i,j \in I$ and $t \in T$, 
\[ (i, t) \; \omega^+_b \; (j,t) \quad \iff \quad i \; \pi_b \; j \quad \mbox{ and }\]
\item for any $i \in I$, 
\[  (i, t^+) \; \omega^+_b \; (i,t^0). \] 
\end{enumerate}
Let $\Omega^+_b$ be the set of all  linear orders on $I\times T$  which satisfy these two conditions. 

When a given BRADSO policy is invoked at a branch $b\in B$ (for some or all of the positions), (i) the relative priority order of cadets 
with identical willingness to serve the increased cost 
remain the same as the baseline priority order $\pi_b$, and (ii) any cadet 
has higher claims for a position at branch $b$ with the increased cost  $t^+$ compared to her claims for the same position with the base cost $t^0$.

How much of an advantage a BRADSO policy grants to a cadet in securing a position at branch $b$
due to her willingness to serve the increased cost $t^+$ differs between distinct
elements of $\Omega^+_b$. 
Given two BRADSO policies $\omega^+_b, \nu^+_b \in \Omega^+_b$, the policy $\nu^+_b$ 
\textbf{has weakly more effective BRADSO} than the policy $\omega^+_b$ if,
\[ \mbox{ for any } i,j \in I, \qquad (i, t^+) \; \omega^+_b \; (j,t^0) \; \implies \;  (i, t^+) \; \nu^+_b \; (j,t^0).
\]
That is, the boost received under $\nu^+_b$ (for the units the BRADSO policy is invoked) is at least as 
much as the boost received under $\omega^+_b$ for any individual when
$\nu^+_b$ has weakly more effective BRADSO than $\omega^+_b$.

\subsection{Examples of BRADSO Policies: Ultimate and Tiered} \label{subsec:BradsoPolicies}

Given a branch $b\in B$ and a baseline priority order $\pi_b \in\Pi$, define the \textbf{ultimate BRADSO policy} $\overline{\omega}_b^+ \in \Omega^+_b$ 
as the BRADSO policy where willingness to serve the increased cost $t^+$ overrides any differences in cadet ranking under
branch-$b$ baseline priority order $\pi_b$. That is, for  any pair of cadet $i,j \in I$, 
\[ (i,t^+) \; \overline{\omega}_b^+ \; (j,t^0).
\]

For the Classes of 2006-2019, USMA implemented the ultimate BRADSO policy. During these years, USMA capped the positions that could be assigned the increased cost $t^+$ at 25 percent of total positions within each branch. 
For any branch $b\in B$, 
cadets who were willing to serve at the increased cost for branch $b$ had higher priority for the 
$q_b^+$ BRADSO-eligible positions than all cadets who were not willing to serve at the increased cost for branch $b$.\medskip

Given a branch $b\in B$ and a baseline priority order $\pi_b \in\Pi$, partition cadets into $n$ tiers $I_b^1, I_b^2, \ldots, I_b^n$
so that, for any  two tiers $\ell, m \in \{1,\ldots,n\}$ and pair of cadets $i,j \in I$,  
    \[ \left. \begin{array}{l}
  \ell < m,\\   
  i \in I_b^{\ell}, \; \mbox{ and}\\ 
  j \in I_b^m  \end{array}  \right\}
\implies \; i \; \pi_b \; j.
\]
Under a \textbf{tiered BRADSO policy\/} $\omega^+_b$, for any  tier $\ell \in \{1,\ldots,n\}$ and three cadets $i, j, k\in I$, 
    \[ \left. \begin{array}{l}
 i \; \pi_b \; k ,\\   
j \; \pi_b \; k, \; \mbox{ and}\\ 
  i,j \in I_b^{\ell}  \end{array}  \right\}
\quad \implies \qquad  \Bigg((k, t^+) \; \omega^+_b \; (i,t^0) \quad \iff \quad  (k, t^+) \; \omega^+_b \; (j,t^0)\Bigg). 
\]
That is, under a tiered BRADSO policy, given two cadets $i,j\in I$ in the same tier and  
a third cadet $k\in I$ with lower  $\pi_b$-priority than both $i$ and $j$,  cadet $k$ can gain priority over cadet $i$ 
through willingness to serve at the increased cost $t^+$ if and only if  cadet $k$ can gain priority over cadet $j$ 
through willingness to serve at the increased cost $t^+$.

For the Classes of 2020 and 2021, tiered BRADSO policies were used.  
In both years, cadets were prioritized by each branch into one of three tiers, which we 
denote high, middle and low.\footnote{Branch rating categories are known to cadets and finalized before cadets submit their preferences for branches.}  
In 2020, when a cadet
expressed a willingness to serve the increased cost $t^+$, it only resulted in higher priority among
cadets who had the same categorical branch rating. For example, a middle tier cadet who
was willing to serve with increased cost would not obtain higher
priority than a high tier cadet who was unwilling to serve with increased
cost. Therefore,  under the 2020 policy, the willingness to 
serve overrides any difference in cadet ranking under $\pi_b$ only among cadets in the same
tier.

Relative to the 2020 policy, the USMA BRADSO policy for the Class of 2021 granted cadets more advantage 
in securing a position at branch $b$. Specifically, if a cadet in the Class of 2021 expressed a willingness 
to serve $t^+$, then she had higher priority over all other cadets if she was in the medium or high tier categories. 
Low tier cadets who expressed a willingness to serve $t^+$ 
only received higher priority among other low tier cadets. 
Formally, the ultimate BRADSO policy is weakly more effective than the 2021 BRADSO policy, which is 
weakly more effective than the 2020 BRADSO policy. 


\subsection{Formulation through the Matching with Contracts Model}
To introduce the outcome of an economy and some of the mechanisms analyzed in the paper, 
the following formulation through the \textit{matching with contracts} model by \cite{hatfield/milgrom:05}
will be helpful. 

For any $i \in I$,  $b \in B$, and $t \in T$, 
the triple $x=(i,b,t)$ is called a \textbf{contract}. It represents
a bilateral match between cadet $i$ and branch $b$ at the cost of $t$. 
Let 
\[ \calx = I \times B \times T \] 
denote the set of  all contracts.  
Given a contract $x \in  \calx$, let $\i(x)$ denote the cadet, $\b(x)$ denote the branch, and $\t(x)$ denote the cost of the contract $x$.  
That is, $x = \big(\i(x), \b(x), \t(x)\big).$

For any  cadet $i\in I$,  let 
\[ \calx_i = \{x \in \calx : \i(x) = i\} \] 
denote the set of contracts that involve cadet $i$. 
Similarly, for any branch $b \in B$, let  
\[ \calx_b = \{x \in \calx : \b(x) = b\} \] 
denote the set of contracts that involve branch $b$. 
Observe that for any cadet $i\in I$,  her preferences $\succ_i \, \in \calq$ originally defined over $B \times T \cup \{\emptyset\}$ 
can be redefined over $\calx_i \cup \{\emptyset\}$ (i.e. her contracts and remaining unmatched) by simply 
interpreting a branch-cost pair $(b,t) \in B\times T$ in the original domain as a contract between cadet $i$ and branch $b$ at cost $t$ in the new domain.  

\subsection{Allocations, Mechanisms, and their Desiderata} \label{sec-axioms}	

An \textbf{allocation} is a (possibly empty) set of contracts $X \subset  \calx$, such that
\[
\begin{array}{ll}
\mbox{(1)\; for any } i \in I, \qquad   & |\{x \in X : \i(x) = i\}| \leq 1,\\ 

\mbox{(2)\; for any } b \in B, \qquad   & |\{x \in X : \b(x) = b\}| \leq q_b, \quad   \mbox{ and }\\ 

\mbox{(3)\; for any } b \in B, \qquad  & |\{x \in X : \b(x) = b  \mbox{ and } \t(x) = t^+\}| \leq q^+_b.  
\end{array}
\]
That is, under an allocation  $X$, no individual can appear in more than one contract,  no branch $b$ can appear in more contracts 
than the number of its positions $q_b$, and no branch $b$ can appear in more than  $q_b^+$ contracts with the increased cost $t^+$. 
Let $\cala$ denote the set of all allocations. 

For a given allocation $X \in \cala$ and cadet $i\in I$, the \textbf{assignment} $X_i$ of cadet $i$ under allocation $X$ is defined as
\[ X_i =  \left\{ \begin{array}{cl}
         (b,t)  & \mbox{ if } (i,b,t) \in X\\
        \emptyset & \mbox{ if } X \cap \calx_i = \emptyset. \end{array} \right.
\]
For the latter case, i.e. if $X_i = \emptyset$, we say that cadet $i$ in \textbf{unmatched} under $X$. 

For a given allocation $X \in \cala$ and cadet $i\in I$, with a slight abuse of the notation,\footnote{The abuse of notation is due to the fact that
while the argument of the function $\b(.)$ is previously introduced as a contract, here it is an assignment. Since a cadet and an assignment 
uniquely defines a (possibly empty) contract, the notational abuse is innocuous.} 
let $\b(X_i)$ be defined as
\[ \b(X_i) =  \left\{ \begin{array}{cl}
         b  & \mbox{ if } (i,b,t) \in X\\
        \emptyset & \mbox{ if } X \cap \calx_i = \emptyset. \end{array} \right.
\]

	
A \textbf{mechanism} is a strategy space $\cals_i$ for each cadet $i \in I$	along with an outcome function 
\[ \varphi : \prod_{i\in I} \cals_i \rightarrow \cala\]
that selects an allocation for each strategy profile.  Let $\cals = \prod_{i\in I} \cals_i$.

Given a mechanism $\big(\cals, \varphi\big)$, the resulting \textbf{assignment function} $\varphi_i: \cals \rightarrow B\times T \cup \{\emptyset\}$
for cadet $i\in I$ is defined as follows: For any $s \in \cals$ and $X = \varphi(s)$,
\[   \varphi_i(s) = X_i. \]



A \textbf{direct mechanism} is a mechanism where  $\cals_i = \calq$ for each cadet $i \in I$. \medskip

We next formulate the desiderata for allocations and mechanisms. 
Our first three axioms are basic, and standard in the literature.
\medskip

\begin{definition} \label{IR}
An allocation $X\in \cala$  satisfies \textbf{individual rationality} if, for any $i\in I$, 
\[ X_i  \succ_i  \emptyset.
\]
A mechanism  $\big(\cals, \varphi\big)$ satisfies \textbf{individual rationality} if, 
the allocation $\varphi(s)$  satisfies individual rationality for any strategy profile $s \in \cals$. 
\end{definition}

\begin{definition} \label{non-wastefulness}
An allocation $X\in \cala$  satisfies satisfies \textbf{non-wastefulness} if for any  $b \in B$ and $i\in I$, 
\[ \left.  \begin{array}{c}
 \big|\{x \in X : \b(x) = b\}\big| <  q_b \, , \; \mbox{ and}\\ 
X_i = \emptyset \end{array}  \right\}
\implies \;  \emptyset \; \succ_i \; (b,t^0).
\]
A mechanism $\big(\cals, \varphi\big)$ satisfies \textbf{non-wastefulness} if,	
the allocation $\varphi(s)$  satisfies non-wastefulness for any strategy profile $s \in \cals$. 
\end{definition}


\begin{definition}
An allocation $X\in \cala$   
\textbf{has no priority reversals} if, for any $i,j \in I$, and $b\in B$
    \[ \left. \begin{array}{c}
  \b(X_j) = b, \; \mbox{ and}\\ 
  X_j  \succ_i  X_i \end{array}  \right\}
\implies \; j \; \pi_b \; i.
\]
A mechanism $\big(\cals, \varphi\big)$  \textbf{has no priority reversals} if,
the allocation $\varphi(s)$  satisfies elimination of priority reversals for any strategy profile $s \in \cals$. 
\end{definition}
   
This condition states that if cadet $j$ is assigned branch
$b$ at any cost  and cadet $i$ prefers cadet $j$'s assignment to her own,
then $j$ must have higher baseline priority than $i$.\footnote{This condition is  identical to the fairness condition
defined by \cite{sonmez/switzer:13}.}  
If instead cadet $i$ strictly prefers cadet $j$'s assignment even though
cadet $j$ has lower baseline priority than cadet $i$, then there is a priority reversal.
When an allocation or mechanism satisfies this axiom, we also say it \textbf{lacks priority reversals}.

Our next axiom formulates how the BRADSO policy is to be implemented. 

\begin{definition} \label{BRADSO-policy}
An allocation  $X \in \cala$ satisfies \textbf{enforcement of the BRADSO policy} if,   for any  $b \in B$, and $i,j \in I$,
\[ \begin{array}{llll}
& (1) & \left. \begin{array}{c}
  X_i = (b,t^+), \; \mbox{ and}\\ 
    (b,t^0) \succ_j X_j  \end{array}  \right\}     \quad  &\implies \quad (i,t^+) \; \omega_b^+ \; (j,t^0), \; \mbox{ and}\\ 
    \mbox{} &&\\
& (2) & \left. \begin{array}{c}
        X_j = (b,t^0),\\ 
      (b,t^+) \succ_i X_i,  \; \mbox{ and}\\
       (i,t^+) \; \omega_b^+ \; (j,t^0) \end{array}  \right\}     \quad  &\implies \quad 
       \big|\big\{i'\in I : X_{i'}=(b,t^+)\big\}\big|  = q^+_b.  
\end{array}\]
A mechanism $\big(\cals, \varphi\big)$ satisfies \textbf{enforcement of the BRADSO policy} if	
the allocation $\varphi(s)$  satisfies enforcement of the BRADSO policy for any strategy profile $s \in \cals$. 
\end{definition}

Here the first condition states that if a cadet $i$ (by invoking the BRADSO policy) receives an assignment $(b,t^+)$
at the expense of another cadet $j$ who would rather receive an assignment of $(b,t^0)$, then it must be the case that
the  increased cost contract of cadet $i$ has higher priority under the BRADSO policy $\omega_b^+$ than the base cost contract
of cadet $j$. 
The second condition, on the other hand, states that if the BRADSO policy is not invoked 
for a cadet $i$ who would rather receive an assignment of $(b,t^+)$ and who has 
higher priority under the BRADSO policy $\omega_b^+$ than the base cost contract
of another cadet $j$ with an assignment of $(b,t^0)$, then it must be the case that the upper limit for BRADSO-eligible positions  at branch $b$ is already reached. 

Our last condition is the highly sought-after incentive compatibility property for direct mechanisms.
\begin{definition} \label{strategy-proofness}
A direct mechanism $\varphi$ is \textbf{strategy-proof} if, for any $\succ \, \in \calq^{|I|}$, any $i \in I$, and any $\succ'_i \, \in \calq$, 
\[ \varphi_i(\succ) \, \succeq_i \, \varphi_i(\succ_{-i}, \succ'_i).
\]
\end{definition}

\section{BRADSO Program for Improved Retention}\label{sec:bradso}

Prior to the Class of 2006, USMA cadets were assigned positions at Army branches 
using a \textit{serial dictatorship\/} that is induced by
a cadet performance ranking known as the \textit{order of merit list (OML)\/}.
Cadets submitted their preferences over
the set of branches, and the highest-OML cadet was assigned her most-preferred branch, 
the second highest-OML cadet was assigned her most-preferred branch among branches with remaining positions, and so on.   
Let us refer to this mechanism as $\varphi^{OML}$.  


In response to declining junior officer retention rates during the late 1990s and early 2000s, the U.S. Army offered a menu of retention incentives to cadets at USMA and ROTC through the \textit{Officer Career Satisfaction Program\/}, first implemented in 2006 \citep{colarusso/lyle/wardynski:10}.
The most popular incentive, which involved a reform of the branching mechanism, was the \textit{branch of choice\/}, or \textit{BRADSO\/} program. 
Under this program, for a given percentage of the positions in any branch $b\in B$,
cadets who are willing to extend their  \textit{Active Duty Service Obligation (ADSO)\/} 
by three years if assigned to  branch $b$ are given higher priority.\footnote{The Officer Career Satisfaction Program also gave cadets the opportunity to receive their post of choice (PADSO) and the guaranteed option to attend graduate school (GRADSO) in exchange for extending their ADSO by three years. Neither PADSO nor GRADSO influenced the branching mechanism.} 
To infer which cadets are willing to serve the additional three years of ADSO for any given branch $b$,
the strategy space of the new mechanism was also modified by requesting cadets to report the set of branches
they are willing to serve the additional ADSO. Hence, the strategy space of each cadet under the modified mechanism is $\calp \times 2^{B}$.

It is important to emphasize that the modified mechanism is not a direct mechanism. Rather than merely submitting their
preferences over branch-cost pairs,  cadets instead submit their preferences over branches alone and ``signal'' their willingness
to serve the increased cost at any branch. 
The structure of the strategy space under the modified mechanism
has two important implications in relation to the axioms we introduce in Section \ref{sec-axioms}. 
First, our primary incentive compatibility axiom, strategy-proofness, is only defined for direct mechanisms. 
Hence, it is not well-defined for the Army's modified mechanism. 
Second, while the remaining four axioms are all well-defined for any mechanism regardless of their strategy spaces, they all depend
on cadet preferences over branch-cost pairs, which is private information.   
Under a direct mechanism, this private information is solicited from cadets, and hence this private information
becomes available to the central planner. As a result, verifying these axioms becomes a straightforward task 
under the ``submitted'' preferences. Moreover if the direct mechanism is strategy-proof, the central planner  has a 
formal basis to assume that the submitted preferences are truthful. 
The mechanism adopted by the USMA for the class of  2006, however, is not a direct mechanism. 
Therefore, verification of these axioms may be less clear under the modified mechanism. 
This distinction, at least partially, contributed the Army's decision to maintain the USMA-2006 mechanism for
over a decade. 
Before formally introducing this mechanism, we first formulate axioms that are both well-defined and possible to verify 
under a simpler strategy space.

\subsection{Quasi-Direct Mechanisms and their Desiderata}

A \textbf{quasi-direct mechanism} is a mechanism where the strategy space is $\cals_i = \calp \times 2^B$ for each cadet $i \in I$. \medskip

We next formulate three axioms for quasi-direct mechanisms; axioms  which play important role in Army's decision 
to reform its branching process both for the USMA and the ROTC for the Class of 2021. \medskip

Our first axiom on quasi-direct mechanisms formulates the goal of charging the increased cost only to cadets
for whom the BRADSO policy has been pivotal in securing a branch.

\begin{definition} \label{BRADSO-IC}
A quasi-direct mechanism $\varphi$  satisfies \textbf{BRADSO-incentive compatibility} (or 
\textbf{BRADSO-IC}) if, for any $s = \big(P_j, B_j\big)_{j\in I} \in (\calp \times 2^B)^{|I|}$, 
$i \in I$, and  $b \in B$, 
\[  \varphi_i(s) = (b, t^+) \; \implies \;  \varphi_i\big((P_i, B_i\setminus\{b\}),\; s_{-i}\big) \not= (b, t^0). 
\]
\end{definition}
That is, any cadet $i\in I$ who receives a position at branch $b$ at the increased cost $t^+$ under $\varphi$ 
should not be able to profit by receiving a position at the same branch at the cheaper base cost $t^0$
by dropping branch $b$ from the set of branches $B_i$ for which she has indicated willingness to serve the increased cost $t^+$.
Alternatively, a cadet should never be charged BRADSO for a branch merely because of his/her willingness to serve the increased cost.

Our next axiom formulates the idea that the willingness to serve the increased cost $t^+$ at a branch should never serve
the sole purpose of enabling an assignment in this branch at the base cost $t^0$. 

\begin{definition} \label{strategic-BRADSO}
A quasi-direct mechanism $\varphi$  satisfies \textbf{elimination of strategic BRADSO}  if, for any $s = \big(P_j, B_j\big)_{j\in I} \in (\calp \times 2^B)^{|I|}$, 
$i \in I$, and  $b \in B$, 
\[  \varphi_i(s) = (b, t^0) \; \implies \;  \varphi_i\big((P_i, B_i\setminus\{b\}),\; s_{-i}\big) = (b, t^0). 
\]
\end{definition}
That is, any cadet $i\in I$ who receives a position at branch $b$ at the base cost $t^0$ under $\varphi$ 
should still do so  upon
dropping branch $b$ from the set of branches $B_i$ for which she has indicated willingness to serve the increased cost $t^+$
(in case $b\in B_i$).\footnote{This statement holds vacuously if $b\not\in B_i$.}
Whenever this axiom fails for a cadet $i\in I$ at a branch $b\in B$,  
cadet $i$ has an opportunity to strategically indicate a willingness to serve the increased cost $t^+$ at branch $b$
and receive a position at  this branch at the base cost $t^0$ which is otherwise beyond reach in the absence of this strategy.  

Our last axiom relaxes the lack of priority reversals formulated in Section \ref{sec-axioms}
by removing any dependence on cadet preference information on branch-cost pairs not solicited by the mechanism. 

\begin{definition} \label{detectable-priorityreversal}
A quasi-direct mechanism $\varphi$  has \textbf{no detectable priority reversals}  if, 
for any $s = \big(P_j, B_j\big)_{j\in I} \in (\calp \times 2^B)^{|I|}$, $b \in B$, and
$i,j \in I$,   
    \[ \left. \begin{array}{c}
  \varphi_j(s) = (b,t^0), \; \mbox{ and}\\ 
    \varphi_i(s) =  (b,t^+) \quad \mbox{ or } \quad b \; P_i \; \b\big(\varphi_i(s)\big)
  \end{array}  \right\}
\implies \; j \; \pi_b \; i.
\]
\end{definition}
This condition requires that whenever
a cadet $j\in I$ is assigned a position at a  branch $b\in B$ at the cheaper base cost $t^0$, 
while another cadet $i\in I$ receives a visibly less desired assignment by 
\begin{itemize}
\item[(i)] either receiving a position at the same branch at the increased cost $t^+$ or 
\item[(ii)] by receiving  a position at a strictly less preferred (and possibly empty) branch based on cadet $i$'s submitted preferences $P_i$ on $B\cup\{\emptyset\}$, 
\end{itemize}
cadet $j$ must have higher baseline priority under branch $b$ than cadet $i$.  \medskip

The distinction between our axiom on the lack of priority reversals and its weaker version on the lack of detectable priority reversals
is subtle.  When a mechanism  has priority reversals,  thus failing the stronger of the two axioms,
there is a cadet $i\in I$ who strictly prefers the assignment of another cadet $j\in I\setminus\{i\}$ despite having higher claims for this position. 
The key difference is that verification of  this anomaly may require knowing the preferences
$\succ_i \, \in \calq$ of cadet $i$ over branch-cost pairs, which is potentially private information that may not be always available (even to the central planner).  
Verification is particularly challenging if
the mechanism is not a direct mechanism. 
In contrast,  when a quasi-direct mechanism has detectable priority reversals, thus failing the weaker of the two axioms, 
there is a cadet $i\in I$ who strictly prefers the assignment of another cadet $j \in I\setminus\{i\}$ no matter what cadet $i$'s preferences $\succ_i \, \in \calq$ over 
branch-cost pairs  are provided that they are consistent with her submitted preferences $P_i \in \calp$ over branches alone. 
In that sense, all detectable priority reversals can be verified under a quasi-direct mechanism, 
but the same is not true for all priority reversals.




\subsection{USMA-2006 Mechanism}

We are ready to introduce the quasi-direct mechanism the  Army has adopted at USMA starting with the Class of 2006 to implement its 
BRADSO program. Since it is a quasi-direct mechanism, the strategy space for this mechanism is given as 
\[ \cals^{2006} = \big(\calp \times 2^{B} \big)^{|I|},
\]
and the following construction is useful to introduce its outcome function:  

Given an OML $\pi$ and a strategy profile $s= (P_i, B_i)_{i\in I} \in \cals^{2006}$, for any branch $b\in B$
construct the following adjusted priority order $\pi^+_b \in \Pi$ on the set of cadets $I$.
For any pair of cadets $i, j \in I$,
\begin{enumerate}
\item $b\in B_i$ and $b\in B_j \quad \implies \quad i \; \pi^+_b \; j \; \iff i \; \pi \; j$,
\item $b\not\in B_i$ and $b\not\in B_j \quad \implies \quad i \; \pi^+_b \; j \; \iff i \; \pi \; j$, \mbox{ and} 
\item $b\in B_i$ and $b\not\in B_j \quad \implies \quad i \; \pi^+_b \; j$.
\end{enumerate}	
Under the adjusted priority order $\pi^+_b$, 
any pair of cadets are rank ordered through the OML $\pi$
if they have indicated the same willingness to serve for branch $b$, and 
through the ultimate BRADSO policy $\overline{\omega}^+_b$ (which 
gives higher priority to the cadet who has indicated to serve the increases cost) otherwise.  

Given an OML $\pi$ and  a strategy profile $s= (P_i, B_i)_{i\in I} \in \cals^{2006}$,
the outcome $\varphi^{2006}(s)$ of the 
\textbf{USMA-2006 mechanism\/} is obtained with the following sequential procedure:
\\
\begin{quote}
\textbf{\textit{Branch assignment\/}}: At any step $\ell \geq 1$ of the procedure, the highest $\pi$-priority cadet $i$ who is not tentatively on hold for a position at any branch
applies to her highest-ranked acceptable branch $b$ under her submitted branch preferences $P_i$ that 
has not rejected her from earlier steps.\footnote{The USMA-2006 mechanism can also 
be implemented with a variant of the algorithm where each cadet who is not tentatively holding a position simultaneously apply to
her next choice branch among branches that has not rejected her application.}  
 
Branch $b$ considers cadet $i$ together with all cadets it has been tentatively holding both for 
its $q^0_b$ primary positions and also for its $q^+_b$ BRADSO-eligible positions, and 
\begin{enumerate}
\item it tentatively holds (up to) $q^0_b$ highest  $\pi$-priority applicants for one of its $q_b^0$ primary positions, 
\item among the remaining applicants it tentatively holds  (up to) $q^+_b$ highest  $\pi^+_b$-priority applicants for one of its $q_b^+$ 
BRADSO-eligible positions, and
\item it rejects any remaining applicant. 
\end{enumerate}
The procedure terminates when no applicant is rejected.   Any cadet who is not tentatively on hold
at any brach remains unmatched, and all tentative branch assignments are finalized. 

\textbf{\textit{Cost assignment\/}}: For any branch $b\in B$,
\begin{enumerate}
\item any cadet $i\in I$ who is assigned one of the $q^0_b$ primary positions at branch $b$ is charged the base cost $t^0$, and
\item any cadet  $i\in I$ who is assigned one of the $q^+_b$  BRADSO-eligible positions is charged 
\begin{enumerate}
\item the increased cost $t^+$ if $b\in B_i$, and 
\item the base cost $t^0$ if $b\not\in B_i$.
\end{enumerate} 
\end{enumerate}
\end{quote} 

\subsection{Shortcomings of the USMA-2006 Mechanism}
 
While a natural extension of its predecessor $\varphi^{OML}$, \cite{sonmez/switzer:13} show 
that the USMA-2006 mechanism has a number of shortcomings. These are largely due to the inability of its strategy space
to capture cadet preferences over branch-cost pairs. 
In particular, they have shown that  the USMA-2006 mechanism fails  BRADSO-IC and has 
priority reversals even at its Nash equilibrium outcomes. 
As a remedy, 
\cite{sonmez/switzer:13} proposed the \textit{cumulative offer mechanism\/} (presented in Section \ref{sec:2020reform}) implemented with
the ultimate BRADSO policy reflecting the Army's BRADSO policy at the time. 

As a direct mechanism, the cumulative offer mechanism requires cadets to submit their preferences
over branch-cost pairs (rather than their preferences over branches alone together with a set of branches
for which cadets indicate their willingness to serve the increased cost $t^+$ to receive preferential treatment for their BRADSO-eligible positions). 
This change in the strategy space was initially seen at the Army as unnecessary
 due to three main reasons:
\begin{enumerate}
\item While in theory the USMA-2006 mechanism has BRADSO-IC failures and detectable priority reversals,
these issues have been relatively rare in practice. For example, each year on average 22 cadets have been affected
by BRADSO-IC failures and 20 cadets have been affected by detectable priority reversals 
under the USMA-2006 mechanism across the Classes of 2014-2019 (These facts are described in further detail below in \fig{Figure \ref{fig:failures}}).  
\item Any potential BRADSO-IC failure or detectable priority reversal can be manually corrected ex-post, 
since each only involves a cadet needlessly paying the increased cost at her assigned branch. 
An ex-post manual reduction of the cost to the base cost $t^0$ completely resolves the issue.   
\item Even though the USMA-2006 mechanism  allows for additional priority reversals which may alter a cadet's branch assignment 
and consequently  cannot be manually corrected ex-post, the 
verification of any such theoretical failure relies on cadet preferences over branch-cost pairs.  Since USMA-2006 is a quasi-direct mechanism, information
on cadet preferences over branch-cost pairs is not available.
\end{enumerate}

In summary, any possible failure of the properties above under the USMA-2006 mechanism can either be manually corrected ex-post or cannot
be verified based on the existing data.
In large part for these reasons, the USMA-2006 mechanism was maintained by the Army for fourteen years
until the Class of 2020. 
At this point, the introduction of a new program aimed at improved talent assignment 
triggered an adjustment in the mechanism, which we describe next. 

\section{Talent-Based Branching Program for Improved Talent Alignment}\label{sec:usma2020}

The Army began piloting the Talent-Based Branching (TBB) program with the USMA Class of 2013 
with the aim of matching cadets to branches which better fit their talents \citep{colarusso2016starting}. A substantial component of TBB is an opportunity for branches to interview and rate cadets into three tiers.  Prior to the Class of 2020, these rating categories did not influence baseline branch priorities at USMA.  Ratings could only indirectly influence a cadet's branch assignments either by causing some cadets to adjust their preferences for branches or by convincing the Army to make an ex-post adjustment to a cadet's branch assignment after executing the branching assignment mechanism.

In July 2019, the Army decided to incorporate branch rating categories into 
baseline branch priorities beginning with the USMA Class of 2020.
Just as the introduction of the BRADSO program triggered a reform in the branching mechanism, the full integration of the TBB program with the
branching process resulted in another adjustment.
The Army  replaced the USMA-2006 mechanism with another quasi-direct mechanism based on the individual-proposing
deferred acceptance algorithm, where branches have heterogeneous baseline priorities over cadets according to the tiered BRADSO policy described in Section \ref{subsec:BradsoPolicies}.

 

A key distinction between the USMA-2006 mechanism and the USMA-2020 mechanism was that, even though the Army continued to cap the number of BRADSO-eligible positions at 25 percent of the total number of positions within each branch,
the Army used the adjusted priority ranking of cadets mainly intended for the BRADSO-eligible positions also for the regular positions.
Through this practice the matching aspect of the branching process was transformed into a standard priority-based assignment problem, which in turn
made it possible for the Army to use the  individual-proposing  deferred acceptance algorithm to determine the branch assignments.
The cost assignments were then subsequently determined based on submitted cadet willingness to serve the increased cost $t^+$. 
Importantly, the Army charged the increased cost to willing cadets in reverse-priority order, stopping when 25 percent of 
cadets assigned to the branch had been charged the increased cost. For example, if 100 cadets were assigned to a branch and 50 
of the cadets volunteered for the increased cost $t^+$, the Army would charge the increased cost to the 25 lowest priority cadets  of the 50 willing to 
serve $t^+$.\footnote{USMA leadership described this mechanism to cadets during two separate briefings.}

\subsection{USMA-2020 Mechanism} \label{USMA-2020}

We next formally introduce and analyze the USMA-2020 mechanism. 
As in the case of USMA-2006 mechanism,  the USMA-2020 mechanism is also a quasi-direct mechanism. 
Hence, the strategy space $\cals^{2020}_i$ for each cadet $i\in I$  under the USMA-2020 mechanism is 
\[ \cals^{2020}_i = \calp \times 2^{B}. 
\]

For the rest of this section, fix a  a profile of baseline branch priorities $(\pi_b)_{b\in B} \in \Pi^{|B|}$ and a
profile of BRADSO policies $\big(\omega^+_b\big)_{b\in B} \in  \prod_{b\in B}\Omega^+_b$. 

Given a strategy profile $s = (P_i, B_i)_{i\in I}$, for any branch $b\in B$ construct the following adjusted
priority order $\pi^+_b \in \Pi$ on the set of cadets $I$.
For any $i, j \in I$,
\begin{enumerate}
\item $b\in B_i$ and $b\in B_j \quad \implies \quad i \; \pi^+_b \; j \; \iff i \; \pi_b \; j$,
\item $b\not\in B_i$ and $b\not\in B_j \quad \implies \quad i \; \pi^+_b \; j \; \iff i \; \pi_b \; j$, \mbox{ and} 
\item $b\in B_i$ and $b\not\in B_j \quad \implies \quad i \; \pi^+_b \; j \; \iff (i,t^+) \; \omega^+_b \; (j,t^0)$.
\end{enumerate}	
Under the priority order $\pi^+_b$, 
any two cadets are rank ordered using the baseline priority order $\pi_b$
if they have indicated the same willingness to serve for branch $b$, and using the BRADSO policy $\omega^+_b$ otherwise.\footnote{When (i) the baseline priority order $\pi_b$ is fixed as OML at each branch $b\in B$ 
and (ii) the BRADSO policy $\omega^+_b$ is fixed as the ultimate BRADSO policy
${\overline{\omega}}^+_b$ at each branch $b\in B$, this construction gives the same adjusted priority order constructed for the USMA-2006 mechanism.}

For any strategy profile $s = (P_i, B_i)_{i\in I}$, 	let $\mu$ be the outcome
of the  \textit{individual-proposing deferred acceptance algorithm\/} \citep{gale/shapley:62} for submitted cadet preferences $(P_i)_{i\in I}$ and constructed
branch priorities $\big(\pi^+_b\big)_{b\in B}$.\footnote{See Appendix \ref{sec:da} for the individual-proposing deferred acceptance algorithm.}

For any strategy profile $s = (P_i, B_i)_{i\in I}$, the outcome  $\varphi^{2020}(s)$ of the \textbf{USMA-2020 mechanism} is given as follows.  	
For any cadet $i \in I$, 
\[ \varphi_i^{2020}(s) =   \left\{ \begin{array}{cl}  
\emptyset & \mbox{if } \; \mu(i)=\emptyset, \\
 \big(\mu(i), t^0\big)   & \mbox{if } \; \mu(i)\not\in B_i  \mbox{ or }   \big|\big\{j\in I : \mu(j)=\mu(i), \; \mu(j)\in B_j, \mbox{ and } i \; \pi_{\mu(i)} \;j\big\}\big| \geq q^+_{\mu(i)},\\
 \big(\mu(i), t^+\big)   & \mbox{if } \; \mu(i)\in B_i  \mbox{ and }  \big|\big\{j\in I  : \mu(j)=\mu(i), \; \mu(j)\in B_j, \mbox{ and } i \; \pi_{\mu(i)} \;j\big\}\big| < q^+_{\mu(i)}.
\end{array} \right.
\]	
In the USMA-2020 mechanism, each cadet $i\in I$ is asked to submit a preference relation $P_i \in \calp$ along with a 
(possibly empty) set of branches 
$B_i\in 2^B$ for which she indicates her willing to serve the increased cost $t^+$ to receive preferential admission. 
A priority order $\pi_b^+$ of cadets is constructed for each branch $b$ by adjusting  the baseline priority order $\pi_b$  using the
BRADSO policy $\omega_b^+$ whenever a pair of cadets  submitted different willingness  to serve the increased cost $t^+$  at branch $b$. 
Cadets' branch assignments are determined by the individual-proposing deferred acceptance algorithm  using the submitted 
profile of cadet preferences $(P_i)_{i\in I}$
and the profile of adjusted priority rankings  $(\pi^+_b)_{b\in B}$. 
A cadet pays the base cost for her branch assignment  if either she has not declared willingness to pay the increased cost for her assigned branch or 
the increased cost capacity for the branch is already filled with cadets who have lower baseline priorities. 
With the exception of those who remain unmatched,  all other cadets pay the increased cost for their branch assignments.

\subsection{Shortcomings of the USMA-2020 Mechanism} \label{sec:shortcomings2020}

Example \ref{knifeedge} in Section \ref{subsec:NE}  shows that the USMA-2020 mechanism fails both BRADSO-IC and
elimination of strategic BRADSO,  and Example \ref{Bayesian} in Section \ref{subsec:NE} shows that it can admit
detectable priority reversals even under its Bayesian Nash equilibrium outcomes. 
Before formally presenting these examples in the next section, we first  describe how these 
failures already surfaced at the USMA in Fall 2019, 
paving the way for our collaboration with the Army. 

Before a formal analysis of the USMA-2020 mechanism was carried out by our team,  USMA leadership already recognized
the possibility of detectable priority reversals under the USMA-2020 mechanism due to either failure of BRADSO-IC
or presence of strategic BRADSO. 
For example, in a typical year, the number of cadets willing to BRADSO for traditionally oversubscribed branches like Military Intelligence 
greatly exceeded 25 percent of the branch's allocations. 
Therefore, by volunteering for BRADSO for an oversubscribed branch, some cadets could receive a priority upgrade even though they may not be charged for it, making detectable priority reversals a theoretical possibility.
Moreover, unlike the detectable priority reversals under the USMA-2006 mechanism, some of these detectable priority reversals
can affect cadet branch assignments, thereby making manual ex-post adjustments infeasible.

Failures of BRADSO-IC, elimination of strategic BRADSO,  or presence of detectable priority reversals, 
especially when not manually corrected ex-post, 
could erode cadets' trust in the Army's branching process. Consider, for example, a comment from a cadet survey administered to the USMA Class of 2020:
\footnote{The survey was administered to the Class of 2020 immediately before they submitted their preferences for branches under the USMA-2020 mechanism.
The response rate to this survey was 98\%. Appendix \ref{survey} contains specific questions and results.} 
 
\begin{quote}
    ``\textit{I believe this system fundamentally does not trust cadets to make the best choice for themselves. It makes it so that we cannot choose what we want and have to play games to avoid force branching}.''
\end{quote}
A mechanism that erodes trust is unlikely to persist in the US Army, where trust is an essential characteristic of the profession and the foundation of the organization's talent management strategy.\footnote{For example, in \textit{The Army Profession}, the US Army's Training and Doctrine Command identifies trust as an essential characteristic that defines the Army as a profession \citep[US][]{army:19}.   The Army's People Strategy describes one of the Army's strategic outcomes as building a professional Army that retains the trust and confidence of the American people and its members \citep[US][Training and Doctrine Command]{army:19b}.} Perhaps unsurprisingly, when considering potential reforms to the USMA-2020 mechanism, the manager of the Talent-Based Branching program stated the the Army prefers a mechanism that incentivizes honest preference submissions.\footnote{Lieutenant Colonel Riley Post, the Talent-Based Branching Program Manager, said ``cadets should be honest when submitting preferences for branches, instead of gaming the system'' in a statement in West Point's official newspaper \citep{garcia:20}.}

To address these types of concerns, USMA leadership decided to execute a simulation using cadet preliminary preferences to inform cadets of the potential cutoffs for each branch.\footnote{Cadets in the Class of 2020 submitted preliminary preferences one month before submitting final preferences. USMA ran the USMA-2020 mechanism on these preliminary preferences to derive results for the simulation, which USMA provided to cadets 6 days prior to the deadline for submitting final preferences.}

As emphasized in the following quote from a U.S. Army news article on the new mechanism, the goal of this simulation was to improve transparency and help cadets to optimize their submitted strategies  \citep{oconnor:19}: 
\begin{quote}
 ``\textit{We're going to tell all the cadets, we're going to show all of them, here's when the branch would have went out, here's the bucket you're in, here's the branch you would have received if this were for real. You have six days to go ahead and redo your preferences and look at if you want to BRADSO or not.'' Sunsdahl said. ``I think it's good to be transparent. I just don't know what 21-year-olds will do with that information}.''
\end{quote}

The same quote, however, also indicates that USMA leadership recognized the challenges in cadets optimizing their strategies under the USMA-2020 mechanism. These concerns were well-justified. For example, considering the relative popularity of a branch like Military Intelligence, it could be advantageous for some cadets to volunteer to BRADSO for it even if they would normally not be willing to serve at the increased cost. Relatedly, other cadets who were willing to BRADSO would also have to determine whether volunteering to BRADSO was a good strategy if the simulation suggested they were likely to secure the branch even without the increased cost. This latter point was particularly salient among cadets interested in the Engineer branch. For example, most cadets who were willing to BRADSO for the Engineer branch were placed in the high priority tier, but results from the simulation indicated the branch was very likely to extend contracts to medium priority cadets by the Engineer branch. As a result, cadets who volunteered to BRADSO for Engineer who were also placed
in the high priority tier by the branch, faced a high probability of being charged BRADSOs under the USMA-2020 mechanism even though it was unlikely these cadets needed to BRADSO to branch Engineer.

Several open-ended survey comments from USMA cadets in the Class of 2020 mirrored USMA leadership's concern that continued use of the USMA-2020 mechanism would erode trust in the branching process.
We present three additional comments articulating concerns related to the lack of BRADSO-IC, the presence of  strategic BRADSO, and the
difficulty of navigating a system with both shortcomings:

\begin{itemize}
    \item[1)] ``\textit{Volunteering for BRADSO should only move you ahead of others if you are actually charged for BRADSO. 
    By doing this, each branch will receive the most qualified people. Otherwise people who are lower in class rank 
    will receive a branch over people that have a higher class rank which does not benefit the branch. Although those who 
    BRADSO may be willing to serve longer, if they aren't charged then they can still leave after their 5 year commitment 
    so it makes more sense to take the cadets with a higher OML.}"
    \item[2)]    ``\textit{I think it is still a little hard to comprehend how the branching process works. For example, I do not know if I put a BRADSO for my preferred branch that happens to be very competitive, am I at a significantly lower chance of getting my second preferred if it happens to be something like engineers? Do I have to BRADSO now if I want engineers??? Am I screwing myself over by going for this competitive branch now that every one is going to try to beat the system????}''
    \item[3)] ``\textit{Releasing the simulation just created chaos and panicked cadets into adding a BRADSO who otherwise wouldn't have.''}
\end{itemize}

\subsection{USMA-2006 and USMA-2020 Mechanism in the Field}

In this section, we use administrative data on cadet rankings, branch priorities, and capacities to investigate the performance
of the USMA-2006 and USMA-2020 mechanisms.  The data cover the West Point Classes of 2014 through 2021.  \fig{Table \ref{fig:tableCapacity}} lists the capacity for each branch, the number of cadets who list the branch as their top choice, and the number of cadets who expressed a willingness to BRADSO for each branch for the Classes of 2020 and 2021. For the Class of 2020, 1,089 cadets participated in the branching process for
17 different branches.  For the Class of 2021, 994 cadets participated in the branching process for 18 different branches.\footnote{We successfully replicated the branch assignment for 99.2\% of cadets in the Classes of 2014 through 2021. See \fig{Appendix \ref{subsec:dataappendix}} for details on our replication rates for each class.}  

\begin{figure}[htp]
    \begin{center}
    \caption{\textbf{Comparison of Outcomes of the USMA-2006 and USMA-2020 Mechanisms\\}}
       \label{fig:failures}
    \includegraphics[scale=1.0]{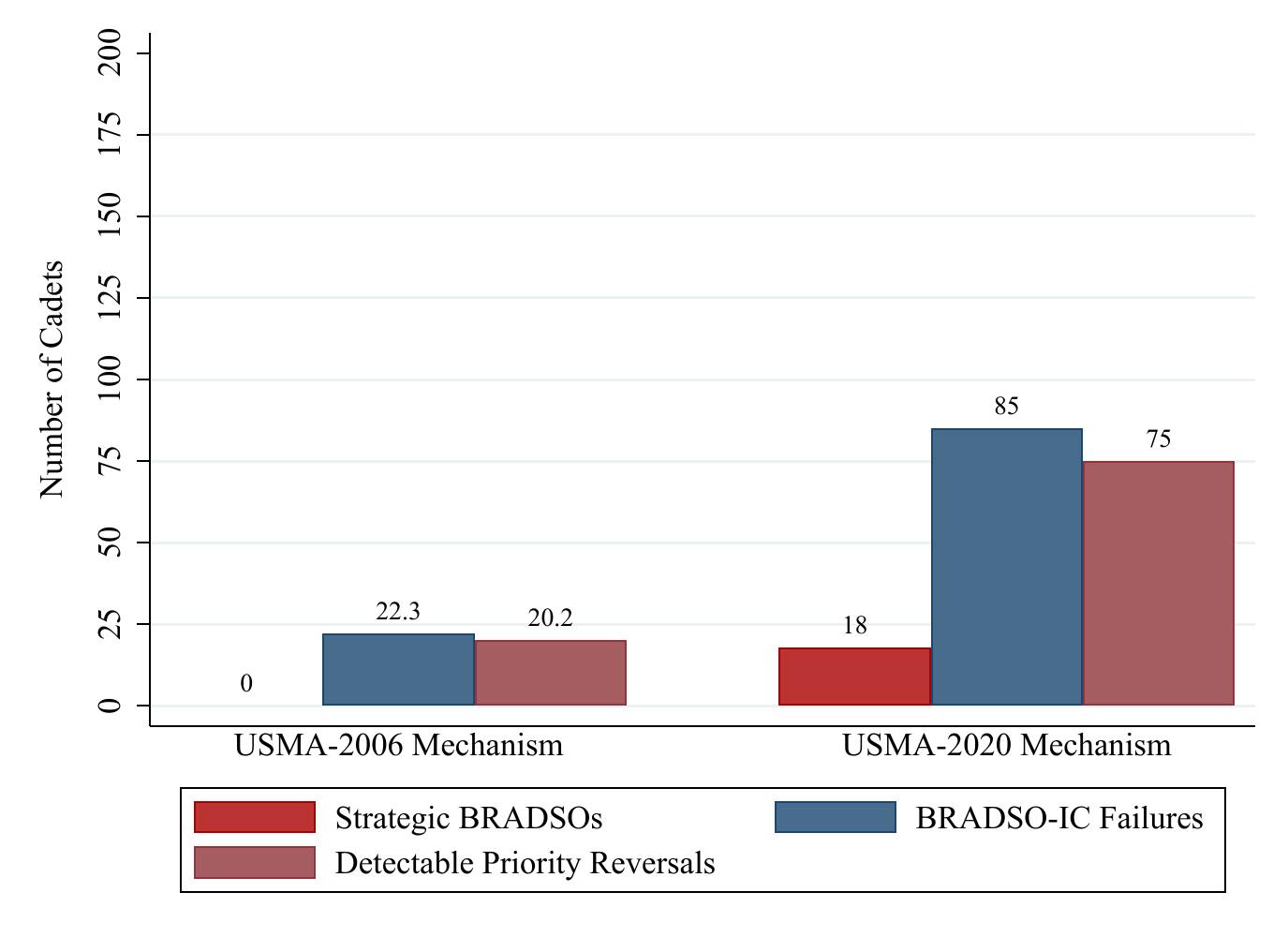}
    \end{center}
    \textbf{Notes.}  This figure reports Strategic BRADSOs, BRADSO-IC Failures, and Detectable Priority Reversals under the USMA-2006  and USMA-2020 Mechanisms.   The leftmost three columns correspond to outcomes under the USMA-2006 Mechanism averaging over six classes who participated in the mechanism from 2014-2019.  The rightmost three columns correspond to outcomes under USMA-2020 Mechanism for the Class of 2020.   A cadet has a Strategic BRADSO if she is assigned a branch at base cost and would have still received that position at base cost if she did not indicate a willingness to BRADSO at the branch. 
A cadet has a BRADSO-IC failure if she is assigned a branch at increased cost, but would receive that branch at base cost if she did not indicate a willingness to BRADSO at the branch.
A cadet has an detectable Priority Reversal if she is assigned a branch at base cost and another cadet with higher priority either receives that branch with BRADSO or is assigned a strictly less preferred branch. 
\end{figure}

\fig{Figure \ref{fig:failures}} tabulates the incidence of BRADSO-IC failures, strategic BRADSO, and 
detectable priority reversals among USMA cadets across the USMA-2006 and USMA-2020 mechanism.
For the USMA-2006 mechanism, we report the average across the Class of 2014 through Class of 2019.
Nearly four times as many cadets are part of BRADSO-ICs from the Class of 2020 (where the USMA-2020 mechanism was used) than earlier Classes from 2014 to 2019 (where USMA-2006 mechanism was used).
\fig{Figure \ref{fig:failures}} shows about 22 cadets were part of BRADSO-IC failures under the USMA-2006 mechanism, while
85 cadets were part of BRADSO-IC failures under the USMA-2020 mechanism.  
Parallel to the incidences on BRADSO-IC failures, \fig{Figure \ref{fig:failures}} shows that
nearly four times as many cadets are part of detectable priority reversals under the USMA-2020 mechanism than under the USMA-2006 mechanism (75 versus 20).
It is  not possible to have a strategic BRADSOs under the USMA-2006 mechanism. 
\fig{Figure \ref{fig:failures}} shows that 18 cadets in the Class of 2020 were part of strategic 
BRADSOs under the USMA-2020 mechanism.  
Importantly, these instances are not possible to remedy ex-post since that would require a change in branch assignments
(rather than merely foregoing a BRADSO charge).

\section{Single Branch Analysis} \label{sec:singlebranch}

As with the USMA-2006 mechanism, truthful revelation of branch preferences is not a dominant strategy under the USMA-2020 mechanism,
thereby making its analysis challenging. 
Fortunately, focusing on a simpler version of the model with a single branch is sufficient to illustrate and analyze the main challenges
of the USMA-2020 mechanism. 
Focusing on this simpler model also offers a clear path to overcome these shortcomings, a path which is extended in Section \ref{sec:2020reform}
to the model in its full generality with multiple branches.  

When there is a single branch $b \in B$, there are only two preferences  for any cadet $i \in I$.
The base cost contract  $(i,b,t^0)$ is by assumption preferred by cadet $i$ to both its 
increased cost version $(i,b,t^+)$ and also to remaining unmatched. Therefore, the only variation in cadet $i$'s preferences
depends on whether the increased cost contact $(i,b,t^+)$ is preferred to remaining unmatched. For any cadet $i \in I$, $|\calq| =2$
When there is a single branch $b\in B$, since 
\begin{itemize}
\item indicating willingness  to serve the increased cost $t^+$ 
under a quasi-direct mechanism can be naturally mapped to the preference relation  where the  increased cost contact $(i,b,t^+)$ is acceptable, whereas 
\item not doing so 
can be naturally mapped to the preference relation  where the  increased cost contact $(i,b,t^+)$ is unacceptable, 
\end{itemize}
any quasi-direct mechanism  can be interpreted as a direct mechanism. 
Therefore, unlike the general version of the model, the axioms of BRADSO-IC and elimination of strategic BRADSO are  
well-defined for direct mechanisms when there is a single branch, and moreover they are both implied by strategy-proofness.\footnote{BRADSO-IC and elimination of strategic BRADSO together are equivalent to strategy-proofness when there is a single branch. Strategy-proofness of a single branch,  called non-manipulability via contractual terms also plays an important role in the analysis of \cite{hatfield/kominers/westkamp:21}.}  

\subsection{Single-Branch Mechanism {\boldmath $\phi^{BR}$} and Its Characterization}

We next introduce a single-branch direct mechanism that is key for our analysis of the USMA-2020 mechanism.
The main feature of this mechanism is its iterative subroutine (in Step 2), which determines how many BRADSO-eligible positions 
are assigned at the increased cost and which cadets receive these positions.   
\smallskip

\begin{quote}
\noindent \textbf{Mechanism} {\boldmath $\phi^{BR}$} 

\noindent For any given profile of cadet preferences $\succ = (\succ_i)_{i\in I} \in \calq^{|I|}$, 
construct the allocation $\phi^{BR}(\succ)$ as follows:\smallskip

\noindent \textbf{Step 0.} Let $I^0 \subset I$ be the set of  $q^0_b$ highest $\pi_b$-priority cadets in $I$. 
For each cadet $i \in I^0$, finalize the assignment of cadet $i$ as $\phi_i^{BR}(\succ) = (b,t^0)$.  \smallskip

\noindent \textbf{Step 1.} Let $I^1 \subset I\setminus I^0$ be the set of  $q^+_b$ highest $\pi_b$-priority cadets in $I\setminus I^0$. 
\textit{Tentatively\/} assign each cadet in $I^1$  a position to at the base cost $t^0$. 
Relabel the set of cadets in $I^1$ so that cadet $i^1 \in I^1$ has the lowest $\pi_b$-priority in $I^1$,  
cadet $i^2 \in I^1$ has the second-lowest $\pi_b$-priority in $I^1$, $\ldots$, and  
cadet $i^{q^+_b} \in I^1$ has the highest $\pi_b$-priority in $I^1$. 
Also relabel the lowest  $\pi_b$-priority cadet in $I^0$ as $i^{q^+_b + 1}$.
\smallskip

\noindent \textbf{Step 2.} This step determines how many positions are assigned at the increased cost $t^+$. \smallskip

\noindent \textbf{Step 2.0.} 
Let $J^0 \subset I \setminus (I^0 \cup I^1)$ be the set of cadets in $I \setminus (I^0 \cup I^1)$ who declared 
the position at the increased cost  $t^+$ as acceptable:
\[ J^0 = \{j \in I\setminus (I^0 \cup I^1) :  (b,t^+) \; \succ_j \; \emptyset\}.
\]
If \[\big|\big\{j \in J^0 : (j, t^+) \; \omega^+_b \; (i^1, t^0) \big\}\big| = 0,\] then finalize Step 2 and proceed to Step 3.
In this case no position will be assigned at the increased cost $t^+$. 

Otherwise, if 
\[\big|\big\{j \in J^0 : (j, t^+) \; \omega^+_b \; (i^1, t^0) \big\}\big|  \geq 1,\] 
then proceed to Step 2.1.  \smallskip

\noindent \textbf{Step 2.}{\boldmath $\ell$.} {\boldmath $(\ell = 1,\ldots, q^+_b)$} 
Let 
\[J^{\ell} = \left\{ \begin{array}{cl}
         J^{\ell -1}  & \mbox{ if } \; \emptyset \, \succ_{i^{\ell}} \, (b,t^+)\\
       J^{\ell -1} \cup\{i^{\ell}\} & \mbox{ if } \; (b,t^+) \succ_{i^{\ell}} \, \emptyset. \end{array} \right.
\]
If \[\big|\big\{j \in J^{\ell} : (j, t^+) \; \omega^+_b \; (i^{\ell +1}, t^0) \big\}\big| = \ell,\] then finalize Step 2 
and proceed to Step 3.\footnote{Since  $J^{\ell} \supseteq J^{\ell -1}$ by construction, the fact that the procedure has reached Step 2.$\ell$ 
implies that the inequality  $\big|\big\{j \in J^{\ell} : (j, t^+) \; \omega^+_b \; (i^{\ell +1}, t^0) \big\}\big| \geq \ell$ must hold.}
In this case $\ell$ positions will be assigned at the increased cost $t^+$. 

Otherwise, if 
\[\big|\big\{j \in J^{\ell} : (j, t^+) \; \omega^+_b \; (i^{\ell +1}, t^0) \big\}\big|  \geq \ell +1,\] 
then proceed to Step 2.$(\ell+1)$, unless $\ell = q^+_b$, in which  case finalize Step 2 and proceed to Step 3.\smallskip 

\noindent \textbf{Step 3.} Let Step 2.$n$ be the final sub-step of Step 2  leading to Step 3. $\{i^1,\ldots,i^n\}\subset I^1$ is the set of cadets in $I^1$ who each
lose their tentative assignment $(b,t^0)$. For each cadet $i \in I^1\setminus \{i^1,\ldots,i^n\}$, 
finalize the assignment of cadet $i$ as $\phi_i^{BR}(\succ) = (b,t^0)$.

For each cadet $i \in J^{n}$ with one of the $n$ highest $\pi_b$-priorities in $J^{n}$, 
finalize the assignment of cadet $i$ as $\phi_i^{BR}(\succ) = (b,t^+)$.
Finalize the assignment of any remaining cadet as $\emptyset$.\medskip
\end{quote}

The key step in the procedure is Step 2 where it is determined how many of the $q^+_b$ positions are  to be awarded at the increased cost $t^+$. 
To determine this number, the BRADSO policy $\omega^+_b$ is used to check 
\begin{itemize}
\item[(1)] whether there is at least one cadet with a lower baseline priority $\pi_b$ than cadet $i^1$, who is willing to serve the increased cost $t^+$ and  
whose increased cost contract has higher priority under the BRADSO  policy $\omega_b^+$ than the base cost contract of cadet $i^1$; 
\item[(2)] whether there are at least two cadets each with a lower baseline priority $\pi_b$ than cadet $i^2$, who are each willing to serve the increased cost $t^+$ and  
whose increased cost contracts have higher priority under the BRADSO  policy $\omega_b^+$ than the base cost contract of cadet $i^2$; \\
\mbox{} $\vdots$\\
\item[($q^+_b$)] whether there are at least $q^+_b$ cadets each with a lower baseline priority $\pi_b$ than cadet $i^{q^+_b}$, who are each willing to serve the increased cost $t^+$ and  whose increased cost contracts have higher priority under the BRADSO  policy $\omega_b^+$ than the base cost contract of cadet $i^{q^+_b}$.
\end{itemize}
Once the number of positions awarded  through increased cost $t^+$ contracts is determined in this way, all other positions are 
assigned to the highest baseline priority cadets  as base cost contracts.  The increased cost contracts are awarded 
to the remaining highest baseline priority cadets who are willing to serve the increased cost $t^+$.\smallskip 

\begin{example} \textbf{(Mechanics of Mechanism $\phi^{BR}$)} \label{ex-mechanics}
There is a single branch $b$ with $q^0_b = 3$ and $q^+_b = 3$. There are eight cadets, with their set given as $I=\{i^1, i^2, i^3, i^4, i^5, i^6,  j^1, j^2\}$.  
The baseline priority order $\pi_b$ is given as
\[  i^6 \; \pi_b \;  i^5 \; \pi_b \;  i^4 \; \pi_b \;  i^3 \; \pi_b \;  i^2 \; \pi_b \;  i^1 \; \pi_b \;  j^1 \; \pi_b \;  j^2,
\]
and the BRADSO policy is the ultimate BRADSO policy $\overline{\omega}^+_b$. 
Cadet preferences are given as
\begin{eqnarray*}
(b,t^0) \; \succ_i \; (b,t^+) \; \succ_i \; \emptyset \qquad  && \mbox{for any } i \in \{i^1,i^3,i^5,j^1\}, \; \mbox{ and} \\
(b,t^0) \; \succ_i \; \emptyset  \; \succ_i \;  (b,t^+) \qquad && \mbox{for any }i\in \{i^2,i^4,i^6,j^2\}.
\end{eqnarray*}
We next run the procedure for the mechanism $\phi^{BR}$. \smallskip

\noindent \textbf{\textit{Step 0\/}}: There are three regular positions. The three highest $\pi_b$-priority cadets in the set $I$ are $i^6$, $i^5$, and $i^4$.
Let $I^0=\{i^4,i^5,i^6\}$, and finalize
the assignments of cadets in $I^0$ as $\phi^{BR}_{i^6}(\succ) = \phi^{BR}_{i^5}(\succ) = \phi^{BR}_{i^4}(\succ) = (b,t^0)$. \smallskip

\noindent \textbf{\textit{Step 1\/}}: There are three BRADSO-eligible positions. Three highest $\pi_b$-priority cadets in the set $I\setminus I^0$ are $i^3$, $i^2$, and $i^1$.
Let $I^1 = \{i^1,i^2,i^3\}$, and the tentative assignment of each cadet in $I^1$ is $(b,t^0)$.  There is no need to relabel the cadets since cadet $i^1$ is already the 
lowest $\pi_b$-priority cadet in $I^1$,  cadet $i^2$ is the 
second lowest $\pi_b$-priority cadet in $I^1$, and cadet $i^3$ is the highest $\pi_b$-priority cadet in $I^1$.

\noindent \textbf{\textit{Step 2.0\/}}: The set of cadets in $I\setminus(I^0 \cup I^1) = \{j^1,j^2\}$ for whom the assignment $(b,t^+)$ is acceptable is
$J^0 = \{j^1\}$. Since
\[ \underbrace{\big|\big\{j\in J^0 : (j,t^+) \; \overline{\omega}^+_b \; (i^1,t^0) \big\}\big|}_{=|J^0|=|\{j^1\}|=1}  \geq 1,
\]
we proceed to Step 2.1. \smallskip

\noindent \textbf{\textit{Step 2.1\/}}: Since $(b,t^+) \, \succ_{i^1} \, \emptyset$, we have  $J^1 = J^0 \cup \{i^1\} = \{i^1,j^1\}$.  Since
\[ \underbrace{\big|\big\{j\in J^1 : (j,t^+) \; \overline{\omega}^+_b \; (i^2,t^0) \big\}\big|}_{=|J^1|=|\{i^1,j^1\}|=2}  \geq 2,
\]
we proceed to Step 2.2. \smallskip

\noindent \textbf{\textit{Step 2.2\/}}: Since $\emptyset \, \succ_{i^2} (b,t^+)$, we have  $J^2 = J^1 = \{i^1,j^1\}$.  Since
\[ \underbrace{\big|\big\{j\in J^2 : (j,t^+) \; \overline{\omega}^+_b \; (i^3,t^0) \big\}\big|}_{=|J^2|=|\{i^1,j^1\}|=2}  = 2,
\]
we finalize Step 2 and proceed to Step 2.3. \smallskip

\noindent \textbf{\textit{Step 3\/}}: Step $2.2$ is the last sub-step of Step 2. Therefore two lowest $\pi_b$-priority cadets
in $I^1$, i.e cadets $i^1$ and $i^2$, lose their tentative assignments of $(b,t^0)$. In contrast, the only remaining cadet in the set 
$I^1 \setminus \{i^1,i^2\}$,  i.e cadet $i^3$ maintains her tentative assignment, which is finalized as  $\phi^{BR}_{i^3}(\succ) = (b,t^0)$.

The two highest priority cadets in $J^2$ are $i^1$ and $j^1$. Their assignments are finalized as  $\phi^{BR}_{i^1}(\succ) = \phi^{BR}_{j^1}(\succ) = (b,t^+)$.
Assignments of the remaining cadets $i^2$ and $j^2$ are finalized as $\emptyset$. The final allocation is:
\[ \phi^{BR}(\succ) = \left( \begin{array}{cccccccc}
i^1 & i^2 & i^3 & i^4 & i^5 & i^6& j^1 & j^2 \\
(b,t^+) & \emptyset & (b,t^0) & (b,t^0) & (b,t^0) & (b,t^0) & (b,t^+) & \emptyset
\end{array} \right).
\]
\mbox{} \hfill $\blacksquare$
\end{example}

Our first result shows that when there is a single branch the direct mechanism $\phi^{BR}$ is the only 
mechanism that satisfies our main desiderata. 

\begin{theorem} \label{thm:singlebranchcharacterization}
Suppose there is a single branch $b$. Fix a baseline priority order $\pi_b \in \Pi$ and a BRADSO policy $\omega^+_b \in \Omega^+_b$. 
A direct mechanism $\varphi$ satisfies
\begin{enumerate}
\item individual rationality, 
\item non-wastefulness, 
\item enforcement of the BRADSO policy, 
\item BRADSO-IC, and
\item has no priority reversals, 
\end{enumerate}
if and only if $\varphi = \phi^{BR}$. 
\end{theorem}

\subsection{Equilibrium Outcomes under the USMA-2020 Mechanism} \label{subsec:NE}

While the USMA-2020 mechanism is not a direct mechanism in general, when there is a single branch it can be
interpreted a direct mechanism. In this case, for any cadet $i \in I$ the first part of the strategy space $\cals_i = \calp \times 2^B$
becomes redundant, and the second part simply solicits whether branch $b$ is acceptable by cadet $i$ or not (analogous to a direct mechanism). 

Our next result shows that
when there is a single branch the truthful outcome of the direct mechanism $\phi^{BR}$ is 
the same as the unique Nash equilibrium outcome of the mechanism $\varphi^{2020}$. 

\begin{proposition} \label{prop:2021NashEqm}
Suppose there is a single branch $b$.  
Fix a baseline priority order $\pi_b \in \Pi$, a BRADSO policy $\omega^+_b \in \Omega^+_b$, and a preference profile 
$\succ \, \in \calq^{|I|}$. 
Then the strategic-form game induced by the mechanism $(\cals^{2020},\varphi^{2020})$ has a unique Nash equilibrium outcome that is
equal to the allocation $\phi^{BR}(\succ)$.\footnote{Using the terminology of the 
\textit{implementation theory\/}, this result can be alternatively stated as
follows: When there is a single branch, the mechanism $(\cals^{2020},\varphi^{2020})$ implements the allocation rule   $\phi^{BR}$ in 
Nash equilibrium. See \cite{maskin/sjostrom:02} and \cite{jackson:01} for surveys of implementation theory.}  
\end{proposition}

Caution is needed when interpreting Proposition \ref{prop:2021NashEqm}; 
if interpreted literally, this result can be misleading. 
What is  more consequential for Proposition \ref{prop:2021NashEqm}  is not the result itself, but rather
its proof which constructs  the equilibrium strategies of cadets.  The proof provides insight
into why the failure of BRADSO-IC, the presence of strategic BRADSO, and the presence of detectable priority reversals are
all common phenomena under the real-life implementation of the USMA-2020 mechanism
(despite the outcome equivalence suggested by Proposition \ref{prop:2021NashEqm}).

Given the byzantine structure of the Nash equilibrium strategies even with a single branch, it is perhaps not surprising that
reaching such a well-behaved Nash equilibrium is highly unlikely to be observed under the USMA-2020 mechanism. 
The following example illustrates  the knife-edge structure of the Nash equilibrium  strategies under the USMA-2020 mechanism. 

\begin{example} \textbf{(Knife-Edge Nash Equilibrium Strategies)} \label{knifeedge}

To illustrate how challenging it is for the cadets to figure out their best responses under the USMA-2020 mechanism, 
we present two scenarios.  The scenarios differ from each other minimally, but cadet best responses differ dramatically. 
Our first scenario is same as the one we presented in Example \ref{ex-mechanics}. \smallskip

\noindent \textbf{\textit{Scenario 1\/}}: There is a single branch $b$ with $q^0_b = 3$ and $q^+_b = 3$. 
There are eight cadets, $I=\{i^1, i^2, i^3, i^4, i^5, i^6,  j^1, j^2\}$.  
The baseline priority order $\pi_b$ is given as
\[  i^6 \; \pi_b \;  i^5 \; \pi_b \;  i^4 \; \pi_b \;  i^3 \; \pi_b \;  i^2 \; \pi_b \;  i^1 \; \pi_b \;  j^1 \; \pi_b \;  j^2 \quad \mbox{and}
\]
and the BRADSO policy is the ultimate BRADSO policy $\overline{\omega}^+_b$. 
Cadet preferences are 
\begin{eqnarray*}
(b,t^0) \; \succ_i \; (b,t^+) \; \succ_i \; \emptyset \qquad  && \mbox{for any } i \in \{i^1,i^3,i^5,j^1\}, \; \mbox{ and} \\
(b,t^0) \; \succ_i \; \emptyset  \; \succ_i \;  (b,t^+) \qquad && \mbox{for any }i\in \{i^2,i^4,i^6,j^2\}.
\end{eqnarray*}
Let $s^*$ be a Nash equilibrium strategy for Scenario 1 under the USMA-2020 mechanism. 
Recall that when there is a single branch $b$, the strategy space for each cadet $i\in I$ is simply $\cals_i =\{b,\emptyset\}$. 
We construct the Nash equilibrium strategies in several phases.\smallskip 

\textit{Phase 1\/}: Consider cadets $i^1$ and $j^1$, each of whom prefers the increased-cost assignment $(b,t^+)$ to remaining unmatched. 
Since there are six positions altogether and there are five higher $\pi_b$-priority cadets than either of these two cadets, 
at most one of them can receive a position (at any cost) unless each of them submit a strategy of $b$. 
And if one of them submit a strategy of $\emptyset$, the other one has a best response strategy of $b$ assuring a position
at the increased cost rather than remaining unmatched. Hence, $s^*_{i^1}=s^*_{j^1}=b$ at any Nash equilibrium.    

\textit{Phase 2\/}: Consider cadet $j^2$ who prefers remaining unmatched to the increased-cost assignment $(b,t^+)$. 
Since she is the lowest $\pi_b$-priority cadet, she cannot receive an assignment of $(b,t^0)$ regardless of her strategy. 
In contrast, she can guarantee remaining unmatched with a strategy of $s_{j^2} = \emptyset$. While this does not at this
point rule out a strategy of  $s_{j^2} = \emptyset$ at Nash equilibrium (just yet), it means $\varphi^{2020}_{j^2}(s^*)=\emptyset$. 

\textit{Phase 3\/}: Consider cadet $i^2$ who prefers remaining unmatched to the increased-cost assignment $(b,t^+)$. 
She is the fifth highest $\pi_b$-priority cadet, so she secures a position if she submits a strategy of $s_{i^2} = b$, 
but the position will have to be at the increased price $t^+$, since the lowest $\pi_b$-priority cadet $j^2$ is remaining unmatched  from Phase 2,
and therefore  there cannot be three cadets with lower $\pi_b$-priority who receive an assignment of $(b,t^+)$.
But since cadet $j^2$  prefers remaining unmatched to the increased-cost assignment $(b,t^+)$, she cannot receive an assignment
of $(b,t^+)$ at Nash equilibria. 
Hence, her Nash equilibrium strategy is $s^*_{i^2}=\emptyset$, and her Nash equilibrium assignment is $\varphi^{2020}_{i^2}(s^*)=\emptyset$. 

\textit{Phase 4\/}: Consider the remaining cadets $i^3$, $i^4$, $i^5$ and $i^6$. 
Since cadets $i^2$ and $j^2$ have to remain unmatched (from Phases 2 and 3) at Nash equilibria,
they each receive a position at Nash equilibrium. Since only the two cadets $i^1$ and $j^1$ from Phases 1-3 have Nash equilibrium
strategies of $b$, the lowest $\pi_b$-priority cadet of the four cadets $i^3$, $i^4$, $i^5$, $i^6$ who submit a strategy of $b$
receives an assignment of $(b,t^+)$. But this cannot happen at Nash equilibria since that particular cadet can instead submit a strategy of $\emptyset$
receiving a more preferred assignment of $(b,t^0)$. Hence, $s^*_i = \emptyset$ and $\varphi^{2020}_{i}(s^*)= (b,t^0)$ for any $i\in \{i^3,i^4,i^5,i^6\}$. 

The unique Nash equilibrium strategy $s^*$ and its Nash equilibrium outcome $\varphi^{2020}(s^*)$ for Scenario 1 are given as:
\[  \begin{array}{lcccccccc}
\mbox{Cadet} & i^1 & i^2 & i^3 & i^4 & i^5 & i^6& j^1 & j^2 \\
\hline\mbox{Nash equilibrium strategy} & b & \emptyset & \emptyset &\emptyset &\emptyset & \emptyset & b &\emptyset\\
\mbox{Nash equilibrium assignment} & (b,t^+) & \emptyset & (b,t^0) & (b,t^0) & (b,t^0) & (b,t^0) & (b,t^+) & \emptyset
\end{array}  \smallskip 
\]
Scenario 1 involves  \textit{BRADSO-IC\/} failures for cadets $i^3$ and $i^5$ whose Nash equilibrium strategies
force them into hiding their willingness  to serve the increased cost $t^+$.  Any deviation from her Nash equilibrium strategy
by truthfully declaring her willingness to serve the increased cost $t^+$ will result in an detectable priority reversal for cadet $i^5$.

\smallskip

\textbf{\textit{Scenario 2\/}}: This scenario differs from Scenario 1 in only the preferences of the lowest $\pi_b$-priority cadet $j^2$
and nothing else. Thus,  cadet preferences for this scenario are given as:
\begin{eqnarray*}
(b,t^0) \; \succ'_i \; (b,t^+) \; \succ'_i \; \emptyset \qquad  && \mbox{for any } i \in \{i^1,i^3,i^5,j^1,j^2\}, \; \mbox{ and} \\
(b,t^0) \; \succ'_i \; \emptyset  \; \succ'_i \;  (b,t^+) \qquad && \mbox{for any }i\in \{i^2,i^4,i^6\}.
\end{eqnarray*}
Let $s'$ be a Nash equilibrium strategy for Scenario 2 under the USMA-2020 mechanism.  \smallskip 

\textit{Phase 1\/}: Identical to Phase 1 for Scenario 1, and thus $s'_{i^1}=s'_{j^1}=b$ at any Nash equilibrium.   

\textit{Phase 2\/}: Consider cadet $i^2$ who prefers remaining unmatched to the increased-cost assignment $(b,t^+)$, 
and cadets $i^3$ and $j^2$, each of whom prefers the increased-cost assignment $(b,t^+)$ to remaining unmatched. 
Since (i) there are six positions altogether,  (ii) three cadets with higher $\pi_b$-priority  than  each one of $i^2,i^3,$ and $j^2$,
and (iii) $s'_{i^1}=s'_{j^1}=b$ from Phase 1, at most one of  the cadets $i^2,i^3,j^2$ can receive an assignment of $(b,t^0)$ if any. 
Therefore, submitting a strategy of $s_{i^3} = \emptyset$ is a best response for cadet $i^3$ only if both cadets $i^2$ and $j^3$
also submit a strategy of $\emptyset$ each. But this cannot happen in Nash equilibria, since it gives cadet $j^2$ a profitable
deviation by submitting a strategy of $s_{j^2} = b$ and jumping ahead of cadets $i^2$ and $i^3$ securing her a position. 
Hence $s'_{i^3} = b$ and $\varphi^{2020}_{i^3}(s')= (b,t^+)$.  When cadet $i^3$ joins 
the  two cadets from Phase 1 each also submitting a strategy of $b$,  this assures that exactly three positions
will be assigned at the increased cost $t^+$.  Therefore a strategy of f $s_{i^2} = b$ assures 
assures cadet $i^2$ an assignment of $(b,t^+)$, which cannot happen at Nash equilibrium. 
Therefore, $s'_{i^2} = \emptyset$ and $\varphi^{2020}_{i^2}(s')= \emptyset$. 
This not only assures that $\varphi^{2020}_{i^3}(s')= \varphi^{2020}_{i^1}(s') =  \varphi^{2020}_{j^1}(s')=(b,t^+)$,  but
it also means that $s'_{j^2} = b$ at Nash equilibrium, for otherwise with two lower $\pi_b$-priority cadets with strategies
of $\emptyset$, cadet $i^3$ would have an incentive to deviate himself and receiving the position at the base cost rather than the increased cost. 
  
\textit{Phase 3\/}:  Consider the remaining cadets $i^4$, $i^5$ and $i^6$. 
Of all lower $\pi_b$-priority cadets, only the cadet $i^2$ and has  Nash equilibrium
strategies of $\emptyset$ from Phases 1 and 2.  Therefore
the lowest $\pi_b$-priority cadet of the three cadets  $i^4$, $i^5$, $i^6$ who submit a strategy of $\emptyset$
receives an assignment of $\emptyset$. But this cannot happen at Nash equilibria since that particular cadet can instead submit a strategy of $b$
and receive a more preferred assignment of $(b,t^0)$ since three lower $\pi_b$-priority cadets already receive an assignment of $(b,t^+)$ each from
Phase 2. Therefore, regardless of their preferences $s'_{i^4} = s'_{i^5} = s'_{i^6} = b$, and 
$\varphi^{2020}_{i^4}(s')= \varphi^{2020}_{i^5}(s') = \varphi^{2020}_{i^6}(s') (b,t^0)$. 

The unique Nash equilibrium strategy $s'$ and its Nash equilibrium outcome $\varphi^{2020}(s')$ for Scenario 2 are given as:
\[  \begin{array}{lcccccccc}
\mbox{Cadet} & i^1 & i^2 & i^3 & i^4 & i^5 & i^6& j^1 & j^2 \\ \hline
\mbox{Nash equilibrium strategy} & b & \emptyset & b & b & b & b & b & b\\
\mbox{Nash equilibrium assignment} & (b,t^+) & \emptyset & (b,t^+) & (b,t^0) & (b,t^0) & (b,t^0) & (b,t^+) & \emptyset
\end{array}  \smallskip 
\]
Not only does the Nash equilibrium strategies of cadets $i^4$ and $i^6$ involve
 strategic BRADSO in Scenario 2 and they have to  declare willingness  to serve the increased cost $t^+$ even though
 under their true preferences they do not, but any deviation from this Nash equilibrium strategy
by declaring their unwillingness to serve the increased cost $t^+$  will result in detectable priority reversals for both cadets. 

Another key insight from this example is the dramatic difference between the Nash equilibrium strategies due to one minor
change in the underlying economy, a preference change in the lowest base priority cadet. 
This minor change only affects the assignment of cadet $i^3$ by changing it from $(b,t^0)$ to $(b,t^+)$. 
It also changes the Nash equilibrium strategy of not only  cadet $i^3$, and also all other higher $\pi_b$-priority cadets $i^4, i^5,$ and $i^6$. 
Moreover, in addition to BRADSO-IC failures and the presence of strategic BRADSO
under Nash equilibria,  any deviation from these strategies result in detectable priority reversals.
The fragility of our equilibrium strategies provides us intuition on the prevalence of these phenomena under the USMA-2020 mechanism. 
\mbox{} \hfill $\blacksquare$
\end{example}

Example \ref{knifeedge} shows that while the failure of BRADSO-IC and the presence of strategic BRADSO
can be observed at Nash equilibria  of the USMA-2020 mechanism,  
the presence of detectable priority reversals is out-of-equilibrium behavior under complete information
when there is a single branch.  Our next example shows that if the complete information assumption is relaxed  
there can also be detectable priority reversals in the
Bayesian equilibria of the USMA-2020 mechanism.

\medskip

\begin{example} \textbf{(Detectable Priority Reversals at Bayesian Equilibria)}  \label{Bayesian}

Suppose there is a single branch $b$ with $q^0_b=q^+_b=1$ and three cadets $i_1, i_2,$ and $i_3$. 
The baseline priority order $\pi_b$ is such that
\[  i_1 \; \pi_b \; i_2 \; \pi_b \; i_3,  
\]
and the BRADSO policy $\omega_b^+$ is the ultimate BRADSO policy $\overline{\omega}^+_b$.

Each cadet has a utility function that is drawn from a distribution with the following two elements, $u$ and $v$, where:
\[ u(b,t^0) = 10, \; u(\emptyset) = 8, \; u(b,t^+)=0, \quad \mbox{ and } \quad v(b,t^0) = 10, \; v(b,t^+) = 8, \; v(\emptyset)=0.
\]
Let us refer to cadets with a utility function $u(.)$ as type 1 and cadets with a utility function $v(.)$ as type 2. 
All cadets have a utility of 10 for their first choice assignment of $(b,t^0)$,  
a utility of 8 for their second choice assignment, and  a utility of 0 for their last choice assignment. 
For type 1 cadets, the second choice is remaining unmatched whereas for type 2 cadets the second choice
is receiving a position at the increased cost $t^+$. 
Suppose each cadet can be of the either type with a probability of 50 percent, and they are all expected utility maximizers.  

The unique Bayesian Nash equilibrium $s^*$ under the incomplete information game induced by the
USMA-2020 mechanism is, for any cadet $i \in \{i_1,i_2,i_3\}$,
\[  s^*_i = \left\{ \begin{array}{cl}
\emptyset & \mbox{ if cadet } i \mbox{ is of type 1,\;  and}\\
b & \mbox{ if cadet } i \mbox{ is of type 2.}   \end{array}
\right.
\]
That is, truth-telling is the unique  Bayesian Nash equilibrium strategy for each cadet. 
However, this unique Bayesian Nash equilibrium strategy results in  detectable priority reversals whenever either
\begin{enumerate}
\item  cadet $i_1$ is of type 1 and cadets $i_2, i_3$ are of type 2, or
\item  cadet $i_1$ is of type 2 and cadets $i_2, i_3$ and are of type 1. 
\end{enumerate}
While cadet $i_2$ receives a position at the base cost $t^0$ in both cases, 
the  highest baseline priority cadet $i_1$ remains unassigned in the first case
and receives a position at the increased cost $t^+$ in the second case. 
\mbox{} \hfill $\blacksquare$
\end{example}








\section{Multi-Branch Analysis and the 2020 Reform} \label{sec:2020reform}



To resolve the problems with the USMA-2020 mechanism, most notably its failure of BRADSO-IC, the possibility of strategic BRADSO, and the 
resulting detectable priority reversals, the Army established a partnership with a team of market designers to design the USMA-2021 mechanism.
Critical to achieving these objectives was the Army's decision to permit cadets in the Class of 2021 to submit preferences over branch-cost pairs. This decision was aided by evidence from a 
cadet survey that mitigated concerns that rating branch-cost pairs would be overly complex or unnecessary. 
Indeed, some of the cadets indicated the need for a system that would allow
them to rank order branch-cost pairs.  One cadet wrote:
\begin{quote}
``
[$\ldots$] \textit{I believe that DMI (Department of Military Instruction) could elicit a new type of ranking list. 
Within my proposed system, people could add to the list of 17 branches BRADSO slots and rank them within that list. 
For example: AV (Aviation) $>$ IN (Infantry) $>$ AV:B (Aviation with BRADSO). While this may be a transmutation of the ``alternate system,'' 
I believe many cadets could utilize this system as it is the case that people view branch without ADSO and BRADSO slots are considered almost different things}.''
\end{quote}
More generally, the survey revealed that more than twice as many cadets prefer a mechanism that allows them to submit preferences over branch-cost pairs relative to a mechanism that requires them to submit preferences over branches and BRADSOs separately as in the USMA-2006 and USMA-2020 mechanism.\footnote{A question on the survey asked cadets whether they prefer a mechanism that allows them to submit preferences over branch-cost pairs, like the COM-BRADSO mechanism, or a mechanism that requires them to submit preferences over branches alone while separately indicating willingness to BRADSO for each branch, like the USMA-2006 and USMA-2020 mechanism.  \fig{Appendix \ref{survey}}  shows that
50 percent of respondents preferred the mechanism that permitted ranking branch-cost pairs, 21 percent preferred the mechanism without the option to rank branch-cost pairs, 24 percent were indifferent, and 5 percent did not understand.}

\subsection{COM-BRADSO Mechanism and its Characterization} \label{subsec:COSM-2021}

Unlike its predecessors USMA-2006 and USMA-2020 mechanisms, 
using the contract terminology is more convenient to describe the \textit{COM-BRADSO\/} mechanism,
adopted for the USMA and the ROTC Class of 2021.
We need the following additional terminology. 

A set of contracts $X \subset \calx_b$ is \textbf{viable} for branch $b\in B$, if  for any $i \in I$, 
\[ (i,b,t^+) \in X \; \implies \; (i,b,t^0) \in X.
\] 
That is,  the increased cost contract of a cadet can be available in a viable set of contracts only together with its base cost version. 

Given a baseline priority order $\pi_b$, let $\omega^0_b$ be the resulting  \textbf{native priority order} on $I\times T$
that is uniquely defined by the following two conditions:
\begin{enumerate}
\item for any $i,j \in I$ and $t,s \in T$, 
\[ (i, t) \; \omega^0_b \; (j,s) \quad \iff \quad i \; \pi_b \; j, \quad \mbox{ and }\]
\item for any $i \in I$, 
\[  (i, t^0) \; \omega^0_b \; (i,t^+). \] 
\end{enumerate}
The native linear order $\omega^0_b$ simply mirrors the baseline priority order $\pi_b$, and 
prioritizes  cadet-cost pairs in $I \times T$ as the cadet of the pair is prioritized under
the baseline priority order $\pi_b$, while giving higher priority to
the base cost $t^0$ over the increased cost $t^+$ for any given cadet. \medskip

Under the COM-BRADSO mechanism, each  branch $b\in B$ relies on the following choice rule to select a set of contracts from 
any set of contracts viable for branch $b$. 

\begin{quote}
 
\noindent \textbf{Choice Rule}  {\boldmath $\calc^{BR}_b$}

\noindent For any set of contracts $X  \subset \calx_b$ that is viable for branch $b$, 

\noindent \textbf{Step 1.} If there are less than $q^0_b$ contracts in $X$ with distinct cadets, then
choose all contracts in $X$ with the base cost $t_0$ and terminate the procedure. 
In this case $\calc^{BR}_b(X) = \big\{x\in X : \t(x)=t^0\big\}$. 

Otherwise, let $X_1$ be the set of $q^0_b$ highest $w^0_b$-priority contracts in $X$ with distinct cadets.\footnote{Since $X$ is
viable and $\omega_b^0$ is the native priority order, all contracts in $X_1$ has the base cost $t^0$.} 
Pick contracts in $X_1$ and proceed to Step 2. \smallskip

\noindent \textbf{Step 2.}  The set of contracts under consideration for this step is  
\[ Y=  \Big\{x\in X\setminus X_1 : \big(\i(x),b,t^0\big)\not\in X_1\Big\}.\]
 
If there are less than $q^+_b$ contracts in $Y$ with distinct cadets, then
pick all contracts in $Y$  with the base cost $t^0$ and terminate the procedure. 
In this case 
 $\calc^{BR}_b(X) = X_1 \cup  \big\{x\in Y : \t(x)=t^0\big\}$. 

Otherwise, let $X_2$ be the set of $q^+_b$ highest $w^+_b$-priority contracts in $Y$ with distinct cadets. 
Pick contracts in $X_2$ and terminate the procedure. 
In this case 
 $\calc^{BR}_b(X) = X_1 \cup X_2$. \medskip
\end{quote}

\noindent Intuitively, the choice rule  $\calc^{BR}_b$  relies on the native priority order $\omega^0_b$ for the first $q^0_b$ positions, 
and on the BRADSO policy  $\omega^+_b$ for the last $q^+_b$ positions. 

Observe that all increased cost contracts are selected in Step 2 of the choice rule $\calc^{BR}_b$. 
Therefore, an increase in the BRADSO cap means using the native priority order $\omega^0_b$ for fewer positions
and the  BRADSO policy  $\omega^+_b$ for more positions, thereby weakly increasing  the number of increased-cost contracts selected 
by the choice rule $\calc^{BR}_b$. 
Moreover, since the increased-cost contracts receive weakly higher priorities when the BRADSO policy becomes more effective at branch $b$, 
such a change in the BRADSO policy also weakly increases the number of increased-cost contracts selected by the choice rule $\calc^{BR}_b$.
We state these two observations in the following result. 
\begin{proposition} \label{prop:BRADSOcomparativestatics}
For any branch $b\in B$ and  set of contracts $X\subset \calx_b$ viable for branch $b$,
\begin{enumerate}
\item the higher the BRADSO cap $q^+_b$ is the weakly higher is the number of increased cost contracts accepted under $\calc^{BR}_b(X)$, and
\item the more effective the BRADSO policy $\omega^+_b$ is the weakly higher is the number of increased cost contracts accepted under $\calc^{BR}_b(X)$.
\end{enumerate}
\end{proposition}

We  are ready to introduce the mechanism central to the Army's 2021 Branching reform. 
For a given list of BRADSO policies $(\omega^+_b)_b\in B$, 
let $\calc^{BR} = (\calc^{BR}_b)_{b\in B}$ denote the list of branch-specific choice rules defined above. 
COM-BRADSO mechanism is a direct mechanism where each cadet reports her preferences over $B\times T \cup \{\emptyset\}$. 
Therefore, the strategy space for each cadet $i \in I$ is
\[ \cals^{COM-BR}_i = \calq.
\]
The outcome function $\phi^{COM-BR}$ for the COM-BRADSO mechanism is given through the following procedure. 

\begin{quote}

\noindent \textbf{Cumulative Offer Mechanism under}  {\boldmath $\calc^{BR}$}\smallskip

\noindent Fix a linear order of cadets $\pi \in \Pi$.\footnote{By Kominers and S\"{o}nmez (2016), 
the outcome is independent of this linear order. Nevertheless, one natural linear order is OML, which is
also used in the construction of branch priorities.} 
For a given profile of cadet preferences $\succ = (\succ_i)_{i\in I} \in \calq^{|I|}$, 
cadets propose their acceptable contracts to branches in a sequence of steps $\ell = 1, 2, \ldots $:\smallskip 

\noindent \textbf{Step 1.}  Let $i_1\in I$ be the highest $\pi$-ranked cadet who has an acceptable contract. 
Cadet $i_1 \in I$ proposes her most preferred contract $x_1 \in \calx_{i_1}$ to branch $\b(x_1)$. 
Branch $\b(x_1)$ holds $x_1$ if $x_1 \in  \calc^{BR}_{\b(x_1)}\big(\{x_1\}\big)$ and rejects $x_1$ otherwise. Set 
$A^2_{\b(x_1)} = \{x_1\}$ and set $A^2_{b'}=\emptyset$ for each $b'\in B\setminus\{\b(x_1)\}$; 
these are the sets of contracts available to branches at the beginning of step 2.\smallskip

\noindent \textbf{Step} {\boldmath $\ell$.} 
Let $i_{\ell} \in I$ be the  highest $\pi$-ranked cadet  for whom no contract is currently held by any branch,
and let $x_{\ell} \in \calx_{i_{\ell}}$ be her most preferred acceptable contract that has not yet been rejected.  
Cadet $i_{\ell}$ proposes contract   $x_{\ell}$ to branch $\b(x_{\ell})$. 
Branch $\b(x_{\ell})$ holds the contracts in $\calc^{BR}_{\b(x_{\ell})}\big(A^{\ell}_{\b(x_{\ell})}\cup \{x_{\ell}\}\big)$ 
and rejects all other contracts in $A^{\ell}_{\b(x_{\ell})}\cup \{x_{\ell}\}$. 
Set  $A^{\ell +1}_{\b(x_{\ell})} = A^{\ell}_{\b(x_{\ell})}\cup \{x_{\ell}\}$ and 
set $A^{\ell +1}_{b'} = A^{\ell}_{b'}$ for each $b'\in B\setminus\{\b(x_{\ell})\}$; 
these are the sets of contracts available to branches at the beginning of step $\ell +1$.\smallskip

The procedure terminates at a step when either no cadet remains with an acceptable contract that has not been rejected, or
when no contract is rejected. All the contracts on hold in the final step are finalized as the outcome $\phi^{COM-BR}(\succ)$ of the 
COSM-2021 mechanism.  \smallskip
\end{quote}

\noindent COM-BRADSO mechanism is a generalization of the COSM mechanism proposed by \cite{sonmez/switzer:13}
for the case of  the ultimate BRADSO policy $\overline{\omega}^+_b$, and a
special case of the cumulative offer mechanism for the matching with slot-specific priorities
model by \cite{kominers/sonmez:16}. 

Our final and main theoretical result shows COM-BRADSO is the only mechanism that satisfies all our desiderata. 

\begin{theorem} \label{cosm}
Fix a profile of baseline priority orders $(\pi_b)_{b\in B} \in \Pi$ and a profile of BRADSO policies $\big(\omega^+_b\big)_{b\in B} \in \prod_{b\in B}\Omega^+_b$. 
A direct mechanism $\varphi$ satisfies
\begin{enumerate} 
\item \textit{individual rationality\/},  
\item \textit{non-wastefulness\/},  
\item \textit{enforcement of the BRADSO policy\/},  
\item \textit{strategy-proofness\/}, and
\item \textit{has no priority reversals\/}, 
\end{enumerate}
if and only if $\varphi$ is the COM-BRADSO mechanism $\phi^{COM-BR}$. 
\end{theorem}

Apart from singling out the COM-BRADSO mechanism as the unique mechanism that satisfies our desiderata, 
to the best of our knowledge Theorem \ref{cosm} is the first joint characterization of an allocation mechanism (i.e. the cumulative offer
process) together with a specific choice rule $\calc^{BR}_b$ for each branch $b\in B$.\footnote{ Characterizations 
of the cumulative offer process are available in the literature  for choice rules that are fixed or that satisfy various criteria. See, for example, \cite{hatfield/kominers/westkamp:21} and \cite{hirata/kasuya:17}.    
Similarly, there are characterization of choice rules in single institution environments which do not involve the cumulative offer process (see, e.g., \cite{echenique/yenmez:15} and \cite{imamura:20}). }
In our application, in addition to the standard axioms of individual rationality,  
non-wastefulness, lack of priority reversals, 
and strategy-proofness, the axiom of enforcement of the BRADSO policy (which directly
formulates the Army policy) uniquely identifies the cumulative offer process under a specific choice rule  $\calc^{BR}_b$.

We finalize our theoretical analysis with a straightforward observation. 
Since (i) a quasi-direct mechanism becomes a direct mechanism with only a single branch,  and 
(ii) strategy-proofness implies BRADSO-IC in this environment, 
Theorems \ref{thm:singlebranchcharacterization} and \ref{cosm} immediately imply the following result. 

\begin{corollary} \label{corollary:phi=cosm}
Suppose there is a single branch $b$. Fix a baseline priority order $\pi_b \in \Pi$ and a BRADSO policy $\omega^+_b \in \Omega^+_b$. 
Then, for any preference profile $\succ \, \in \calq^{|I|}$, 
\[  \phi^{COM-BR}(\succ)  = \phi^{BR}(\succ).  
\]
\end{corollary}

\subsection{Field Evidence on COM-BRADSO}

\begin{figure}[htp]
    \begin{center}
    \caption{\textbf{BRADSO Ranking Relative to Non-BRADSO Ranking by Class of 2021}\\}
    \label{fig:nonconsecutive_bradso}
    \includegraphics[scale = 1.0]{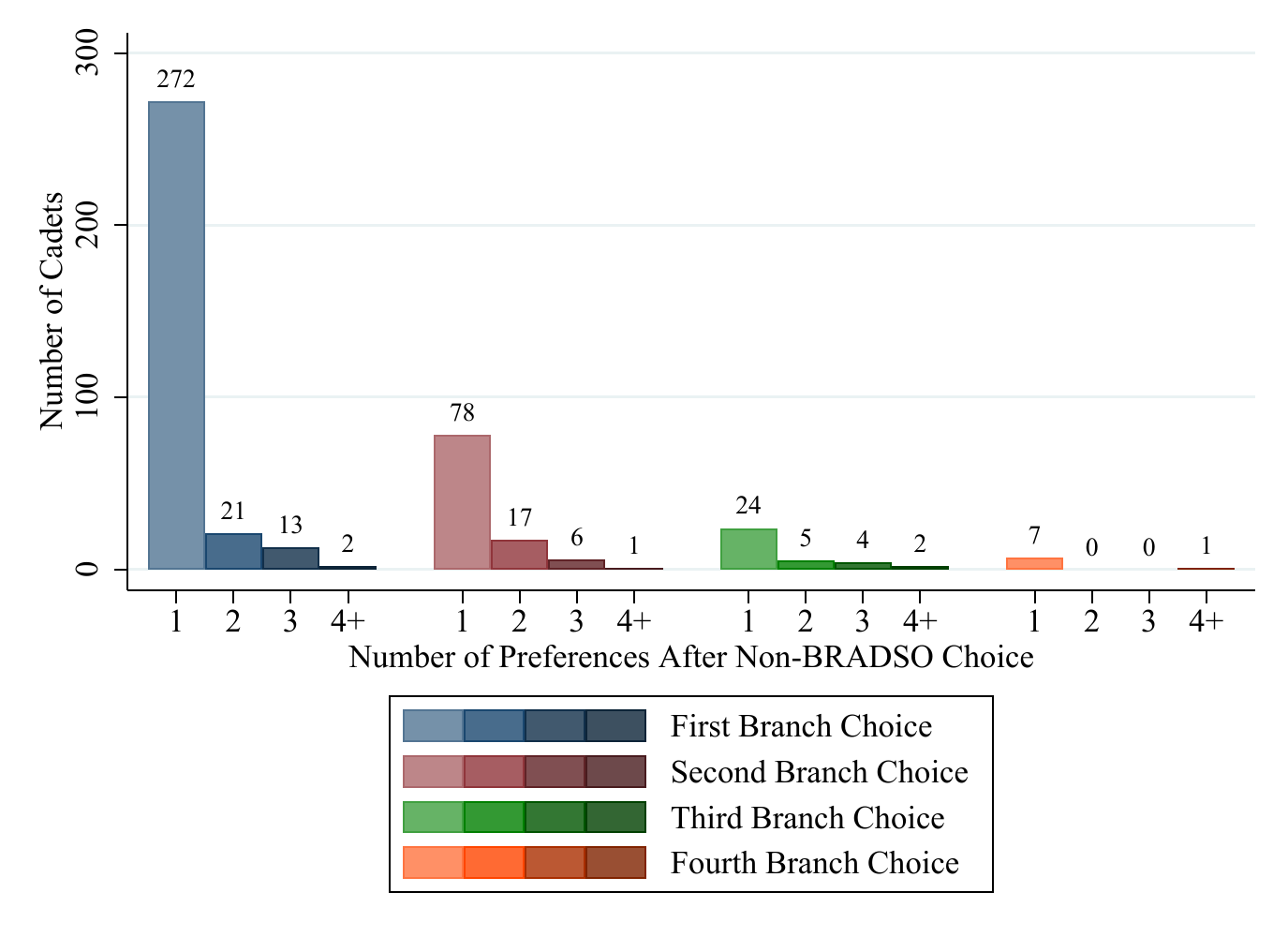}
    \end{center}
\textbf{Notes.} This figure reports where in the preference list a branch is ranked with BRADSO relative to where it is ranked without BRADSO.  A value of 1 (2 or 3) indicates that the branch is ranked with BRADSO immediately after (two places or three places after, respectively) the branch is ranked at base cost.  4+ means that the a branch is ranked with BRADSO four or more choices after the branch is ranked at base cost.
\end{figure}

Relative to USMA-2020, a key benefit of the COM-BRADSO mechanism is the ability for cadets to submit preferences over branch-cost pairs. As described above, survey results from the Class of 2020 indicated that about half of the cadets prefer a mechanism that permits them to submit preferences over branch-cost pairs over a mechanism that does not. Preference data from the Class of 2021 confirms that this flexibility was used by cadets.  \fig{Figure \ref{fig:nonconsecutive_bradso}} provides details on the extent to which cadets did not rank a branch with increased cost immediately after
the branch at base cost. For each of 994 cadet first branch choices, 272 cadets rank that branch with BRADSO as their second choice and 36 cadets rank that 
branch with BRADSO as their third choice or lower.  These 36 cadets would not have been able to express this preference under the message space of a quasi-direct mechanism 
like the USMA-2006 mechanism
or the USMA-2020 mechanism.  When we consider the next branch on a cadet's rank order list, cadets also value the flexibility of the new mechanism.  
For the branch that appears next on the rank order list, 78 cadets rank that branch with BRADSO as their immediate next highest choice and 24 cadets rank that branch
with BRADSO two or more places below on their rank order list.  These 24 cadets also would not have been able to express this preference under a quasi-direct mechanism.  

The fact that COM-BRADSO is a strategy-proof mechanism which elicits rankings over branch-price pairs allows us to compare outcomes under the USMA-2006 and USMA-2020
mechanisms with knowledge of the underlying branch-price preference relationship. 
In \fig{Figure \ref{fig:failures}}, we could only measure detectable priority reversals and not all priority reversals.  To measure all priority
reversals, 
we use preferences over branch-price pairs under COM-BRADSO to construct a truthful strategy denoted $s_i = (P_i,B_i)$ under a quasi-direct mechanism by using the branch rank ordering for $P_i$ and assuming that if a cadet ever expresses a willingness to BRADSO at a branch, then the cadet is willing to BRADSO under $B_i$.   Taking this constructed strategy as input, we then simulate the USMA-2006 and USMA-2020 mechanism
using the branch capacities and priorities from the Class of 2021.   Under the USMA-2006 mechanism simulation,
there are 29 priority reversals and 20 are detectable priority reversals.  Under the USMA-2020 mechanism simulation, there are 204 priority reversals and 197 are detectable priority reversals.  This
suggests that the detectable priority reversals in practice likely constitute the major of priority reversals for the Classes of 2014-2019, which used the USMA-2006 mechanism,
and for the Class of 2020, which used the USMA-2020 mechanism.

\begin{figure}[htp]
    \begin{center}
    \caption{\textbf{USMA-2006 and USMA-2020 Mechanism Performance under Truthful Strategies Simulated from Preference Data from Class of 2021}\\}
    \label{fig:2021sim}
    \includegraphics[scale = 1.0]{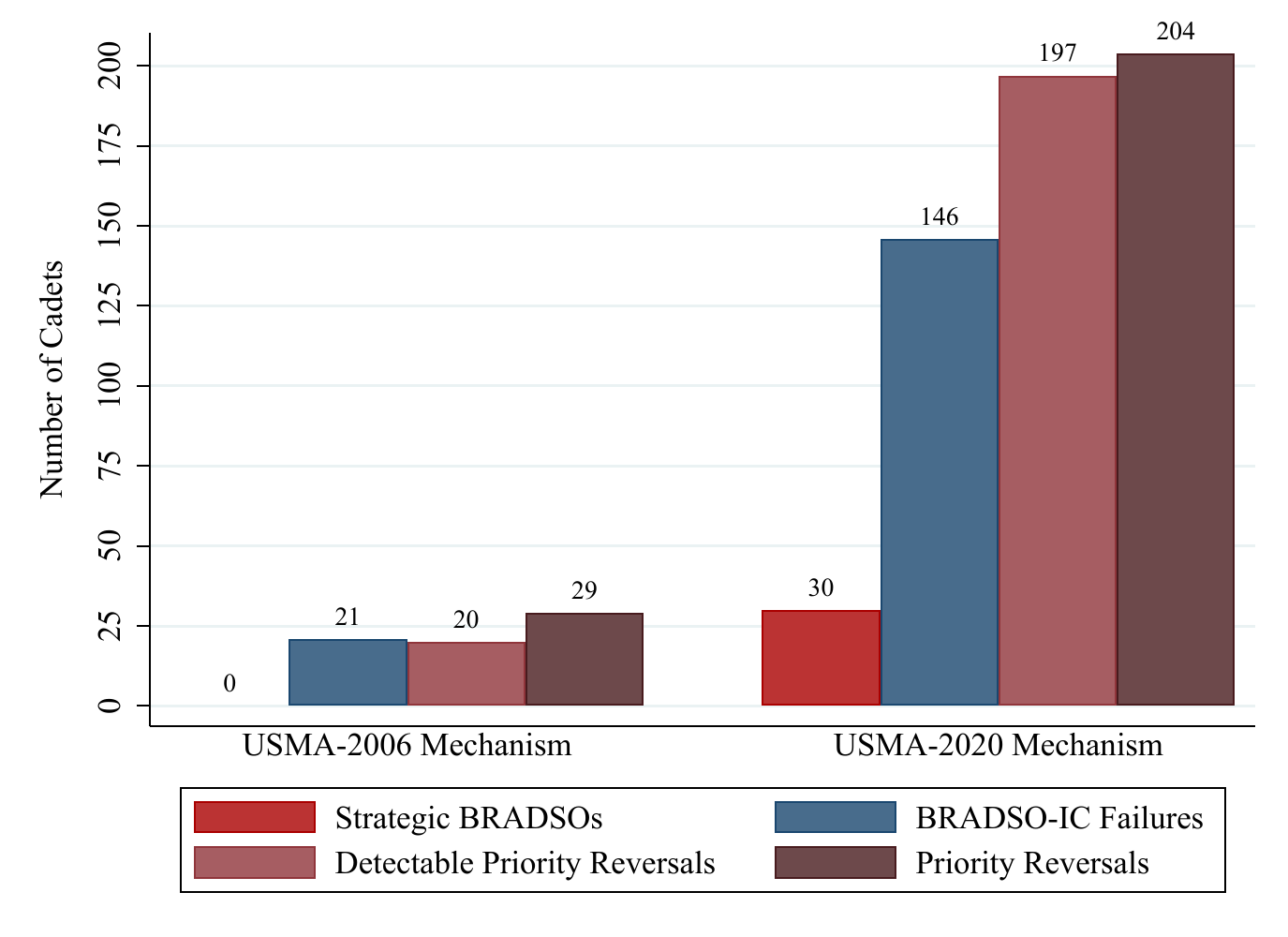}
    \end{center}
    \textbf{Notes.} 
    USMA used the strategy-proof COM-BRADSO mechanism for the Class of 2021. 
This figure uses data  from the Class of 2021 on cadet preferences, branch priorities, and branch capacities to simulate the outcomes of the mechanisms USMA-2006 and
USMA-2020. Since the strategy spaces of the mechanisms USMA-2006 and USMA-2020 differ from that of the mechanism COM-BRADSO, cadet strategies 
that correspond to truthful branch-preferences and BRADSO willingness are 
are simulated from cadet preferences over branch-cost pairs under the COM-BRADSO mechanism. 
    Truthful strategies for the mechanisms USMA-2006 and USMA-2020 are constructed from Class of 2021 preferences by assuming that a preference indicating willingness to BRADSO at a branch means the cadet's strategy under the USMA-2006 and USMA-2020 mechanisms has her willing to BRADSO.   Strategic BRADSOs, BRADSO-IC Failures, and Detectable Priority Reversals are defined in \fig{Figure \ref{fig:failures}}.  To compute Priority Reversals, we compare a cadet's outcome in the USMA-2006 or USMA-2020 mechanism to a cadet's preference submitted under the COM-BRADSO mechanism. If a cadet prefers a higher ranked choice and has higher priority over a cadet who is assigned that choice, then the cadet is part of a Priority Reversal.  
\end{figure}

Using truthful strategies to evaluate the USMA-2006 and USMA-2020 mechanism, \fig{Figure \ref{fig:2021sim}} shows that there are nearly seven times as many BRADSO-IC failures under the USMA-2020 mechanism compared to the USMA-2006 mechanism (146 vs. 21) and seven times as many priority reversals under the USMA-2020 mechanism compared to the USMA-2006 mechanism (204 vs. 29).  This pattern of 
behavior suggests that the comparison reported in \fig{Figure \ref{fig:failures}} potentially understates the dramatic increase in BRADSO-IC failures and priority reversals stemming from
the adoption of the USMA-2020 mechanism because that comparison was based on strategies in the mechanism and not underlying cadet preferences.

\begin{figure}[htp]
\begin{center}
\caption{\textbf{USMA-2020 Mechanism Performance Under Indicative and Final Strategies}}
\label{fig:usma2020indicative}
    \includegraphics[scale = 1.0]{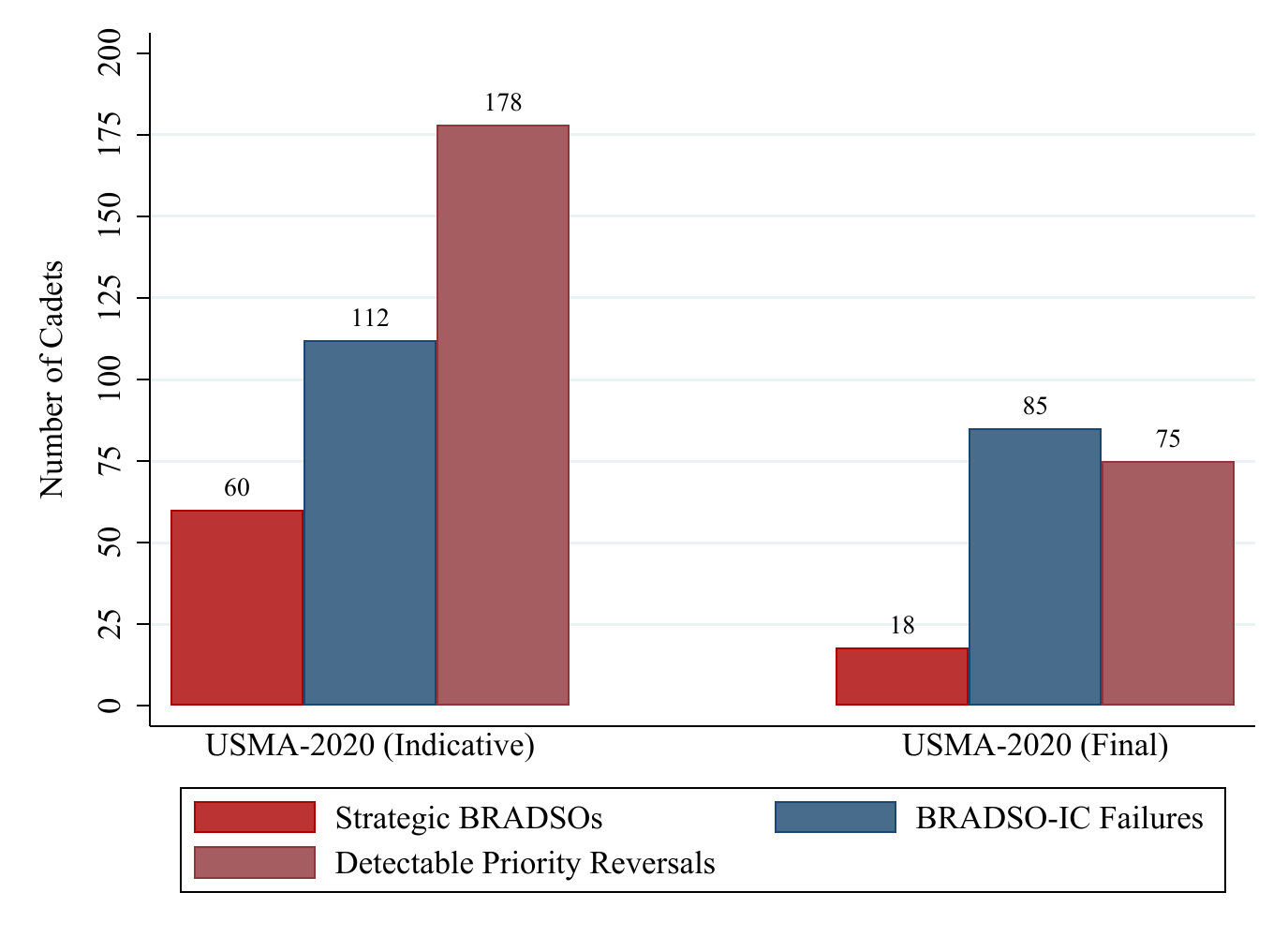}
\end{center}
\textbf{Notes.} This figure reports on the number of Strategic BRADSOs, BRADSO-IC failures, Detectable Priority Reversals, and Priority Reversals under indicative strategies submitted
in a dry-run of the USMA-2020 mechanism and final strategies of the USMA-2020 mechanism for the Class of 2020. 
\end{figure}

One reason the comparison between USMA-2006 and USMA-2020 in \fig{Figure \ref{fig:failures}} 
is not as striking as the comparison in \fig{Figure \ref{fig:2021sim}} is that, 
as we have presented in Section \ref{sec:shortcomings2020}, many cadets 
were well-aware of the necessity to strategically make their BRADSO choices under the USMA-2020 mechanism.  
Our analysis in Section \ref{sec:singlebranch} illustrates the perverse incentives in the USMA-2020
mechanism.  For the Class of 2020, a dry-run of the mechanism where cadets submitted indicative rankings of branches and learned about their assignment  took place.
After observing their dry-run assignment, cadets were allowed to submit a final set of rankings under USMA-2020, 
and therefore had the opportunity to revise their strategies in response
to this feedback.   \fig{Figure \ref{fig:usma2020indicative}} tabulates strategic BRADSOs, BRADSO-IC failures, and detectable priority
reversals under indicative and final preferences.  Final preferences result in fewer strategic BRADSOs, BRADSO-IC failures, and detectable
priority reversals.  This pattern is consistent with some cadets responding to the dry-run by ranking branch choices in response to these issues.

\begin{figure}[htp]
\begin{center}
\caption{\textbf{USMA-2020 Mechanism Performance under Truthful Strategies Simulated from Indicative and Final Preference Data from Class of 2021}\\}
\label{fig:usma2021indicative}
    \includegraphics[scale = 0.9]{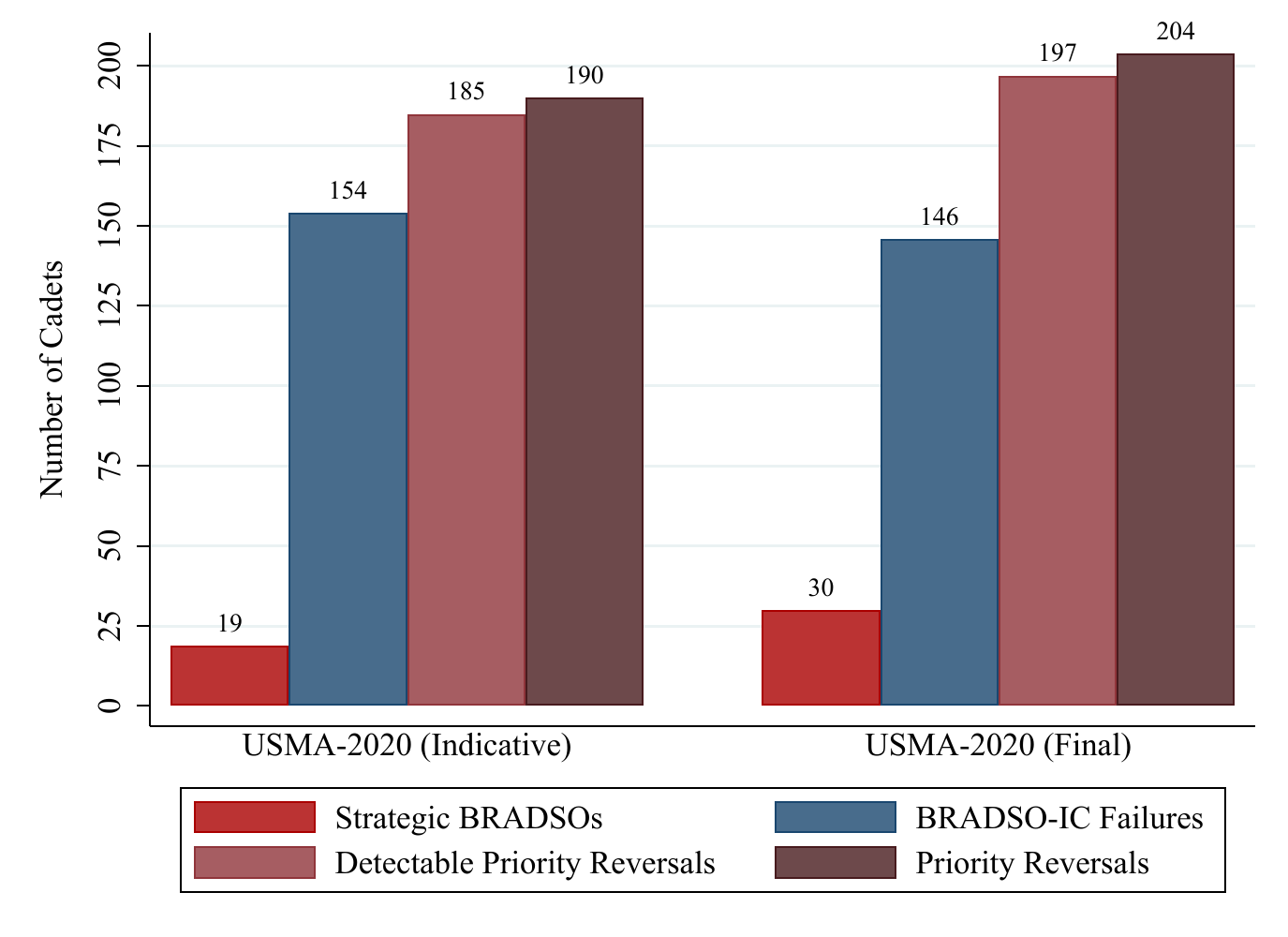}
\end{center}

\textbf{Notes.} 
USMA used the strategy-proof COM-BRADSO mechanism for the Class of 2021. 
This figure uses data  from the indicative and final rounds from the Class of 2021 on cadet preferences, branch priorities, and branch capacities to simulate the outcome of the 
USMA-2020 mechanism. Since the strategy space of the mechanism USMA-2020 differs from that of the mechanism COM-BRADSO, cadet strategies 
that correspond to truthful branch-preferences and BRADSO willingness are 
are simulated from cadet preferences over branch-cost pairs under the COM-BRADSO mechanism. 
Truthful strategies are constructed from Class of 2021 preferences 
by assuming that a preference indicating willingness to BRADSO at a branch means the cadet's strategy under the USMA-2006 and USMA-2020 mechanisms has her willing to BRADSO.  USMA-2020 (Indicative) reports outcomes using strategies constructed from preferences submitted in the dry-run of COM-BRADSO.
USMA-2020 (Final) reports outcomes using strategies constructed from preferences submitted in the final run of COM-BRADSO.
\end{figure}

In general, cadets form their preferences over branches over time as they acquire
more information about branches and their own tastes.  Therefore, the change documented in \fig{Figure \ref{fig:usma2020indicative}} may simply
reflect general preference formation from acquiring information about branches, and not revisions to preferences in response to the specific mechanism.  We briefly investigate this possibility by looking at the presence of strategic BRADSOs,
BRADSO-IC failures, and priority reversals using data on the indicative and final preferences from the Class of 2021.  This class participated in the 
strategy-proof COM-BRADSO mechanism.  We take indicative and final cadet preferences under COM-BRADSO and construct truthful strategies, following
the approach described above, for the USMA-2020 mechanism.  \fig{Figure \ref{fig:usma2021indicative}} shows that with preferences
constructed from a strategy-proof mechanism, there are only modest differences in strategic BRADSOs, BRADSO-IC failures, and priority reversals between the indicative and final rounds.
This comparison supports our claim that revisions of rank order lists in response to a dry-run of the USMA-2020 mechanism might understate
the issues this mechanism created, and why these issues became so pronounced with the USMA-2020 mechanism relative to the USMA-2006 mechanism. 

\subsection{Trade-off Between Talent Alignment vs. Retention}

As presented in Section \ref{subsec:COSM-2021}, the COM-BRADSO mechanism is based on 
\begin{enumerate}
\item the \textit{cumulative offer process\/}, and 
\item the choice rule $\calc^{BR}_b$ at any branch $b\in B$, which is a function of 
\begin{enumerate}
\item the baseline priority order $\pi_b$, 
\item the BRADSO cap $q_b^+$, and
\item the BRADSO policy $\omega^+_b$.
\end{enumerate}
\end{enumerate}
The flexibility of the COM-BRADSO mechanism to
accommodate branch-specific priorities under any BRADSO policy and BRADSO cap allows 
policy makers to determine the appropriate balance between branch priorities and the enforcement of contracts 
with increased costs. This was critical during the reform of the USMA-2020 mechanism where Army and USMA leadership had several discussions about the potential BRADSO policy for the Class of 2021. As described in the excerpt below from a news article describing an interview with the Talent-Based Branching Program Manager, selecting the specific BRADSO policy presented the Army with a trade-off between retention and talent alignment \citep{garcia:20}: 
\begin{quote}
A key question the Army considered when designing this year's mechanism was how much influence to give cadets who are willing to BRADSO. 
    If every cadet who volunteers to BRADSO can gain priority, or ``jump'' above, every cadet who did not volunteer to BRADSO, 
    then that could improve Army retention through more cadets serving an additional three years, but it could also result in 
    more cadets being assigned to branches that do not prefer them.
    \end{quote} 
The comparative static results in Proposition \ref{prop:BRADSOcomparativestatics} in Section \ref{subsec:COSM-2021} motivate our 
empirical analysis of different BRADSO policies.  While the results on the BRADSO collected given in  Proposition \ref{prop:BRADSOcomparativestatics} hold for a given
branch, in theory they may not hold in aggregate across all branches under COM-BRADSO.\footnote{The fact that
a global comparative static result does not hold in matching models with slot-specific priorities has been explored in other work,
including \cite{dur_boston} and \cite{dur/pathak/sonmez:20}. Both papers contains examples showing that how a comparative static across all branches need not hold.  However, the two papers also show empirically that these theoretical cases do not apply in their applications.  See, also, \cite{pathak/sonmez/unver/yenmez:20a}.}  However, 
as we show next,  the comparative static properties do hold in our  simulations with the Class of 2021 data for several BRADSO policies.

\begin{figure}[htp]
    \begin{center}
    \caption{\textbf{Number of BRADSOs Charged Across BRADSO Policies and Cap Sizes}\\}
    \label{fig:bradsocap}
    \includegraphics[scale = 0.9]{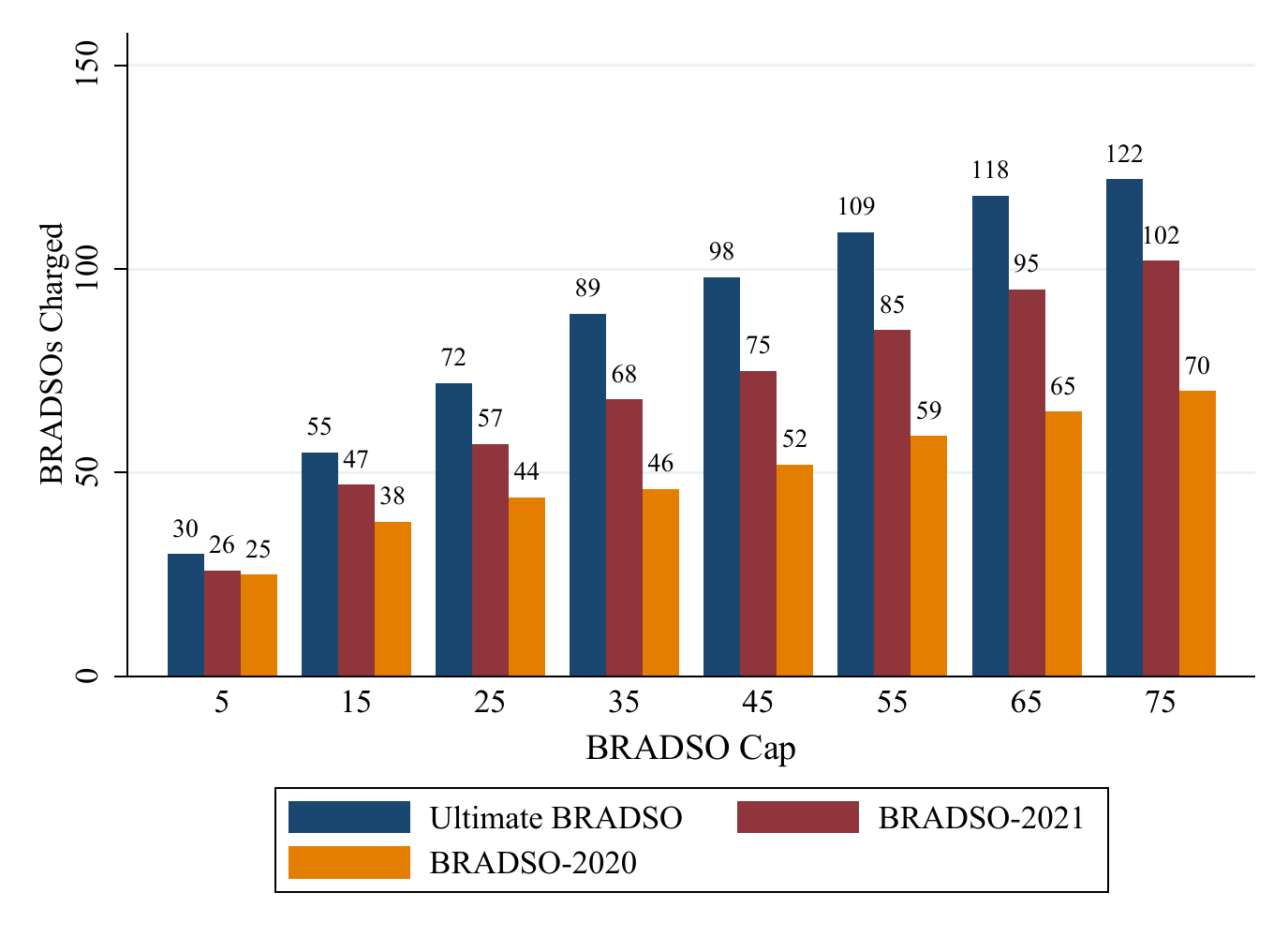}
    \end{center}
    \textbf{Notes.} This figure reports on the number of BRADSOs charged for three BRADSO policies: Ultimate BRADSO, BRADSO-2020, and BRADSO-2021 using data from the Class of 2021.  The BRADSO cap ranges from 5\% to 75\% of slots at each branch.  Each outcome is computed by running COM-BRADSO given stated cadet preferences under different BRADSO policies and cap sizes.
\end{figure}

The Army considered three BRADSO policies: the ultimate BRADSO policy and two tiered BRADSO policies.  Under BRADSO-2020, a cadet who expressed a willingness to
sign a BRADSO contract only obtained priority over other cadets who had the same categorical branch rating.  Under BRADSO-2021, a cadet who expressed a willingness to 
sign a BRADSO contract obtained higher priority over all other cadets if she was in the medium or high category.  To illustrate the trade-off between talent alignment and retention, \fig{Figure \ref{fig:bradsocap}} uses preferences from the Class of 2021 and re-runs the COM-BRADSO mechanism under these three BRADSO policies for different levels of BRADSO cap $q_b^+$, where  $q_b^+$ is expressed as a percentage of $q_b$, the total number of positions for branch $b$. 

To measure the effects of BRADSO policies on BRADSOs collected, \fig{Figure \ref{fig:bradsocap}} shows how the number of BRADSOs charged increases with $q_b^+$ and with the closeness of the BRADSO policy to the ultimate BRADSO policy. That is, for a given $q_b^+$ the BRADSO-2021 policy results in more BRADSOs charged than BRADSO-2020 policy, but fewer BRADSOs charged than the ultimate BRADSO policy.  When the BRADSO cap is small, there is relatively
little difference between BRADSO policies.  For example, when the BRADSO cap is 15\% of slots, 55 BRADSOs are charged under 
the ultimate BRADSO, 47 BRADSOs are charged under
BRADSO-2021, and 38 BRADSOs are charged under BRADSO-2020.  When the BRADSO cap is larger, the BRADSO policy has a larger effect on BRADSOs collected.
When the BRADSO cap is 65\%, 118 BRADSOs are charged under the ultimate BRADSO, 95 BRADSOs are charged under BRADSO-2020,
and 65 BRADSOs are charged under BRADSO-2021.  


The ability to run this analysis on the effects of BRADSO policies is a side-benefit of a strategy-proof mechanism, like COM-BRADSO.  At the request of the Army, we had conducted a similar analysis using data from the Class of 2020, but this analysis required stronger assumptions on cadet 
preferences.\footnote{Because cadets in the Class of 2020 did not submit preferences over branch-cost pairs, we assumed that all BRADSOs are consecutive, and also considered
different assumptions on the prevalence of non-consecutive BRADSOs.  These assumptions are not needed when cadets can rank branch-price pairs in a strategy-proof mechanism.}
As a result of this analysis, the Army decided to adopt the BRADSO-2021 policy and  increase the BRADSO cap, $q_b^+$, from 25 to 35 percent.  These are both policies
that increase the power of BRADSO.  However, USMA decided against adopting the ultimate BRADSO policy because branches remained opposed to giving more BRADSO power to low tier cadets.  
  
\section{Conclusion}\label{sec:conclusion}
In July 2019, the US Army implemented sweeping changes to the Army's Talent-Based Branching Program by adopting the USMA-2020 mechanism for the West Point, or USMA, Class of 2020.   The impetus for this change was to give Army branches greater influence and to ultimately assign cadets to better fitting branches.  
However, the USMA-2020 mechanism retained the same restricted strategy space as the previous USMA-2006 mechanism.  The
performance of the USMA-2020 mechanism made several underlying issues more apparent.

Our paper describes these reforms and shows how they facilitated the
adoption of a cumulative offer mechanism for the Class of 2021.  Our main result is that
the cumulative offer mechanism with a particular choice function is the only mechanism that
satisfies intuitive criteria, all formulating the Army's objectives. 
We also formally and empirically study the USMA-2020 mechanism.  That investigation
provides insights into the perverse incentives in this mechanism and why
these challenges became difficult to ignore for the Class of 2020.

When it was first formulated in \cite{sonmez/switzer:13},  cadet-branch matching became the first real-life application of the matching with contracts framework
with a non-trivial role for the contractual terms.   Our work builds on foundational theory by \cite{kelso/crawford:82},  \cite{hatfield/milgrom:05}, and \cite{hatfield/kojima:10} and applied theory papers by \cite{sonmez/switzer:13} and \cite{sonmez:13}.  This sequence of papers opened the door to influence 
mechanisms deployed in the field, and eventually led to the redesign of USMA's mechanism.  In this respect, we contribute to a market design literature where abstract theoretical models, which are often not contemplated in terms of particular applications, go on to have practical applications and ultimately influence real-world mechanisms.  We hope the chronology of the military's reform which links theory to practice follows the model of other market design applications, such as
for the medical match, spectrum auctions, school assignment, kidney exchange, internet advertising, and course assignment.\footnote{For the medical match, see \cite{gale/shapley:62}, \cite{roth:82}, and \cite{roth/peranson:99}.  For package auctions, see \cite{kelso/crawford:82}, \cite{demange/gale/sotomayor:86}, \cite{milgrom:00}, \cite{ausubel/milgrom:03}, \cite{milgrom/segal:17}, and \cite{milgrom/segal:20}.  For school assignment, see \cite{gale/shapley:62}, \cite{balinski/sonmez:99}, \cite{abdulkadiroglu/sonmez:03}, \cite{pathak/sonmez:08}, and \cite{abdulkadiroglu/pathak/roth:09}.  For kidney exchange, see \cite{shapley/scarf:74}, \cite{abdulkadiroglu/sonmez:99}, \cite{roth/sonmez/unver:04} and \cite{kidneyaea}.   For internet advertising, see \cite{shapley/shubik:71}, \cite{edelman/ostrovsky/schwarz:07}, and \cite{varian:06}. For course allocation, see \cite{varian:74}, \cite{sonmez/unver:10}, \cite{budish:11}, \cite{budish/cantillon:12}, and \cite{budish:17}.
}    Moreover, after the adoption of the cumulative offer mechanism at the Israeli Psychology Master's Match \citep{hassidim/romm/shorrer:17}, 
the Army's use of the COM-BRADSO mechanism is, as far as we know, the second field application of matching with contracts.    

While the Army initially resisted reforms to the USMA branching process, the challenges due to failures of certain principles formalized by our axioms led the Army to partner with us to fix these challenges. The Army sought a mechanism that not only promoted retention and talent alignment as USMA-2020 did, but that was also incentive compatible. The desire for incentive compatibility was partly to build cadets' trust in Army labor markets \citep{garcia:20}, and partly to obtain truthful information on cadet preferences. The latter objective is particularly important for Army efforts to understand and address the lack of minority representation in branches like Infantry and Armor, branches that produce a disproportionate share of Army generals \citep{briscoe2013black,kofoed2019effect}.  In that sense, reform shows the practical relevance and power of the matching with contracts framework, as well as the importance of building mechanisms with straightforward incentives to engender trust between organizations and their employees. 

After seeing the value of COM-BRADSO, the Army made two decisions to expand its utilization and to help achieve other policy objectives. First, although the Army originally planned to use the USMA-2020 mechanism to branch more than 3,000 ROTC cadets graduating in 2021, after observing the shortcomings of USMA-2020, the Army changed course and decided to adopt COM-BRADSO for ROTC instead. The decision to use COM-BRADSO for ROTC was in part due to concerns that ROTC's previous branching mechanism generated dead zones that made priority reversals particularly visible, as discussed in \citet{sonmez:13}. Second, for the West Point and ROTC Classes of 2022, the Army has asked us to modify COM-BRADSO to help address shortages of cadets willing to volunteer for the Army's \textit{branch-detail program}.\footnote{Officers who participate in the branch-detail program serve the first three to four years of their Army career in a ``detail'' branch before transitioning to the branch they received through their commissioning source's branching process. The Army faces an imbalance in branch strengths when there is a shortage of cadets willing to branch-detail.} We hope to report on these developments in future work.

Finally, beyond cadet branching, our experience with the cumulative offer mechanism
shows that it could be used in other internal labor markets where participants have preferences over jobs and contract terms.
For example, if the US Army were to permit officers to bid increased service obligations to obtain higher priority for certain jobs, then it could use COM-BRADSO to determine assignments for the 15,000 officers who participate in the Army's officer labor market each year \citep{greenberg/crow/wojtaszek:20}. 
Likewise, the US Air Force Academy uses a centralized process to assign graduates to career fields using
cadet preferences with a linear program \citep{armacost/lowe:05}.   \citet{hatfield/kominers/westkamp:21} and \citet{cowgill2021matching} describe other promising settings
for cumulative offer mechanisms.



\newpage
 \bibliographystyle{economet}
\bibliography{bib_all}

\newpage
\appendix

\section{Proofs}
\label{proofs}

\noindent \textbf{Proof of Theorem \ref{thm:singlebranchcharacterization}}: Suppose there is only one branch $b\in B$, 
and fix  a profile of cadet preferences $\succ \, \in \calq$.
We first show that the direct mechanism $\phi^{BR}$ satisfies the five axioms.

\textbf{\textit{Individual rationality\/}}: This axiom holds immediately under $\phi^{BR}$, since no cadet $i\in I$ is considered for a position at the increased cost $t^+$ unless her submitted preferences is such that $(b,t^+)  \succ_i \emptyset$.

\textbf{\textit{Non-wastefulness\/}}: Since there is only one branch and we already established \textit{individual rationality\/}, we
can focus on cadets who consider a position the base cost acceptable. With this observation, 
\textit{non-wastefulness\/} also holds immediately under $\phi^{BR}$, since all positions are allocated at Steps 0 and 1 at the base cost $t^0$ either 
as a final assignment or a tentative one. Tentative assignments from Step 1 may be altered later on by increasing their cost  to $t^+$ and
possibly changing their recipients, but not by leaving the position unassigned, hence assuring non-wastefulness.

\textbf{\textit{Lack of priority reversals}}:  Under the mechanism $\phi^{BR}$, each of the $q^0_b$ highest $\pi_b$-priority cadets are assigned a position
at the base cost $t^0$ at Step 0, and each of the next $q^+_b$ highest $\pi_b$-priority cadets are tentatively assigned a position at the base cost $t^0$ at Step 1. 
Tentative positions are lost in Step 2 only if there is excess demand from qualified cadets who are willing to pay the increased cost $t^+$, and starting
with the lowest $\pi_b$ priority cadets with tentative assignments. That assures that, for any $i,j \in I$,
\begin{equation} \label{eqn-euj1}
\phi^{BR}_j(\succ) = (b,t^0)  \succ_i \phi^{BR}_i(\succ) \quad \implies \quad j \; \pi_b \; i.
\end{equation} 
Moreover positions at the increased cost $t^+$ are offered to cadets with highest $\pi_b$ priorities among those 
(i) who fail to receive a position at the base cost $t^0$ and (ii) who declare the expensive assignment $(b,t^+)$ as acceptable. 
Therefore, for any $i,j \in I$,
\begin{equation} \label{eqn-euj2}
\phi^{BR}_j(\succ) = (b,t^+)  \succ_i \phi^{BR}_i(\succ) = \emptyset \quad \implies \quad j \; \pi_b \; i.
\end{equation} 
Relations \ref{eqn-euj1} and \ref{eqn-euj2} imply that mechanism $\phi^{BR}$  \textit{has no priority reversals\/}.

\textbf{\textit{BRADSO-IC\/}}:  Fix a cadet $i \in I$. 
For a given profile of preferences for all cadets except cadet $i$,  
whether cadet $i\in I$ receives an assignment of $(b,t^0)$ under the mechanism $\phi^{BR}$ is
independent of cadet $i$'s preferences under the mechanism $\phi^{BR}$: 
Cadets who are among the $q^0_b$ highest $\pi_b$-priority cadets in $I$ always receive an assignment at the base cost $t^0$; 
cadets who are not among the $q$ highest $\pi_b$-priority cadets in $I$ never receive an assignment at the base cost $t^0$; 
and for any cadet $i$ who has one of the highest $q$ but not one of the highest $q^0_b$ priorities, whether she receives  an assignment at the base cost $t^0$
depends on how many lower $\pi_b$-priority cadets are both willing to pay the increase cost $t^+$ and also able to ``jump ahead of'' the cadet $i$ through the
BRADSO policy. Hence if a cadet receives a position under $\phi^{BR}$ at the increased cost $t^+$, changing her reported preferences can only 
result in losing the position altogether. Therefore mechanism $\phi^{BR}$ satisfies \textit{BRADSO-IC\/}. 
 
\textbf{\textit{Enforcement of the BRADSO policy\/}}: The procedure for the mechanism $\phi^{BR}$ initially assigns all positions to the $q_b$ highest 
$\pi_b$-priority cadets at the base cost $t^0$, although the assignments of the $q^+_b$-lowest $\pi_b$-priority  cadets among these awardees 
are only tentative.  Step 2 of the procedure for mechanism $\phi^{BR}$ ensures that, if any cadet $j\in I$ loses her tentative assignment $(b,t^0)$ from Step 1, 
then any cadet $i\in I$ who receives an assignment of $(b,t^+)$ is such that $(i,t^+) \; \omega^+_b \; (j,t^0)$. Therefore, 
\begin{equation} \label{eqn-BRADSO1}
\left. \begin{array}{l}
  \phi^{BR}_i(\succ) = (b,t^+), \; \mbox{ and}\\ 
    (b,t^0) \succ_j \phi^{BR}_j(\succ)  \end{array}  \right\}     \quad  \implies \quad (i,t^+) \; \omega_b^+ \; (j,t^0). 
\end{equation}
Moreover, Step 2 of the same procedure also ensures that, 
for any $\ell \in \{1,\dots, q^+_b\}$, the $\ell^{\footnotesize th}$ lowest $\pi_b$-priority cadet $i^{\ell}$ with a tentative assignment of $(b,t^0)$
cannot maintain this tentative assignment, for as long as  there are  at least $\ell$ lower $\pi_b$-priority cadets  who 
are both willing to pay the increased cost $t^+$ and also able to ``jump ahead of'' the cadet $i^{\ell}$ through the BRADSO policy. 
Therefore,
\begin{equation} \label{eqn-BRADSO2}
\left. \begin{array}{l}
        \phi^{BR}_j(\succ) = (b,t^0),\\ 
      (b,t^+) \succ_i \phi^{BR}_i(\succ),  \; \mbox{ and}\\
       (i,t^+) \; \omega_b^+ \; (j,t^0) \end{array}  \right\}     \quad  \implies \quad 
       \big|\big\{i'\in I : \phi^{BR}_{i'}(\succ)=(b,t^+)\big\}\big|  = q^+_b.   
\end{equation}
Relations \eqref{eqn-BRADSO1} and \eqref{eqn-BRADSO2} imply that mechanism $\phi^{BR}$ satisfies \textit{enforcement of the BRADSO policy\/}. \\

\textbf{\textit{Uniqueness\/}}: We next show that mechanism $\phi^{BR}$ is the only mechanism that satisfies all five axioms. 

Let the direct mechanism $\varphi$ satisfy \textit{individual rationality,  non-wastefulness,
BRADSO-IC, enforcement of the BRADSO policy\/}, and \textit{has no priority reversals}. 
We want to show that $\varphi(\succ) = \phi^{BR}(\succ)$. 

If there are less than or equal to $q$ cadets for whom the assignment $(b,t^0)$ is acceptable under the preference profile $\succ$, 
all such cadets  must receive an assignment of $(b,t^0)$ by 
\textit{individual rationality, non-wastefulness\/}, and \textit{BRADSO-IC\/}. 
Since this is also the case under the allocation $\phi^{BR}(\succ)$, 
the result holds immediately for this case.
 
Therefore, w.l.o.g assume that there are strictly more than $q$ cadets for whom the assignment $(b,t^0)$ is acceptable under the preference profile $\succ$. 
Let $I^0$ be the set of $q^0_b$ highest $\pi_b$-priority cadets in $I$. 
By \textit{non-wastefulness\/}, all positions are assigned under $\varphi(\succ)$.  Since at most $q^+_b$ positions  can be awarded at the 
increased cost $t^+$, at least $q^0_b$ positions has to be allocated at the base cost  $t^0$. 
Therefore,  
\begin{equation} \label{eqn1}
\mbox{for any } i\in I^0, \qquad 
\varphi_i(\succ) = (b,t^0) = \phi^{BR}_i(\succ)
\end{equation}
by \textit{lack of priority reversals\/}. 

Let $I^1$ be the set of $q^+_b$ highest $\pi_b$-priority cadets in $I\setminus I^0$. 
Relabel the cadets in the set $I^1$ so that for any $\ell \in \{1,\ldots, q^+_b\}$, cadet $i^{\ell}$ is the ${\ell}^{\mbox{\footnotesize th}}$-lowest $\pi_b$-priority cadet in $I^1$. 
Let 
\[J^0 = \big\{j \in I\setminus (I^0\cup I^1) : (b,t^+) \succ_j \emptyset\big\}. 
\]
By \textit{individual rationality\/} and the \textit{lack of priority reversals\/},  
\begin{equation} \label{eqn2}
\mbox{for any } i\in I\setminus(I^0\cup I^1 \cup J^0), \qquad 
\varphi_i(\succ) = \emptyset = \phi^{BR}_i(\succ).
\end{equation}
By relations \eqref{eqn1} and \eqref{eqn2},
the only set of cadets whose assignments are yet to be determined under $\varphi(\succ)$ are cadets in $I^1 \cup J^0$. 
Moreover, by the \textit{lack of priority reversals\/}, cadets in $J^0$ can only receive a position at the increased cost $t^+$. 
That is, 
\begin{equation} \label{eqn3}
\mbox{for any } j\in J^0, \qquad 
\varphi_j(\succ) \not=  (b,t^0).
\end{equation}

For the next phase of our proof, we will rely on the sequence of individuals $i^1,\ldots,i^{q^+_b}$ and the
sequence of sets  $J^0, J^1, \ldots$ 
that are constructed for the  Step 2 of the mechanism $\phi^{BR}$. 
Here individual $i^1$ is the $q^{\footnotesize \mbox{th}}$ highest $\pi_b$-priority cadet in set $I$, cadet 
$i^2$ is the $(q-1)^{\footnotesize \mbox{th}}$ highest $\pi_b$-priority cadet in set $I$, and so on. 
The starting element of the second  sequence is  $J^0 = \{j \in I\setminus (I^0\cup I^1) : (b,t^+) \succ_j \emptyset\}$. 
Assuming Step 2.$n$ is the last sub-step of Step 2,
the remaining elements of the latter  sequence  for $n\geq 1$ is given as follows: For any $\ell \in \{1,\ldots, n\}$, 
\[ J^{\ell} = \left\{ \begin{array}{cl}
         J^{\ell -1}  & \mbox{ if } \; \emptyset \,  \succ_{i^{\ell}}  \, (b,t^+)\\
       J^{\ell -1} \cup\{i^{\ell}\} & \mbox{ if } \; (b,t^+) \succ_{i^{\ell}} \, \emptyset \end{array} \right.
\]
We have three cases to consider. \medskip

\noindent \textit{\textbf{Case 1.}\/} $n=0$ \smallskip

For this case, by  the mechanics of the Step 2 of the mechanism $\phi^{BR}$, we have 
\begin{equation} \label{eqn-case1a}
\big|\big\{j \in J^{0} : (j, t^+) \; \omega^+_b \; (i^1, t^0) \big\}\big| = 0.
\end{equation}
Therefore,  by relations \ref{eqn2}, \ref{eqn3}, and  condition (1) of the axiom \textit{enforcement of the BRADSO policy\/}, 
\begin{equation}
\mbox{for any }  i \in I\setminus (I^0 \cup I^1), \qquad \varphi_i(\succ) = \emptyset = \phi^{BR}_i(\succ). 
\end{equation}
Hence by \textit{non-wastefulness\/},
\begin{equation} \label{eqn-case1b}
\mbox{for any }  i \in I^1, \qquad \varphi_i(\succ) \in \big\{(b,t^0),(b,t^+)\big\}. 
\end{equation}
But since $\varphi$ satisfies \textit{individual rationality\/}, relation \eqref{eqn-case1b} implies that $\varphi_i(\succ) = (b,t^0)$
for any $i\in I^1$ with $\emptyset \succ_i (b,t^+)$. 
Furthermore for any $i\in I^1$ with $(b,t^+) \succ_i \emptyset$,  instead reporting 
the fake preference relation $\succ'_i \in \calq$ with $\emptyset \succ'_i (b,t^+)$ would guarantee cadet $i$ an assignment of 
$\varphi_i(\succ_{-i},\succ'_i)=(b,t^0)$ due to the same arguments applied for the economy $(\succ_{-i},\succ'_i)$,
and therefore by \textit{BRADSO-IC\/} these cadets too must receive an assignment of $(b,t^0)$ each. Hence  
\begin{equation} \label{eqn-case1c}
\mbox{ for any } i\in I^1, \qquad \varphi_i(\succ) = (b,t^0) = \phi_i^{BR}(\succ). 
\end{equation}
Relations \eqref{eqn1},  and \eqref{eqn-case1c}  imply $\varphi(\succ) = \phi^{BR}(\succ)$,  completing the proof for Case 1.$\blacksquare$ \medskip

\noindent \textit{\textbf{Case 2.}\/} $n \in \{1,\ldots, q^+_b-1\}$ \smallskip

For this case, by the mechanics of the Step 2 of the mechanism $\phi^{BR}$, we have 
\begin{equation} \label{eqn-case2a}
\mbox{for any } \ell \in \{1,\ldots, n\}, \qquad  \big|\big\{j \in J^{\ell -1} : (j, t^+) \; \omega^+_b \; (i^{\ell}, t^0) \big\}\big| \geq \ell,
\end{equation}
and
\begin{equation} \label{eqn-case2b}
\big|\big\{j \in J^{n} : (j, t^+) \; \omega^+_b \; (i^{n+1}, t^0) \big\}\big| = n.
\end{equation}
Since mechanism $\varphi$ satisfies  condition (2) of the axiom \textit{enforcement of the BRADSO policy\/}, 
the \textit{lack of priority reversals\/} and relation \ref{eqn-case2a} imply
\begin{equation} \label{eqn6}
\mbox{ for any } i\in \{i^1, \dots, i^{n}\}, \qquad \varphi_i(\succ)\not= (b,t^0). 
\end{equation}
Therefore, by \textit{non-wastefulness\/} and relations \eqref{eqn1}, \eqref{eqn2}, \eqref{eqn3}, and \eqref{eqn6}, 
at least  $n$ positions must be assigned at the increased cost $t^+$.

Moreover, since mechanism $\varphi$ satisfies \textit{non-wastefulness, lack of priority reversals\/}, 
and condition (1) of the axiom \textit{enforcement of the BRADSO policy\/},
relation \eqref{eqn-case2b} implies
\begin{equation} \label{eqn7}
\mbox{ for any } i\in \{i^{n+1}, \dots, i^{q^+_b}\}, \qquad \varphi_i(\succ) \in \big\{(b,t^0),(b,t^+)\big\}. 
\end{equation}
But since $\varphi$ satisfies \textit{individual rationality\/}, relation \eqref{eqn7} implies that $\varphi_i(\succ) = (b,t^0)$
for any $i\in\{i^{n+1}, \dots, i^{q^+_b}\}$ with $\emptyset \succ_i (b,t^+)$. 
Furthermore for any $i\in\{i^{n+1}, \dots, i^{q^+_b}\}$ with $(b,t^+) \succ_i \emptyset$,  instead reporting 
the fake preference relation $\succ'_i \in \calq$ with $\emptyset \succ'_i (b,t^+)$ would guarantee cadet $i$ an assignment of 
$\varphi_i(\succ_{-i},\succ'_i)=(b,t^0)$ due to the same arguments applied for the economy $(\succ_{-i},\succ'_i)$,
and therefore by \textit{BRADSO-IC\/} these cadets too must receive an assignment of $(b,t^0)$ each. Hence  
\begin{equation} \label{eqn8}
\mbox{ for any } i\in \{i^{n+1}, \dots, i^{q^+_b}\}, \qquad \varphi_i(\succ) = (b,t^0) = \phi_i^{BR}(\succ). 
\end{equation}
Since we have already shown that at least  $n$ positions must be assigned at an increased cost of $t^+$, 
relation \eqref{eqn8} implies that exactly $n$ positions must be assigned this cost, 
and therefore for any cadet $j \in J^n$ who is one of the $n$ highest $\pi_b$-priority cadets  in $J^n$, 
\begin{equation} \label{eqn9}
\varphi_j(\succ) = (b,t^+) = \phi_i^{BR}(\succ)
\end{equation}
by the \textit{lack of priority reversals\/}. 

Relations \eqref{eqn1}, \eqref{eqn8}, and \eqref{eqn9}  imply $\varphi(\succ) = \phi^{BR}(\succ)$,  completing the proof for Case 2. $\blacksquare$ \medskip

\noindent \textit{\textbf{Case 3.}\/} $n= q^+_b$ \smallskip

For this case, by the mechanics of the Step 2 of the mechanism $\phi^{BR}$, we have 
\begin{equation} \label{eqn-case3}
\mbox{for any } \ell \in \{1,\ldots, q^+_b\}, \qquad  \big|\big\{j \in J^{\ell -1} : (j, t^+) \; \omega^+_b \; (i^{\ell}, t^0) \big\}\big| \geq \ell.
\end{equation}
Since mechanism $\varphi$ satisfies  condition (2) of the axiom \textit{enforcement of the BRADSO policy\/}, 
relation \ref{eqn-case3} implies
\begin{equation} \label{eqn11}
\mbox{ for any } i\in \underbrace{\{i^1, \dots, i^{q^+_b}\}}_{=I^1}, \qquad \varphi_i(\succ)\not= (b,t^0). 
\end{equation}
Therefore, by \textit{non-wastefulness\/} and the \textit{lack of priority reversals}, 
exactly $q^+_b$ positions must be assigned at the increased cost $t^+$.
Hence  for any cadet $j \in J^{q^+_b}$ who is one of the $q^+_b$ highest $\pi_b$-priority cadets  in $J^{q^+_b}$, 
\begin{equation} \label{eqn12}
\varphi_j(\succ) = (b,t^+) = \phi_i^{BR}(\succ)
\end{equation}
by \textit{elimination of priority 
reversals\/}. 

Relations \eqref{eqn1} and \eqref{eqn12}  imply $\varphi(\succ) = \phi^{BR}(\succ)$,  completing the proof for Case 3, thus finalizing
the proof of the theorem. $\blacksquare$ \qed \medskip
 
\noindent \textbf{Proof of Proposition \ref{prop:2021NashEqm}}: Suppose that there is only one branch $b\in B$.  
Fixing  the profile of cadet preferences $\succ \, \in \calq$, the baseline priority order $\pi_b$, and the BRADSO policy
$\omega^+_b$, consider the strategic-form game induced by the USMA-2020 mechanism $(\cals^{2020}, \varphi^{2020})$. 
When there is only one branch,  the first part of the strategy space becomes redundant and the second part  contains only the two elements
$b$ and $\emptyset$. Hence, for any cadet $i\in I$, the strategy space of cadet $i \in I$ under the USMA-2020 mechanism is $\cals^{2020}_i = \{\emptyset, b\}$. 

For a given strategy profile $s \in \cals^{2020}$, construct the  priority order $\pi^+_b(s)$ as follows: 
For any $i, j \in I$,
\begin{enumerate}
\item $\; s_i = s_j \qquad \qquad \qquad \quad \implies \qquad  \qquad i \;\, \pi^+_b(s) \; j \; \iff i \; \pi_b \; j$,
\item $\; s_i = b$ and $s_j = \emptyset \qquad \, \implies \qquad i \;\, \pi^+_b(s) \; j \; \iff (i,t^+) \; \omega^+_b \; (j,t^0)$.\smallskip
\end{enumerate}

Let $I^+(s)$ be the set of $q_b$ highest $\pi^+_b(s)$-priority cadets in $I$. \smallskip
		
For any cadet $i \in I$, the outcome of the USMA-2020 mechanism is given as,
\[ \varphi_i^{2020}(s) =   \left\{ \begin{array}{cl}  
\emptyset & \mbox{if } \;\;  i\not\in I^+(s),\\
 (b, t^0)   & \mbox{if } \;\;   i\in I^+(s) \mbox{ and }   s_i = \emptyset,\\
 (b, t^0)   & \mbox{if } \;\;   i\in I^+(s) \mbox{ and }   s_i = b \; \mbox{ and }  \;  \big|\{j\in I^+(s) : s_j = b \mbox{ and } i \; \pi_b \;j\}\big| \geq q^+_b,\\
 (b, t^+)   & \mbox{if } \;\;   i\in I^+(s) \mbox{ and }  s_i = b \; \mbox{ and } \;  \big|\{j\in I^+(s) : s_j = b \mbox{ and } i \; \pi_b \;j\}\big|  < q^+_b. 
\end{array} \right.
\]	
We first prove a lemma on the structure of Nash equilibrium strategies of the strategic-form game induced by the 
USMA-2020 mechanism $(\cals^{2020},\varphi^{2020})$. 
\begin{lemma} \label{lemma1} 
Let $s^*$  be a Nash equilibrium of the strategic-form game induced by the mechanism $(\cals^{2020},\varphi^{2020})$. Then,
for any $i,j \in I$, 
\[ \varphi^{2020}_j(s^*) \succ_i \varphi^{2020}_i(s^*) \quad \implies \quad j \; \pi_b \; i. 
\]
\end{lemma}

\noindent \textit{Proof of Lemma \ref{lemma1}\/}:   Let $s^*$  be a Nash equilibrium of the strategic-form game 
induced by the USMA-2020 mechanism $(\cals^{2020},\varphi^{2020})$. Contrary to the claim suppose that, there exists $i,j \in I$ such that  
\[ \varphi^{2020}_j(s^*) \succ_i \varphi^{2020}_i(s^*) \quad \mbox{ and } \quad  i \; \pi_b \; j. 
\]
There are three possible cases, where in each case we reach a contradiction by showing that cadet $i$ has a profitable deviation 
by mimicking the strategy of cadet $j$: \smallskip

\noindent \textbf{\textit{Case 1\/}:} $\varphi^{2020}_j(s^*) = (b,t^0)$ and $\varphi^{2020}_i(s^*) = (b,t^+)$. \smallskip

Since by assumption $\varphi^{2020}_i(s^*) = (b,t^+)$, 
\[ s^*_i = b.
\]
Moreover the assumptions $\varphi^{2020}_j(s^*) = (b,t^0)$,\; $\varphi^{2020}_i(s^*) \not= (b,t^0)$,\; and  $i \; \pi_b \; j$ imply
\begin{equation} \label{case1-eqn2}
j \in I^+(s^*) \quad \mbox{ and } \quad   s^*_j = \emptyset.
\end{equation}
But then, relation \eqref{case1-eqn2} and the assumption $i \; \pi_b \; j$ imply that, 
 for the alternative strategy $\hat{s}_i = \emptyset$ for cadet $i$, 
\[ i \in I^+(s_{-i}^*, \hat{s}_i),
\]
and thus 
\[ \varphi_i^{2020}(s_{-i}^*, \hat{s}_i) = (b,t^0) \succ_i \varphi^{2020}_i(s^*),
\] 
contradicting  $s^*$ is a Nash equilibrium strategy.  This completes the proof  for Case 1. $\blacksquare$  \smallskip

\noindent \textbf{\textit{Case 2\/}:} $\varphi^{2020}_j(s^*) = (b,t^0)$ and $\varphi^{2020}_i(s^*) = \emptyset$. \smallskip

Since by assumption $\varphi^{2020}_j(s^*) = (b,t^0)$,\; $\varphi^{2020}_i(s^*) = \emptyset$,\; and  $i \; \pi_b \; j$, we must have
\begin{equation} \label{lemma1case2-eqn1}
j \in I^+(s^*) \quad \mbox{ and } \quad   s^*_j = b \quad \mbox{ and } \quad   \big|\{k\in I^+(s^*) : s^*_k = b \mbox{ and } j \; \pi_b \;k\}\big| \geq q^+_b, 
\end{equation}
and 
\[ s^*_i = \emptyset. 
\]
But then, relation \eqref{lemma1case2-eqn1} and the assumption  $i \; \pi_b \; j$ imply that, 
for the alternative strategy $\hat{s}_i = b$ for cadet $i$, 
\[
i \in I^+(s_{-i}^*, \hat{s}_i) \quad \mbox{ and } \quad   \hat{s}_i = b \quad \mbox{ and } \quad   \big|\{k\in I^+(s_{-i}^*, \hat{s}_i) : s^*_k = b \mbox{ and } i \; \pi_b \;k\}\big| \geq q^+_b, 
\]
and thus 
\[ \varphi_i^{2020}(s_{-i}^*, \hat{s}_i) = (b,t^0) \succ_i \varphi^{2020}_i(s^*),
\] 
contradicting  $s^*$ is a Nash equilibrium strategy.  This completes the proof for Case 2.  $\blacksquare$  \smallskip

\noindent \textbf{\textit{Case 3\/}:} $\varphi^{2020}_j(s^*) = (b,t^+)$ and  $\varphi^{2020}_i(s^*) = \emptyset$. \smallskip

Since by assumption $\varphi^{2020}_j(s^*) = (b,t^+)$, 
\begin{equation} \label{case3-eqn1}
j \in I^+(s^*) \quad \mbox{ and } \quad s^*_j = b.
\end{equation}
Moreover, since $  \varphi^{2020}_i(s^*) = \emptyset$ by assumption, 
\[  i\not\in  I^+(s^*).
\]
Therefore, since  $i \; \pi_b \; j$ by assumption, 
\[ j \in I^+(s^*) \;  \mbox{ and } \; i\not\in  I^+(s^*) \qquad \implies \qquad s^*_i = \emptyset.  
\]
But then, again thanks to assumption  $i \; \pi_b \; j$, the relation \eqref{case3-eqn1} implies that, 
for the alternative strategy $\hat{s}_i = b$ for cadet $i$, 
\[
i \in I^+(s_{-i}^*, \hat{s}_i), 
\]
and thus 
\[ \underbrace{\varphi_i^{2020}(s_{-i}^*, \hat{s}_i)}_{\in \{(b,t^0),(b,t^+)\}} \succ_i \varphi^{2020}_i(s^*),
\] 
contradicting  $s^*$ is a Nash equilibrium strategy,\footnote{Unlike the first two cases, in this case cadet $i$ may even get a better assignment than
cadet $j$ (i.e. cadet $i$ may receive an assignment of $(b,t^0)$) by mimicking cadet $j$'s strategy. }  
completing the proof for Case 3, and concluding the proof of Lemma \ref{lemma1}. $\blacksquare$ \mbox{}\hfill$\diamondsuit$\smallskip 
 
For the next phase of our proof, we rely on the construction in  the  Step 2 of the mechanism $\phi^{BR}$:
Let $I^0$ be the set of $q^0_b$ highest $\pi_b$-priority cadets in $I$, and  
$I^1$ be the set of $q^+_b$ highest $\pi_b$-priority cadets in $I\setminus I^0$. 
Relabel the set of  cadets in $I^1$, so that $i^1$ is the lowest $\pi_b$-priority cadet in $I^1$, 
$i^2$ is the second lowest $\pi_b$-priority cadet in $I^1$,\ldots, and $i^{q^+_b}$ is the highest  $\pi_b$-priority cadet in $I^1$. 
Note that, cadet $i^1$ is the $q^{\footnotesize \mbox{th}}$ highest $\pi_b$-priority cadet in set $I$, cadet 
$i^2$ is the $(q-1)^{\footnotesize \mbox{th}}$ highest $\pi_b$-priority cadet in set $I$, and so on.
Let $J^0 = \{j \in I\setminus (I^0\cup I^1) : (b,t^+) \succ_j \emptyset\}$. 
Assuming Step 2.$n$ is the last sub-step of Step 2  of the mechanism $\phi^{BR}$, 
for any $\ell \in \{1,\ldots, n\}$, let 
\[ J^{\ell} = \left\{ \begin{array}{cl}
         J^{\ell -1}  & \mbox{ if } \; \emptyset \, \succ_{i^{\ell}} \, (b,t^+)\\
       J^{\ell -1} \cup\{i^{\ell}\} & \mbox{ if } \; (b,t^+) \succ_{i^{\ell}} \, \emptyset \end{array} \right.
\]
Recall that, under the mechanism $\phi^{BR}$, exactly $n$ cadets receive an assignment of $(b,t^+)$. 
We will show that, the same is also the case under the Nash equilibria of the strategic-form game 
induced by the USMA-2020 mechanism $(\cals^{2020},\varphi^{2020})$.

Let $s^*$  be a Nash equilibrium of the strategic-form game induced by the USMA-2020 mechanism $(\cals^{2020},\varphi^{2020})$.
We have three cases to consider:

\noindent \textit{\textbf{Case 1}\/}: $n=0$

Since by assumption  $n=0$ in this case, 
\begin{equation} \label{eqm-n=0a}
\big\{j\in J^0 : (j,t^+) \; \omega^+_b \; (i^1,t^0)\big\} = \emptyset. 
\end{equation}
Towards a contradiction, suppose there exists a cadet  $i \in I\setminus(I^0\cup I^1)$ such that $i\in I^+(s^*)$. 
Since cadet $i^1$ is the $q^{\mbox{\footnotesize th}}$ highest $\pi_b$-priority cadet in $I$, the assumption  $i\in I^+(s^*)$ and relation \eqref{eqm-n=0a} imply
\begin{equation}  \label{eqm-n=0b}
i \not\in J^0 \implies \emptyset \; \succ_i (b,t^+). 
\end{equation} 
Moreover, since cadet $i$ is not one of the $q$ highest $\pi_b$-priority cadets in $I$, 
\begin{equation}  \label{eqm-n=0c}
i\in I^+(s^*) \implies s^*_i = b. 
\end{equation}
But this means cadet $i$ can instead submit an alternative strategy $\hat{s}_i = \emptyset$, assuring that she remains unmatched, 
contradicting $s^*$ is a Nash equilibrium.
Therefore, 
\begin{equation}  \label{eqm-n=0d}
\mbox{for any } i\in I\setminus (I^0 \cup I^1), \qquad (i,t^+) \; \omega^+_b \; (i^1,t^0) \implies s^*_i = \emptyset, 
\end{equation}
which in turn implies 
\begin{equation}  \label{eqm-n=0e}
I^+(s^*) = I^0 \cup I^1. 
\end{equation}
Hence all cadets in $I^0 \cup I^1$ receive a position under $\varphi^{2020}(s^*)$.
Next consider the lowest $\pi_b$-priority cadet $i\in I^0 \cup I^1$ such that $\varphi_i^{2020}(s^*) = (b,t^+)$. 
This can only happen if $s^*_i =b$. 
But this means cadet $i$ can instead submit an alternative strategy $\hat{s}_i = \emptyset$, assuring that $\varphi_i^{2020}(s_{-i}^*,\hat{s}_i) = (b,t^0)$
by relation \eqref{eqm-n=0d},  contradicting $s^*$ is a Nash equilibrium.
Hence
\begin{equation}
\mbox{for any } i\in I^0 \cup I^1, \qquad \varphi_i^{2020}(s^*) = (b,t^0) = \phi^{BR}_i(\succ), 
\end{equation}
and therefore $ \varphi^{2020}(s^*) = \phi^{BR}(\succ)$. 

Finally observe that the strategy profile $s'$ where $s'_i = \emptyset$  for any cadet $i\in I$ is a Nash equilibrium, with an outcome 
$\varphi^{2020}(s') = \phi^{BR}(\succ)$, showing that there exists a Nash equilibrium  completing the proof 
for Case 1. $\blacksquare$ \medskip

For any $\ell \in \{1,\ldots, n\}$, let $\overline{J^{\ell}}$ be the set of $\ell$ highest $\pi_b$-priority cadets in the set $J^{\ell}$:
\[ \overline{J^{\ell}} = \Big\{j \in J^{\ell}  : \; \big|\{i \in J^{\ell} : \; i  \; \pi_b \; j\}\big|<\ell\Big\}
\]
Before proceeding with the next two cases, we prove the following lemma that will be helpful for both cases.  

\begin{lemma} \label{lemma2}
Suppose there are $n>0$ positions allocated at the increased cost $t^+$ under the allocation $\phi^{BR}(\succ)$. Then, for any  Nash equilibrium
$s^*$ of the strategic-form game induced by the USMA-2020 mechanism $(\cals^{2020},\varphi^{2020})$ and $\ell \in \{1,\ldots,n\}$,
\begin{enumerate}
\item $\varphi^{2020}_{i^{\ell}}(s^*) = (b,t^+) \quad \iff \quad (b,t^+) \succ_{i^{\ell}} \emptyset$, \; and \smallskip
\item  $\varphi^{2020}_i(s^*) = (b,t^+)$ \quad for any $i\in \overline{J^{\ell}}$.
\end{enumerate}
\end{lemma}

\noindent \textit{Proof of Lemma \ref{lemma2}\/}: Let $s^*$ be a Nash equilibrium 
of the strategic-form game induced by the USMA-2020 mechanism $(\cals^{2020},\varphi^{2020})$. 
First recall that,  
\[ \mbox{for any } j \in I \setminus (I^0 \cup I^1), \qquad  \varphi_j^{2020}(s^*) \in \big\{(b,t^+),\emptyset\big\},
\]
and therefore, since any cadet $j \in I \setminus (I^0 \cup I^1 \cup J^0)$ prefers remaining unmatched to receiving a position at the increased cost $t^+$ 
and she can assure remaining unmatched by submitting the strategy $s_j = \emptyset$,
\begin{equation} \label{lemma2-eqn1}
\mbox{for any } j \in I \setminus (I^0 \cup I^1 \cup J^0),  \qquad \varphi_j^{2020}(s^*) = \emptyset.
\end{equation}
Also, by the mechanics of the Step 2 of the mechanism $\phi^{BR}$,
\begin{equation} \label{lemma2-eqn2}
\mbox{for any } \ell \in \{1,\ldots, n\}, \qquad  \big|\big\{j \in J^{\ell -1} : (j, t^+) \; \omega^+_b \; (i^{\ell}, t^0) \big\}\big| \geq \ell. 
\end{equation}

The proof of the lemma is by induction on $\ell$.  We first prove the result for $\ell = 1$. 

Consider the highest $\pi_b$-priority cadet $j$ in the set  $\big\{j \in J^0 : (j, t^+) \; \omega^+_b \; (i^1, t^0) \big\}$. 
By relation \ref{lemma2-eqn2}, such a cadet exists. 

First assume that $(b,t^+) \succ_{i^1} \emptyset$. 
In this case, $J^1 = J^0 \cup \{i^1\}$ and
cadet $i^1$ is the highest $\pi_b$-priority cadet in $J^1$. Hence $\overline{J^1} = \{i^1\}$ in this case. 
Consider the Nash equilibrium strategies of cadet $i^1$ and cadet $j$. 
If $s^*_{i^1} = \emptyset$, then  by relation \eqref{lemma2-eqn1} her competitor
cadet $j$ can secure himself an assignment of $(b,t^+)$ by reporting a strategy of $s_j =b$, 
which would mean cadet $i^1$ has to remain unassigned, since by Lemma \ref{lemma1} no cadet in $I^0 \cup I^1$  can envy 
the assignment of cadet $i^1$  at Nash equilibria. 
In contrast, reporting a strategy of $s_{i^1}=b$ assures that cadet $i^1$ receives a position, which is preferred at any price to 
remaining unmatched by assumption $(b,t^+) \succ_{i^1} \emptyset$. 
Therefore, $s^*_{i^1} = b$, and  hence
\begin{equation} \label{lemma2-eqn3}
(b,t^+) \succ_{i^1} \emptyset \quad \implies  \quad
\left\{ \begin{array}{ll}
& \varphi^{2020}_{i^1}(s^*) = (b,t^+),  \; \mbox{ and }\\  
& \varphi^{2020}_{i}(s^*) = (b,t^+)  \quad \mbox{ for any } i\in \overline{J^{1}}=\{i^1\}. \end{array} \right.
\end{equation}

Next assume that $\emptyset \succ_{i^1} (b,t^+)$. 
In this case $J^1 = J^0$ and cadet $j$ is the highest $\pi_b$-priority cadet in $J^1$.  Hence $\overline{J^1} = \{j\}$ in this case.
By Lemma \ref{lemma1},  no cadet in $(I^0\cup I^1)\setminus\{i^1\}$  can envy the assignment of cadet $i^1$ at Nash equilibria. 
Therefore, a strategy of $s_{i^1}=b$ means that cadet $i$ receives an assignment of $(b,t^+)$, which is inferior to  
remaining unmatched by assumption. 
Therefore $s^*_{i^1} = \emptyset$. 
Moreover reporting a strategy of $s_j=\emptyset$ means that cadet $j$ remains unmatched, whereas
reporting a  strategy of $s_j=b$ assures that she  receives an assignment of $(b,t^+)$, which is preferred to  
remaining unmatched since $j \in J^0$. Therefore, $s^*_{i^1} = \emptyset$, and  hence
\begin{equation} \label{lemma2-eqn4}
\emptyset \succ_{i^1} (b,t^+) \quad \implies  \quad
\left\{ \begin{array}{ll}
& \varphi^{2020}_{i^1}(s^*) = \emptyset,  \; \mbox{ and }\\  
& \varphi^{2020}_{i}(s^*) = (b,t^+)  \quad \mbox{ for any  } i\in \overline{J^{1}}=\{j\}. \end{array} \right.
\end{equation}
Relations  \eqref{lemma2-eqn3} and  \eqref{lemma2-eqn4} complete the proof for $\ell = 1$. \smallskip

Next assume that the inductive hypothesis holds for $\ell = k<n$. We want to show that the result holds for $\ell = (k+1)$ as well. 

By the inductive hypothesis,
\begin{equation} \label{lemma2-eqn5}
 \mbox{for any } i\in \overline{J^k}, \quad  \varphi^{2020}_{i}(s^*) = (b,t^+).
\end{equation}
By relation \ref{lemma2-eqn2}, there are at least $k+1$ cadets in the set $\big\{j \in J^k : (j, t^+) \; \omega^+_b \; (i^{k+1}, t^0) \big\}$.
Therefore, since there are $k$ cadets in the set $\overline{J^k}$, there is at least one cadet in the set 
\[\big\{j \in J^k : (j, t^+) \; \omega^+_b \; (i^{k+1}, t^0) \big\}\setminus \overline{J^k}.
\]
Let $j$ be the highest $\pi_b$-priority cadet in this set. 

First assume that $(b,t^+) \succ_{i^{k+1}} \emptyset$. 
In this case $J^{k+1} = J^k \cup \{i^{k+1}\}$ and
cadet $i^{k+1}$ is the highest $\pi_b$-priority cadet in $J^{k+1}$. Hence $\overline{J^{k+1}} = \overline{J^k} \cup \{i^{k+1}\}$ in this case. 
Consider the Nash equilibrium strategies of cadet $i^{k+1}$ and cadet $j$. 
If $s^*_{i^{k+1}} = \emptyset$, then by relation \eqref{lemma2-eqn1}
cadet $j$ can secure herself an assignment of $(b,t^+)$ by reporting a strategy of $s_j =b$, 
which would mean cadet $i^{k+1}$ has to remain unassigned, since by Lemma \ref{lemma1} no cadet in $(I^0 \cup I^1)\setminus \{i^1,\ldots, i^k\}$  
can envy the assignment of cadet $i^{k+1}$ at Nash equilibria 
and by relation \eqref{lemma2-eqn5} all cadets in $\overline{J^k}$ receive an assignment of $(b,t^+)$.\footnote{Since 
$\left|(I^0 \cup I^1)\setminus \{i^1,\ldots, i^k\}\right|=(q-k)$
and $\left|\overline{J^k}\right| = k$, this basically means cadets
$i^{k+1}$ and $j$ are competing for a single position.} 
In contrast, reporting a strategy of $s_{i^{k+1}}=b$ assures that cadet $i^{k+1}$ receives a position, which is preferred at any price to 
remaining unmatched by assumption $(b,t^+) \succ_{i^{k+1}} \emptyset$. 
Therefore,  $s^*_{i^{k+1}} = b$, and  hence
\begin{equation} \label{lemma2-eqn6}
(b,t^+) \succ_{i^{k+1}} \emptyset \; \implies  \;
\left\{ \begin{array}{ll}
& \varphi^{2020}_{i^{k+1}}(s^*) = (b,t^+),  \; \mbox{ and }\\  
& \varphi^{2020}_{i}(s^*) = (b,t^+)  \; \mbox{ for any } i\in \overline{J^{k+1}}= \overline{J^k}\cup\{i^{k+1}\}. \end{array} \right.
\end{equation}
Next assume that $\emptyset \succ_{i^{k+1}} (b,t^+)$. 
In this case $J^{k+1} = J^k$ and  $\overline{J^{k+1}} = \overline{J^k} \cup \{j\}$.
By Lemma \ref{lemma1},  no cadet in $I^0\cup I^1 \setminus \{i^1,\ldots, i^k\}$  can envy the assignment of cadet $i^{k+1}$ at Nash equilibria. 
Therefore, since all cadets in $\overline{J^k}$ receive an assignment of $(b,t^+)$  by relation \eqref{lemma2-eqn5},
a strategy of $s_{i^{k+1}}=b$ means that cadet $i^{k+1}$ receives an assignment of $(b,t^+)$, which is inferior to  
remaining unmatched by assumption. Therefore $s^*_{i^{k+1}} = \emptyset$. 
Moreover reporting a strategy of $s_j=\emptyset$  means that cadet $j$ remains unmatched, whereas
reporting a  strategy of $s_j=b$ assures that she  receives an assignment of $(b,t^+)$, which is preferred to  
remaining unmatched since $j \in J^k$. Therefore, $s^*_{i^{k+1}} = \emptyset$, and  hence
\begin{equation} \label{lemma2-eqn7}
\emptyset \succ_{i^{k+1}} (b,t^+) \quad \implies  \quad
\left\{ \begin{array}{ll}
& \varphi^{2020}_{i^{k+1}}(s^*) = \emptyset,  \; \mbox{ and }\\  
& \varphi^{2020}_{i}(s^*) = (b,t^+)  \quad \mbox{ for any  } i\in \overline{J^{k+1}}=\overline{J^k}\cup\{j\}. \end{array} \right.
\end{equation}
Relations  \eqref{lemma2-eqn6} and  \eqref{lemma2-eqn7} complete the proof for $\ell = k+1$, 
and conclude the proof of Lemma \ref{lemma2}. \mbox{}\hfill$\diamondsuit$ \medskip

We are ready to complete prove the theorem for our last two cases:

\noindent \textit{\textbf{Case 2.}\/} $n \in \{1,\ldots, q^+_b-1\}$ \smallskip

For this case, by the mechanics of the Step 2 of the mechanism $\phi^{BR}$, 
\begin{equation} \label{case2-eqn1}
\big|\big\{j \in J^{n} : (j, t^+) \; \omega^+_b \; (i^{n+1}, t^0) \big\}\big| = n.
\end{equation}
Consider cadet $i^{n+1}$. There are $q-(n+1)$ cadets with higher $\pi_b$-priority, and
by relation \eqref{case2-eqn1} there are $n$ cadets in $J^n$ whose increased-cost assignments have higher $\omega^+_b$ priority under the BRADSO policy
than the base-cost assignment for cadet $i^{n+1}$.
For any other cadet $i \in I\setminus \Big(J^n \cup I^0 \cup \big(I^1 \setminus \{i^1,\ldots,i^{n+1}\}\big)\Big)$ with $(i, t^+) \; \omega^+_b \; (i^{n+1}, t^0)$, 
we must have $\emptyset \succ_i (b,t^+)$ since  $J^n \supseteq J^0$. 
Therefore none of these individuals can receive an assignment of $(b,t^+)$ under a Nash equilibrium strategy, and hence 
the number of cadets who can have higher $\pi^+_b(s^*)$-priority than cadet is $i^{n+1}$ is at most 
$q-(n+1)+n = q-1$ under any Nash equilibrium strategy. That is, cadet $i^{n+1} \in I^+(s^*)$ regardless of her submitted strategy, 
and therefore,
\begin{equation}  \label{case2-eqn2}
\varphi^{2020}_{i^{n+1}}(s^*) = (b,t^0), 
\end{equation}
since  her best response $s^*_{i^{n+1}}$ to $s^*_{-i^{n+1}}$ results in an assignment of $(b,t^0)$. 
Moreover, Lemma \ref{lemma1} and relation \eqref{case2-eqn2} imply that,  
for any cadet $i \in I^0 \cup \big(I^1 \setminus  \{i^1,\ldots,i^{n+1}\}\big)$,
\begin{equation}  \label{case2-eqn3}
\varphi^{2020}_i(s^*) = (b,t^0).
\end{equation}
Hence Lemma \ref{lemma2} and relations \eqref{case2-eqn2}, \eqref{case2-eqn3} imply
$\varphi^{2020}(s^*) = \phi^{BR}(\succ)$. 

Finally, the strategy profile $s'$ where $s'_i = b$  for any cadet $i\in J^n$ and $s'_j = \emptyset$
for any cadet  $j \in I\setminus J^n$ is a Nash equilibrium, with an outcome 
$\varphi^{2020}(s') = \phi^{BR}(\succ)$, showing that there exists a Nash equilibrium  completing the proof 
for Case 2. $\blacksquare$ \medskip

\noindent \textit{\textbf{Case 3.}\/} $n =  q^+_b$ \smallskip

Since at most $q^+_b$ positions can be assigned at the increased cost $t^+$, Lemma \ref{lemma1} and Lemma \ref{lemma2} immediately imply
$\varphi^{2020}(s^*) = \phi^{BR}(\succ)$. 

Finally the strategy profile $s'$ where $s'_i = b$  for any cadet $i\in J^{q^+_b} \cup I^0$ and $s'_j = \emptyset$
for any cadet  $j \in I\setminus \big(J^n \cup I^0\big)$ is a Nash equilibrium, with an outcome 
$\varphi^{2020}(s') = \phi^{BR}(\succ)$, showing that there exists a Nash equilibrium  completing the proof  for Case 3, 
and the proof of the proposition. $\blacksquare$ \qed \medskip

\noindent \textbf{Proof of Theorem \ref{cosm}}: Fix $(\pi_b)_{b\in B} \in \Pi^{|B|}$ and $(\omega^+_b)_{b\in B} \in \prod_{b\in B}\Omega^+_b$.  

For any cadet $i\in I$,  branch $b$, and preference $\succ_i \; \in \calq$, 
by assumption we have $(b,t^0) \succ_i (b,t^+)$. 
Therefore, since cadet proposals to branches follow their submitted preferences, 
the set of contracts available to any branch at any stage of the cumulative offer process is \textit{viable\/}. 
That is, whenever the increased cost contact $(i,b,t^+)$ of a cadet $i\in I$ is available for a branch $b\in B$, so is her base cost contract $(i,b,t^0)$.  \smallskip

We first show that the mechanism $\phi^{COM-BR}$ satisfies the five  axioms. 
For the proofs of \textit{individual rationality, non-wastefulness, lack of priority reversals,\/} and
\textit{enforcement of BRADSO policy\/}, fix $\succ \, \in \calq^{|I|}$. 
\smallskip

\textbf{\textit{Individual rationality\/}}: 
No cadet $i\in I$ ever makes a proposal to a branch $b$ at the increased cost $t^+$ 
under the cumulative offer process,  unless
her preferences are such that $(b,t^+)  \succ_i \emptyset$. Hence the mechanism $\phi^{COM-BR}$ satisfies \textit{individual rationality\/}. \smallskip

\textbf{\textit{Non-wastefulness\/}}: For any branch $b\in B$, 
unless there are  already $q$ contracts with distinct cadets on hold, 
it is not possible for 
all contracts of any given cadet to be rejected at any stage of the cumulative offer process under the choice rule $\calc^{BR}_b$.  
Hence the mechanism $\phi^{COM-BR}$  satisfies \textit{non-wastefulness\/}. \smallskip

\textbf{\textit{Lack of priority reversals\/}}:  Suppose that $\phi_j^{COM-BR}(\succ)\succ_i\phi_i^{COM-BR}(\succ)$ for a pair of cadets $i,j\in I.$  Since 
the mechanism $\phi^{COM-BR}$
is \textit{individually rational\/}, $\phi_j^{COM-BR}(\succ)  \not= \emptyset$. Let branch $b\in B$ and cost $t\in\{t^0,t^+\}$ be such that $\phi_j^{COM-BR}(\succ)  = (b,t)$. 
Let $k$ be the final step of the cumulative offer process. 
Since  $\phi_j^{COM-BR}(\succ) \succ_i  \phi_i^{COM-BR}(\succ)$, cadet $i$ has proposed
the contract $(i,b,t)$ to branch $b$ at some step of the cumulative offer  process, which is rejected by branch $b$ (strictly speaking for the first time) 
either immediately or at a later step. 
Since the proposed contracts remain available until the termination  of the procedure under the cumulative offer process,\footnote{It is this feature
of the cumulative offer process that is emphasized in its name.} 
the contract  $(i,b,t)$ is also rejected by branch $b$ at the final Step $k$ of the cumulative offer process.  
In contrast, since $\phi_j^{COM-BR}(\succ)  = (b,t)$, contract $(j,b,t)$ is chosen by branch $b$ at the final step $k$ of the cumulative offer process.  
If the contract $(j,b,t)$ is accepted as one of the first $q^0_b$ positions under the choice rule $\calc^{BR}_b$, then $(j,b,t) \; \omega^0_b \; (i,b,t)$. 
Otherwise, if the contract $(j,b,t)$ is accepted as one of the last $q^+_b$ positions under the choice rule $\calc^{BR}_b$, then $(j,b,t) \; \omega^+_b \; (i,b,t)$.
In either case we have $j \; \pi_b \; i$,  proving that the mechanism $\phi^{COM-BR}$ \textit{has no priority reversals\/}. \smallskip

\textbf{\textit{Enforcement of the BRADSO policy\/}}: First suppose that cadets $i,j\in I$ are such that
$\phi^{COM-BR}_i(\succ) = (b,t^+)$ and  $(b,t^0) \succ_j \phi^{COM-BR}_j(\succ)$.  
The relation $(b,t^0) \succ_j \phi^{COM-BR}_j(\succ)$ implies that cadet $j$ has proposed
the contract $(j,b,t^0)$ to the branch $b$ at some step of the cumulative offer  process, which is rejected by branch $b$ either immediately or at a later step. 
Let $k$ be the final step of the cumulative offer process. 
Since the proposed contracts remain available until the termination  of the procedure under the cumulative offer process, the contract  $(j,b,t^0)$ is also
rejected by branch $b$ at the final Step $k$ of the cumulative offer process. 
More specifically, it is rejected by the choice rule $\calc^{BR}_b$ at the final Step $k$  both for the first $q^0_b$ positions using the 
native priority order $\omega^0_b$ and for the last $q^+_b$ positions using the BRADSO policy $w^+_b$. 
In contrast, contract $(i,b,t)$ is chosen by branch $b$ at the final Step $k$ of the cumulative offer process  using the BRADSO policy $w^+_b$.
Therefore, 
\begin{equation} \label{COSM-BRADSO1}
\left. \begin{array}{l}
  \phi^{COM-BR}_i(\succ) = (b,t^+), \; \mbox{ and}\\ 
    (b,t^0) \succ_j \phi^{COM-BR}_j(\succ)  \end{array}  \right\}     \quad  \implies \quad (i,t^+) \; \omega_b^+ \; (j,t^0). 
\end{equation}

Next suppose that cadets $i,j\in I$ are such that  $\phi^{COM-BR}_j(\succ) = (b,t^0)$, \; $(b,t^+) \succ_i \phi^{COM-BR}_i(\succ)$, \; 
$(i,t^+) \; \omega_b^+ \; (j,t^0)$, and moreover, let cadet $j$ be the lowest $\pi_b$-priority cadet 
with  an assignment of $\phi^{COM-BR}_j(\succ) = (b,t^0)$.   
The relation  $(b,t^+) \succ_i \phi^{COM-BR}_i(\succ)$ implies that cadet $i$ has proposed
the contract $(j,b,t^+)$ to the branch $b$ at some step of the cumulative offer  process, which is rejected by branch $b$ either immediately or at a later step. 
Let $k$ be the final step of the cumulative offer process. 
Since the proposed contracts remain available until the termination  of the procedure under the cumulative offer process, the contract  $(j,b,t^+)$ is also
rejected by branch $b$ at the final Step $k$ of the cumulative offer process. 
More specifically, it is rejected by the choice rule $\calc^{BR}_b$ at the final Step $k$  even for the last $q^+_b$ positions using the BRADSO policy $w^+_b$. 
Therefore, since by assumption we have $(i,t^+) \; \omega_b^+ \; (j,t^0)$, cadet $j$  must have received one of the first $q^0$ positions using the 
native priority order $\omega^0_b$. But since cadet $j$ is the lowest $\pi_b$-priority cadet  with an assignment of $\phi^{COM-BR}_j(\succ) = (b,t^0)$, 
that means no cadet has received any of the last $q^+_b$ positions at the base cost of $t^0$. Therefore, since $\phi^{COM-BR}$ satisfies
\textit{non-wastefulness\/}, 
\begin{equation} \label{COSM-BRADSO2}
\left. \begin{array}{l}
          \phi^{COM-BR}_j(\succ) = (b,t^0),\\ 
      (b,t^+) \succ_i   \phi^{COM-BR}_i(\succ),  \; \mbox{ and}\\
       (i,t^+) \; \omega_b^+ \; (j,t^0) \end{array}  \right\}     \quad  \implies \quad 
       \big|\big\{i'\in I :   \phi^{COM-BR}_{i'}(\succ)=(b,t^+)\big\}\big|  = q^+_b.  \\ 
\end{equation}
Relations \eqref{COSM-BRADSO1} and \eqref{COSM-BRADSO2} imply that mechanism $\phi^{COM-BR}$ satisfies \textit{enforcement of the BRADSO policy\/}. \smallskip

\indent \textbf{\textit{Strategy-proofness\/}}: Our model is a special case of \textit{matching problems with slot-specific priorities\/} 
by \cite{kominers/sonmez:16}. Hence \textit{strategy-poofness\/} of the mechanism $\phi^{COM-BR}$  is a direct corollary of
their Theorem 3, which proves \textit{strategy-proofness\/} of the cumulative offer mechanism more broadly for matching problems with slot-specific priorities. \medskip

\textbf{\textit{Uniqueness\/}}: We prove uniqueness via two lemmata.

\begin{lemma} \label{lemma3} 
Let  $X, Y \in \cala$ be two distinct allocations that satisfy  \textit{individual rationality, non-wastefulness, enforcement of BRADSO policy\/},
and \textit{have no priority reversals\/}. Then there exists a cadet $i \in I$ who receives non-empty and distinct assignments under $X$ and $Y$. 
\end{lemma}

\noindent \textit{Proof of Lemma \ref{lemma3}\/}: The proof is by contradiction.  Fix $\succ \; \in \calq^{|I|}$. 
Let  $X, Y\in \cala$ be two distinct allocations that 
satisfy  \textit{individual rationality, non-wastefulness, enforcement of BRADSO policy\/},
and \textit{have no priority reversals\/}.
 To derive the desired contradiction, suppose that,  for any cadet $i \in I$,
 \begin{equation} \label{lem3-eqn0}
 X_i \not= Y_i \quad  \implies \quad X_i = \emptyset \; \mbox{ or  } \; Y_i =\emptyset. 
\end{equation}

Pick any branch $b\in B$ such that $X_b \not= Y_b$. Let $j\in I$ be the highest $\pi_b$-priority cadet who
is assigned to branch $b$  either under  $X$ or under $Y$ but not both. 
W.l.o.g.,  let cadet  $j$ be assigned to branch $b$ under allocation $X$ but  not under allocation $Y$. By relation \eqref{lem3-eqn0}, 
\[ Y_j = \emptyset.
\]
Since allocation $Y$ satisfies \textit{non-wastefulness\/}, there exists a cadet $k\in I$ who is assigned to branch $b$ under allocation $Y$
but not under allocation $X$. By relation \eqref{lem3-eqn0},  
 \[
X_k = \emptyset,
\]
and therefore, by choice of cadet $j$, cadet $k$ has lower $\pi_b$-priority than cadet $j$.  
Moreover, since allocation $Y$ \textit{has no priority reversals\/} and $Y_j = \emptyset$, we have 
 \begin{equation} \label{lem3-eqn3}
Y_k = (b,t^+),
\end{equation}
and   since allocation $Y$ satisfies (condition 1 of) the axiom \textit{enforcement of BRADSO policy\/}, we have 
 \begin{equation} \label{lem3-eqn4}
(k, t^+) \; \omega^+_b \; (j,t^0). 
\end{equation}
Also relation \eqref{lem3-eqn3} and \textit{individual rationality\/} allocation $Y$ imply
\begin{equation} \label{lem3-eqn5}
 (b,t^+) \, \succ_k \; \emptyset. 
\end{equation}
Define
\[
I^* \equiv \{i\in I \, : \, X_i = (b,t^+)\}. 
\]
Since allocation $X$ satisfies (condition 2 of) the axiom \textit{enforcement of BRADSO policy\/}, the assumption
$X_j \in \{(b,t^0), (b,t^+)\}$ and relation \eqref{lem3-eqn4} imply
\begin{equation} \label{lem3-eqn6}
|I^*| = q^+_b,
\end{equation}
and since allocation $X$ \textit{has no priority reversals\/} and $X_k=\emptyset$, for any $i\in I^*$, 
\begin{equation} \label{lem3-eqn7}
i \; \pi_b \; k.
\end{equation}
But since $Y_k = (b,t^+)$ by relation \eqref{lem3-eqn3} and $|I^*| = q^+_b$ by relation \eqref{lem3-eqn6}, 
there exists a cadet $\ell\in I^*$ with $Y_{\ell} \not= (b,t^+) = X_{\ell}$, and therefore by relation  \eqref{lem3-eqn0} we have,
\begin{equation} \label{lem3-eqn8}
Y_{\ell} = \emptyset. 
\end{equation}
Since $X$ satisfies \textit{individual rationality\/} and $\ell \in I^*$, we have
\[ (b,t^+) \; \succ_{\ell} \; \emptyset,
\]
and therefore relations \eqref{lem3-eqn3}, \eqref{lem3-eqn7}, and \eqref{lem3-eqn8} imply
allocation $Y$ \textit{has  priority reversals\/}, giving us the desired contradiction and completing the proof of Lemma \eqref{lemma3}. 
\mbox{}\hfill$\diamondsuit$ \medskip

\begin{lemma} \label{lemma4} There can be at most one direct mechanism that satisfies 
 \textit{individual rationality, non-wastefulness, enforcement of BRADSO policy, strategy-proofness\/}, 
 and \textit{has no priority reversals\/}. 
\end{lemma}

\noindent \textit{Proof of Lemma \ref{lemma4}\/}: The proof of the lemma is inspired by a technique introduced by \cite{hirata/kasuya:17}.  
Towards a contradiction, suppose there exists two distinct direct mechanisms $\varphi$ and $\psi$ that satisfy  
\textit{individual rationality, non-wastefulness, enforcement of BRADSO policy, strategy-proofness\/}, and \textit{have no priority reversals\/}. 
Let the preference profile $\succ^* \in \calq^{|I|}$ be such that,
\begin{enumerate}
\item $\varphi(\succ^*) \not= \psi(\succ^*)$, and
\item the aggregate number of acceptable contracts between all cadets is minimized 
among all preference profiles    $\widetilde{\succ} \in \calq^{|I|}$ such that  $\varphi(\widetilde{\succ}) \not= \psi(\widetilde{\succ})$. 
\end{enumerate}
Let  $X= \varphi(\succ^*)$ and $Y= \psi(\succ^*)$. By Lemma \ref{lemma3}, there exists a cadet $i\in I$ such that
\begin{enumerate}
\item $X_i \not= \emptyset$, 
\item $Y_i \not= \emptyset$, and
\item $X_i \not= Y_i$. 
\end{enumerate}
Since both allocations $X$ and $Y$ satisfy \textit{individual rationality\/}, 
\[ X_i \; \succ^*_i \; \emptyset \quad \mbox{ and } \;  Y_i \; \succ^*_i \; \emptyset. 
\]
W.l.o.g., assume
\[ X_i \; \succ^*_i \; Y_i \; \succ^*_i \; \emptyset.
\]
Construct the preference relation $\succ'_i \in \calq$ as follows:\smallskip

If $X_i = (b,t^0)$ for some $b\in B$, then 
\[ (b,t^0) \; \succ'_i \; \emptyset \; \succ'_i \; (b',t') \qquad \mbox{ for any } (b',t')\in B\times T \setminus \{(b,t^0)\}.
\]
Otherwise, if 
$X_i = (b,t^+)$ for some $b\in B$, then 
\[ (b,t^0) \; \succ'_i \; (b,t^+) \; \succ'_i \;  \emptyset \; \succ'_i \; (b',t') \qquad \mbox{ for any } (b',t')\in B\times T \setminus \{(b,t^0),(b,t^+)\}.
\]
Since $X_i \succ^*_i  Y_i \,\succ^*_i  \emptyset$ and $(b,t^0)  \succ^*_i (b,t^+)$, the preference relation 
$\succ'_i$ has strictly fewer acceptable contracts for cadet $i$ than the preference relation $\succ^*_i$.

By \textit{strategy-proofness\/} of the mechanism $\psi$, we have
\[ \underbrace{\psi_i(\succ^*_i,\succ^*_{-i})}_{=Y_i} \; \succeq^*_i \;  \psi_i(\succ'_i,\succ^*_{-i}), 
\]
and since no branch-cost pair $(b',t') \in B\times T$ with $Y_i \succ'_i (b',t')$ is acceptable under $\succ'_i$, 
by \textit{individual rationality\/}  of the mechanism $\psi$ we have
\begin{equation} \label{lemma4-eqn1}
 \psi_i(\succ'_i,\succ^*_{-i}) = \emptyset.
\end{equation}
Similarly, by \textit{strategy-proofness\/} of the mechanism $\varphi$, we have
\[ \varphi_i(\succ'_i,\succ^*_{-i}) \; \succeq'_i  \; \underbrace{\varphi_i(\succ^*_i,\succ^*_{-i})}_{=X_i}, 
\] 
which in turn implies 
\begin{equation} \label{lemma4-eqn2}
 \varphi_i(\succ'_i,\succ^*_{-i}) \not= \emptyset.
\end{equation}
But then, by relations  \eqref{lemma4-eqn1} and \eqref{lemma4-eqn2} we have 
\[  \varphi(\succ'_i,\succ^*_{-i}) \not= \psi(\succ'_i,\succ^*_{-i}),
\]
giving us the desired contradiction, 
since between all cadets the preference profile $(\succ'_i,\succ^*_{-i})$ has strictly fewer acceptable contracts than 
the preference profile $\succ^*$. This completes the proof of Lemma \ref{lemma4}. 
\mbox{}\hfill$\diamondsuit$ \medskip

Since we have already shown that $\phi^{COM-BR}$ satisfies all five axioms, Lemma \ref{lemma4} establishes the
uniqueness, concluding the proof of Theorem  \ref{cosm}. 
\qed \medskip

\noindent \textbf{Proof of Corollary \ref{corollary:phi=cosm}}: Since \textit{BRADSO-IC\/} is implied by \textit{strategy-proofness\/}, 
Corollary  \ref{corollary:phi=cosm} is a direct implication of Theorems \ref{thm:singlebranchcharacterization} and \ref{cosm}. \qed 


\newpage


\begin{table}[hp]
	\begin{center}
    \caption{\textbf{Branches and Applications for Classes of 2020 and 2021}}
    \label{fig:tableCapacity}
    \includegraphics[scale = 0.6]{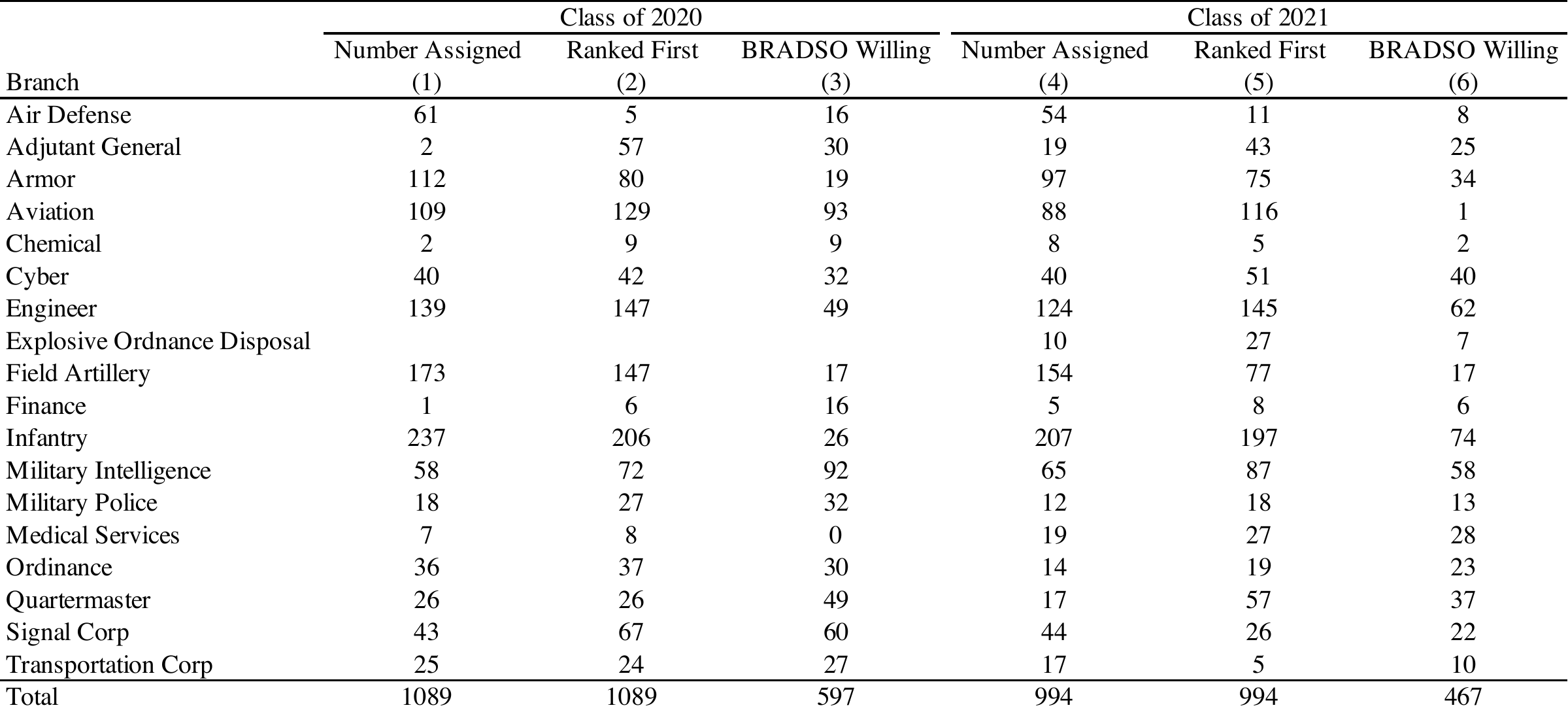}
	\end{center}
    \textbf{Notes.} This table reports information on branches for the Class of 2020 and 2021.  Number Assigned equals the capacity of the branch.  Ranked First is the number of cadets ranking the branch as their highest rank choice.  BRADSO Willing is the number of cadets who rank a BRADSO contract at the branch anywhere on their rank order list. Explosive Ordnance Disposal was not a branch option for the Class of 2020.  
\end{table}

\newpage

\newpage

\newpage

\newpage

\newpage

\section{Online Appendix: Supplementary Material}
\label{subsec:dataappendix}

\subsection{Individual-Proposing Deferred Acceptance}\label{sec:da}

The USMA-2020 mechanism was based on the individual-proposing deferred acceptance
algorithm \citep{gale/shapley:62}.  Given a ranking over branches, 
the individual-proposing deferred acceptance algorithm  (DA)  produces a matching as follows.

\medskip
	\begin{quote}
        \noindent{}{\bf Individual-Proposing Deferred Acceptance Algorithm ($\mathbf{DA}$)}

		\noindent{}{\bf Step 1:}
            Each cadet applies to her most preferred branch.
            Each branch $b$ tentatively assigns
            applicants with the highest priority until
            all cadets are chosen or all $q_{b}$ slots
            as assigned and permanently rejects the rest. If there are no rejections, then stop.
	
		\noindent{}{\bf Step k:}
			Each cadet who was rejected in Step k-1 applies to her next
			preferred branch, if such a branch exists. Branch $b$ tentatively assigns cadets
            with the highest priority until all all cadets are chosen or all $q_b$ slots
            are assigned and permanently rejects the rest. If there are no rejections,
            then stop.\smallskip

            The algorithm terminates when there are no rejections, at which point all tentative
            assignments are finalized.
	\end{quote}

\subsection{Cadet Survey Questions and Answers}
\label{survey}

\renewcommand\thefigure{\thesection.\arabic{figure}}
\renewcommand\thetable{\thesection.\arabic{table}}
\setcounter{figure}{0}
\setcounter{table}{0}

In fall 2020, the Army administered
a survey of cadets.  This survey asked two questions related to assignment mechanisms, one on cadet understanding of USMA-2020 and the other on cadet preferences over assignment mechanisms. This section reports the questions and the distribution of survey responses.

\bigskip

\textbf{Question 1.} \textit{What response below best describes your understanding of the impact of volunteering to BRADSO for a branch in this year's branching process?}
\begin{itemize}
     \item[A.] I am more likely to receive the branch, but I am only charged a BRADSO if I would have failed to receive the branch had I not volunteered to BRADSO. (43.3\% of respondents)
     \item[B.] I am charged a BRADSO if I receive the branch, regardless of whether volunteering to BRADSO helped me receive the branch or not. (9.5\% of respondents)
     \item[C.] I am more likely to receive the branch, but I may not be charged a BRADSO if many cadets who receive the same branch not only rank below me but also volunteer to BRADSO. (38.8\% of respondents)
     \item[D.] I am more likely to receive the branch, but I do not know how the Army determines who is charged a BRADSO. (6.7\% of respondents)
     \item[E.] I am NOT more likely to receive the branch even though I volunteered to BRADSO. (1.8 percent of respondents)
\end{itemize}

38.8\% of cadets answered the correct answer (answer C). 43.3\% of cadets believed that the 2020 mechanism would only charge a BRADSO if required to receive the branch (answer A)

\bigskip

\textbf{Question 2}. \textit{A cadet who is charged a BRADSO is required to serve an additional 3 years on Active Duty. Under the current
mechanism, cadets must rank order all 17 branches and indicate if they are willing to BRADSO for each branch choice. For
example:}
\begin{itemize}
     \item \textbf{Current Mechanism Example}:
   \begin{itemize}
       \item  \textbf{1: AV/BRADSO, 2: EN, 3: CY}
   \end{itemize}
\end{itemize}
\bigskip
Under an alternative mechanism, cadets could indicate if they prefer to receive their second branch choice without a
BRADSO charge more than they prefer to receive their first branch choice with a BRADSO charge. For example:
\begin{itemize}
     \item \textbf{Alternative Mechanism Example}:
   \begin{itemize}
       \item  \textbf{1: AV, 2: EN, 3: AV/BRADSO, 4: CY} 
   \end{itemize}
\end{itemize}
\bigskip
When submitting branch preferences, which mechanism would you prefer?

\begin{itemize}
     \item A. Current Mechanism (21.4\% of respondents)
     \item B. Alternative Mechanism (49.7\% of respondents)
     \item C. Indifferent (24.2\% of respondents)
     \item D. Do Not Understand (4.8\% of respondents)
\end{itemize}

\newpage

\begin{table}[htp]
    \begin{center}
    \caption{\textbf{Mechanism Replication Rate}}
    \includegraphics[scale = 0.9]{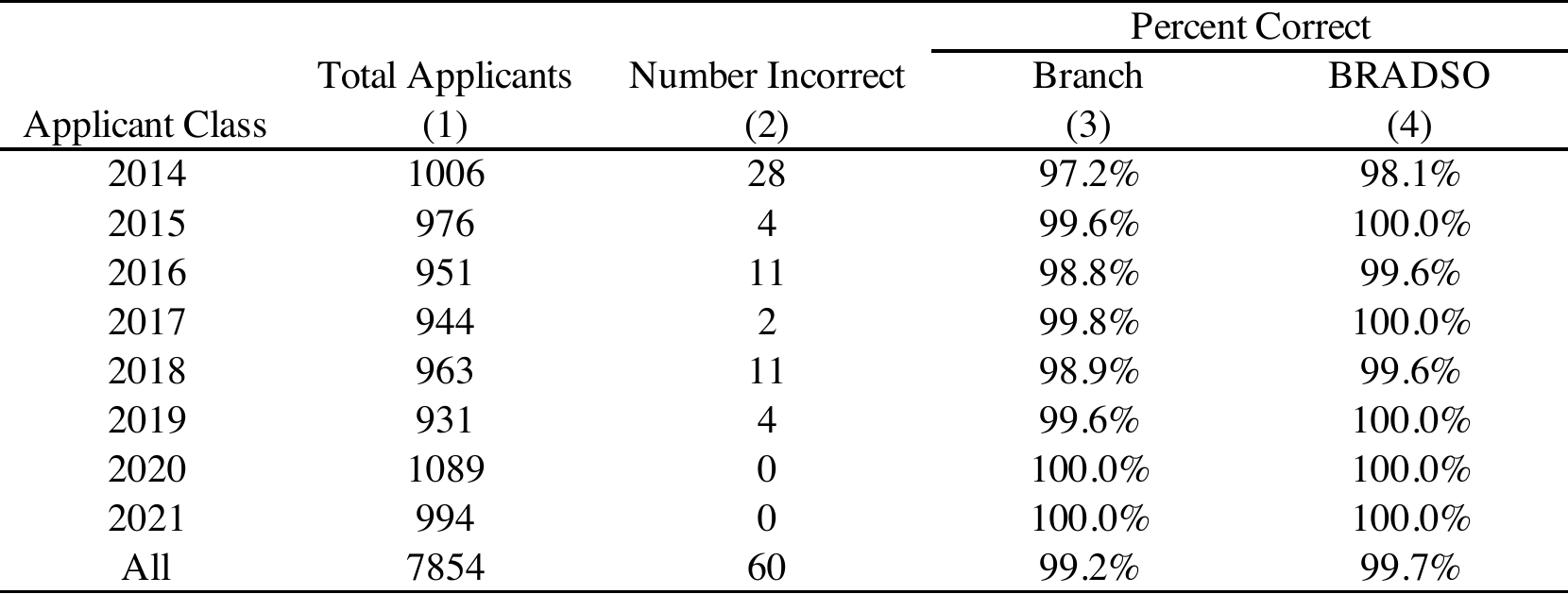}
    \end{center}
    \textbf{Notes.}  This table reports the replication rate of the USMA assignment mechanism across years. The
    USMA-2006 mechanism is used for the Classes of 2014-2019, USMA-2020 mechanism is used for the Class of 2020, and
    the COM-BRADSO mechanism is used for the Class of of 2021.  Number incorrect are the number of cadets who obtain a different
    assignment under our replication.   Branch percent correct is the number of branch assignments that we replicate.
    BRADSO percent correct is the number of BRADSO assignments we replicate.
    \label{fig:tableReplicationRate}
\end{table}

\end{document}